\documentclass{ptephy}

\preprintnumber{XXXX-XXXX} 

\usepackage{cite} 
\usepackage{graphics} 
\usepackage{subfig} 
\usepackage{url} 
\usepackage{multirow}
\usepackage{color}

\newcommand{\numu}{\ensuremath{\nu_{\mu}}}                   
\newcommand{\numubar}{\ensuremath{\overline{\nu}_{\mu}}}                   
\newcommand{\nue}{\ensuremath{\nu_{e}}}                   
\newcommand{\nuebar}{\ensuremath{\overline{\nu}_{e}}}                   
     
\newcommand{\deltacp}{\ensuremath{\delta_{CP}}}

\begin{document}

\title{Physics Potential of a Long Baseline Neutrino Oscillation Experiment
Using J-PARC Neutrino Beam 
and Hyper-Kamiokande}

\newcommand{\noaffiliation}{}
\newcommand{\BERN}{1}
\newcommand{\BOSTON}{2}
\newcommand{\UBC}{3}
\newcommand{\UCDAVIS}{4}
\newcommand{\UCI}{5}
\newcommand{\CSU}{6}
\newcommand{\SACLAY}{7}
\newcommand{\CHONNAM}{8}
\newcommand{\DONGSHIN}{9}
\newcommand{\DUKE}{10}
\newcommand{\DURHAM}{11}
\newcommand{\LLR}{12}
\newcommand{\EDINBURGH}{13}
\newcommand{\ETHZ}{14}
\newcommand{\GENEVA}{15}
\newcommand{\HAWAII}{16}
\newcommand{\IMPERIAL}{17}
\newcommand{\BARI}{18}
\newcommand{\NAPOLI}{19}
\newcommand{\PADOVA}{20}
\newcommand{\ROME}{21}
\newcommand{\INR}{22}
\newcommand{\ISU}{23}
\newcommand{\KEK}{24}
\newcommand{\KOBE}{25}
\newcommand{\KYOTO}{26}
\newcommand{\LNF}{27}
\newcommand{\LANCASTER}{28}
\newcommand{\LIVERPOOL}{29}
\newcommand{\LANL}{30}
\newcommand{\LSU}{31}
\newcommand{\MADRID}{32}
\newcommand{\MSU}{33}
\newcommand{\MIYAGI}{34}
\newcommand{\NAGOYA}{35}
\newcommand{\KMI}{36}
\newcommand{\STELAB}{37}
\newcommand{\NCNR}{38}
\newcommand{\OKAYAMA}{39}
\newcommand{\OCU}{40}
\newcommand{\OXFORD}{41}
\newcommand{\PITTSBURGH}{42}
\newcommand{\REGINA}{43}
\newcommand{\RIO}{44}
\newcommand{\ROCHESTER}{45}
\newcommand{\QMUL}{46}
\newcommand{\RHUL}{47}
\newcommand{\SAOPAULO}{48}
\newcommand{\SHEFFIELD}{49}
\newcommand{\SNU}{50}
\newcommand{\SEOYEONG}{51}
\newcommand{\STONYBROOK}{52}
\newcommand{\RAL}{53}
\newcommand{\SKKU}{54}
\newcommand{\TOHOKU}{55}
\newcommand{\ERI}{56}
\newcommand{\KAMIOKA}{57}
\newcommand{\RCCN}{58}
\newcommand{\IPMU}{59}
\newcommand{\TOKYO}{60}
\newcommand{\TITECH}{61}
\newcommand{\TRIUMF }{62}
\newcommand{\TORONTO}{63}
\newcommand{\WARSAW}{64}
\newcommand{\WARWICK}{65}
\newcommand{\WASHINGTON}{66}
\newcommand{\WINNIPEG}{67}
\newcommand{\VT}{68}
\newcommand{\WROCLAW}{69}
\newcommand{\YORK}{70}

\author{%
\name{K.~Abe}{\KAMIOKA,\IPMU},
\name{H.~Aihara}{\TOKYO,\IPMU},
\name{C.~Andreopoulos}{\LIVERPOOL},
\name{I.~Anghel}{\ISU},
\name{A.~Ariga}{\BERN},
\name{T.~Ariga}{\BERN},
\name{R.~Asfandiyarov}{\GENEVA},
\name{M.~Askins}{\UCDAVIS},
\name{J.J.~Back}{\WARWICK},
\name{P.~Ballett}{\DURHAM}
% ...
\\
\name{(The Hyper-Kamiokande Proto-Collaboration)}{}
}

\address{
\affil{\BERN}{{University of Bern, Albert Einstein Center for Fundamental Physics, Laboratory for High Energy Physics (LHEP), Bern, Switzerland}}
\affil{\BOSTON}{{Boston University, Department of Physics, Boston, Massachusetts, U.S.A.}}
\affil{\UBC}{{University of British Columbia, Department of Physics and Astronomy, Vancouver, British Columbia, Canada }}
\affil{\UCDAVIS}{{University of California, Davis, Department of Physics, Davis, California, U.S.A.}}
\affil{\UCI}{{University of California, Irvine, Department of Physics and Astronomy, Irvine, California, U.S.A.}}
\affil{\CSU}{{California State University, Department of Physics, Carson, California, U.S.A.}}
\affil{\SACLAY}{{IRFU, CEA Saclay, Gif-sur-Yvette, France}}
\affil{\CHONNAM}{{Chonnam National University, Department of Physics, Gwangju, Korea}}
\affil{\DONGSHIN}{{Dongshin University, Department of Physics, Naju, Korea}}
\affil{\DUKE}{{Duke University, Department of Physics, Durham, North Carolina, U.S.A.}}
\affil{\DURHAM}{{University of Durham, Science Laboratories, Durham, United Kingdom}}
\affil{\LLR}{{Ecole Polytechnique, IN2P3-CNRS, Laboratoire Leprince-Ringuet, Palaiseau, France}}
\affil{\EDINBURGH}{{University of Edinburgh, School of Physics and Astronomy, Edinburgh, United Kingdom}}
\affil{\ETHZ}{{ETH Zurich, Institute for Particle Physics, Zurich, Switzerland}}
\affil{\GENEVA}{{University of Geneva, Section de Physique, DPNC, Geneva, Switzerland}}
\affil{\HAWAII}{{University of Hawaii, Department of Physics and Astronomy, Honolulu, Hawaii, U.S.A.}}
\affil{\IMPERIAL}{{Imperial College London, Department of Physics, London, United Kingdom}}
\affil{\BARI}{{INFN Sezione di Bari and Universit\`a e Politecnico di Bari, Dipartimento Interuniversitario di Fisica, Bari, Italy}}
\affil{\NAPOLI}{{INFN Sezione di Napoli and Universit\`a di Napoli, Dipartimento di Fisica, Napoli, Italy}}
\affil{\PADOVA}{{INFN Sezione di Padova and Universit\`a di Padova, Dipartimento di Fisica, Padova, Italy}}
\affil{\ROME}{{INFN Sezione di Roma, Roma, Italy}}
\affil{\INR}{{Institute for Nuclear Research of the Russian Academy of Sciences, Moscow, Russia}}
\affil{\ISU}{{Iowa State University, Department of Physics and Astronomy, Ames, Iowa, U.S.A.}}
\affil{\KEK}{{High Energy Accelerator Research Organization (KEK), Tsukuba, Ibaraki, Japan}}
\affil{\KOBE}{{Kobe University, Department of Physics, Kobe, Japan}}
\affil{\KYOTO}{{Kyoto University, Department of Physics, Kyoto, Japan}}
\affil{\LNF}{{Laboratori Nazionali di Frascati, Frascati, Italy}}
\affil{\LANCASTER}{{Lancaster University, Physics Department, Lancaster, United Kingdom}}
\affil{\LIVERPOOL}{{University of Liverpool, Department of Physics, Liverpool, United Kingdom}}
\affil{\LANL}{{Los Alamos National Laboratory, New Mexico, U.S.A.}}
\affil{\LSU}{{Louisiana State University, Department of Physics and Astronomy, Baton Rouge, Louisiana, U.S.A. }}
\affil{\MADRID}{{University Autonoma Madrid, Department of Theoretical Physics, Madrid, Spain}}
\affil{\MSU}{Michigan State University, Department of Physics and Astronomy,  East Lansing, Michigan, U.S.A.}
\affil{\MIYAGI}{{Miyagi University of Education, Department of Physics, Sendai, Japan}}
\affil{\NAGOYA}{{Nagoya University, Graduate School of Science, Nagoya, Japan}}
\affil{\KMI}{{Nagoya University, Kobayashi-Maskawa Institute for the Origin of Particles and the Universe, Nagoya, Japan}}
\affil{\STELAB}{{Nagoya University, Solar-Terrestrial Environment Laboratory, Nagoya, Japan}}
\affil{\NCNR}{{National Centre for Nuclear Research, Warsaw, Poland}}
\affil{\OKAYAMA}{{Okayama University, Department of Physics, Okayama, Japan}}
\affil{\OCU}{{Osaka City University, Department of Physics, Osaka, Japan}}
\affil{\OXFORD}{{Oxford University, Department of Physics, Oxford, United Kingdom}}
\affil{\PITTSBURGH}{{University of Pittsburgh, Department of Physics and Astronomy, Pittsburgh, Pennsylvania, U.S.A.}}
\affil{\REGINA}{{University of Regina, Department of Physics, Regina, Saskatchewan, Canada}}
\affil{\RIO}{{Pontif{\'\i}cia Universidade Cat{\'o}lica do Rio de Janeiro, Departamento de F\'{\i}sica, Rio de Janeiro, Brazil}}
\affil{\ROCHESTER}{{University of Rochester, Department of Physics and Astronomy, Rochester, New York, U.S.A.}}
\affil{\QMUL}{{Queen Mary University of London, School of Physics and Astronomy, London, United Kingdom}}
\affil{\RHUL}{{Royal Holloway University of London, Department of Physics, Egham, Surrey, United Kingdom}}
\affil{\SAOPAULO}{{Universidade de S\~ao Paulo, Instituto de F\'{\i}sica, S\~ao Paulo, Brazil}}
\affil{\SHEFFIELD}{{University of Sheffield, Department of Physics and Astronomy, Sheffield, United Kingdom}}
\affil{\SNU}{{Seoul National University, Department of Physics, Seoul, Korea}}
\affil{\SEOYEONG}{{Seoyeong University, Department of Fire Safety, Gwangju, Korea }}
\affil{\STONYBROOK}{{State University of New York at Stony Brook, Department of Physics and Astronomy, Stony Brook, New York, U.S.A.}}
\affil{\RAL}{{STFC, Rutherford Appleton Laboratory, Harwell Oxford, and Daresbury Laboratory, Warrington, United Kingdom}}
\affil{\SKKU}{{Sungkyunkwan University, Department of Physics, Suwon, Korea}}
\affil{\TOHOKU}{{Research Center for Neutrino Science, Tohoku University, Sendai, Japan}}
\affil{\ERI}{{University of Tokyo, Earthquake Research Institute, Tokyo, Japan}}
\affil{\KAMIOKA}{{University of Tokyo, Institute for Cosmic Ray Research, Kamioka Observatory, Kamioka, Japan}}
\affil{\RCCN}{{University of Tokyo, Institute for Cosmic Ray Research, Research Center for Cosmic Neutrinos, Kashiwa, Japan}}
\affil{\IPMU}{{University of Tokyo, Kavli Institute for the Physics and Mathematics of the Universe (WPI), Todai Institutes for Advanced Study, Kashiwa, Chiba, Japan}}
\affil{\TOKYO}{{University of Tokyo, Department of Physics, Tokyo, Japan}}
\affil{\TITECH}{{Tokyo Institute of Technology, Department of Physics, Tokyo, Japan}}
\affil{\TRIUMF }{{TRIUMF, Vancouver, British Columbia, Canada}}
\affil{\TORONTO}{{University of Toronto, Department of Physics, Toronto, Ontario, Canada}}
\affil{\WARSAW}{{University of Warsaw, Faculty of Physics, Warsaw, Poland}}
\affil{\WARWICK}{{University of Warwick, Department of Physics, Coventry, United Kingdom}}
\affil{\WASHINGTON}{{University of Washington, Department of Physics, Seattle, Washington, U.S.A.}}
\affil{\WINNIPEG}{{University of Winnipeg, Department of Physics, Winnipeg, Manitoba, Canada}}
\affil{\VT}{{Virginia Tech, Center for Neutrino Physics, Blacksburg, Virginia, U.S.A.}}
\affil{\WROCLAW}{{Wroclaw University, Faculty of Physics and Astronomy, Wroclaw, Poland}}
\affil{\YORK}{{York University, Department of Physics and Astronomy, Toronto, Ontario, Canada}}
\email{masashi@phys.s.u-tokyo.ac.jp}
}

\subjectindex{C03, C04, C32}

\begin{abstract}
Hyper-Kamiokande will be a next generation underground water Cherenkov detector with a total (fiducial) mass of 0.99 (0.56) million metric tons, approximately 20 (25) times larger than that of Super-Kamiokande.
One of the main goals of Hyper-Kamiokande is the study of $CP$ asymmetry in the lepton sector using accelerator neutrino and anti-neutrino beams.

In this paper, the physics potential of a long baseline neutrino experiment using the Hyper-Kamiokande detector and a neutrino beam from the J-PARC proton synchrotron is presented.
The analysis uses the framework and systematic uncertainties derived from the ongoing T2K experiment.
With a total exposure of 7.5~MW $\times$ 10$^7$ sec integrated proton beam power (corresponding to $1.56\times10^{22}$ protons on target with a 30~GeV proton beam) to a $2.5$-degree off-axis neutrino beam,
it is expected that the leptonic $CP$ phase $\deltacp$ can be determined to better than 19 degrees for all possible values of $\deltacp$, and $CP$ violation can be established with a statistical significance of more than $3\,\sigma$ ($5\,\sigma$) for $76\%$ ($58\%$) of the $\,\deltacp$ parameter space.
Using both $\nu_e$ appearance and $\nu_\mu$ disappearance data, the expected 1$\sigma$ uncertainty of $\sin^2\theta_{23}$ is 0.015(0.006) for $\sin^2\theta_{23}=0.5(0.45)$. 
\end{abstract}

\maketitle

\section{Introduction}
The discovery of neutrino oscillations by the Super-Kamiokande (Super-K) 
experiment in 1998~\cite{Fukuda:1998mi}  opened a new window to explore physics beyond the Standard Model (BSM).
Evidence of neutrino oscillations is one of the most convincing experimental proofs known today for the existence of BSM physics at work.
The mixing parameters of neutrinos were found
to be remarkably different from those of quarks, which suggests the presence of an unknown flavor symmetry waiting to be explored.
The extremely small masses of neutrinos compared with those of their charged partners
lead to the preferred scenario of a seesaw mechanism~\cite{Minkowski:1977sc, GellMann:1980vs, Yanagida:1979as, Mohapatra:1979ia}, in which small neutrino masses are 
a reflection of the ultra-high energy scale of BSM physics.

Furthermore, a theoretical framework called leptogenesis points to the intriguing possibility that 
$CP$ asymmetries related to flavor mixing among the three generations of neutrinos may have played an important role 
in creating the observed matter-antimatter asymmetry in the universe~\cite{Fukugita:1986hr}.
This makes a study of the full picture of neutrino masses and mixings and the
measurement of the $CP$ asymmetry
in the neutrino sector among the most important and urgent subjects in today's 
elementary particle physics world.

$CP$ asymmetry in the neutrino sector arising from the presence of the phase which corresponds to the Kobayashi-Maskawa phase~\cite{Kobayashi:1973fv} in the quark sector, can only be seen if all the three
mixing angles governing neutrino oscillations differ from zero.
The Super-K detector has successfully measured all three angles. The angle $\theta_{23}$ was first measured in atmospheric neutrino observations~\cite{Fukuda:1998mi}, $\theta_{12}$ was constrained in solar neutrino observations~\cite{Fukuda:2001nk} (together with another water Cherenkov detector SNO~\cite{Ahmad:2001an}), and the evidence of non-zero $\theta_{13}$ was found by T2K~\cite{Abe:2011sj} which used Super-K as the far detector.
In 2013, T2K established $\nu_\mu \to \nu_e$ oscillation with 7.3\,$\sigma$ 
significance, leading the way towards $CP$ violation measurements in neutrinos~\cite{Abe:2013hdq} in combination with precise measurements of $\theta_{13}$ by reactor neutrino experiments~\cite{An:2013zwz,Abe:2014lus,Ahn:2012nd}.
The highly successful Super-K program indicates that Hyper-Kamiokande (Hyper-K)
is well placed to discover $CP$ violation.

In this paper, the physics potential of a long baseline neutrino experiment using the Hyper-Kamiokande detector and a neutrino beam from the J-PARC proton synchrotron is presented.

\begin{figure}[tbp]
  \centering
    \includegraphics[width=0.8\textwidth]{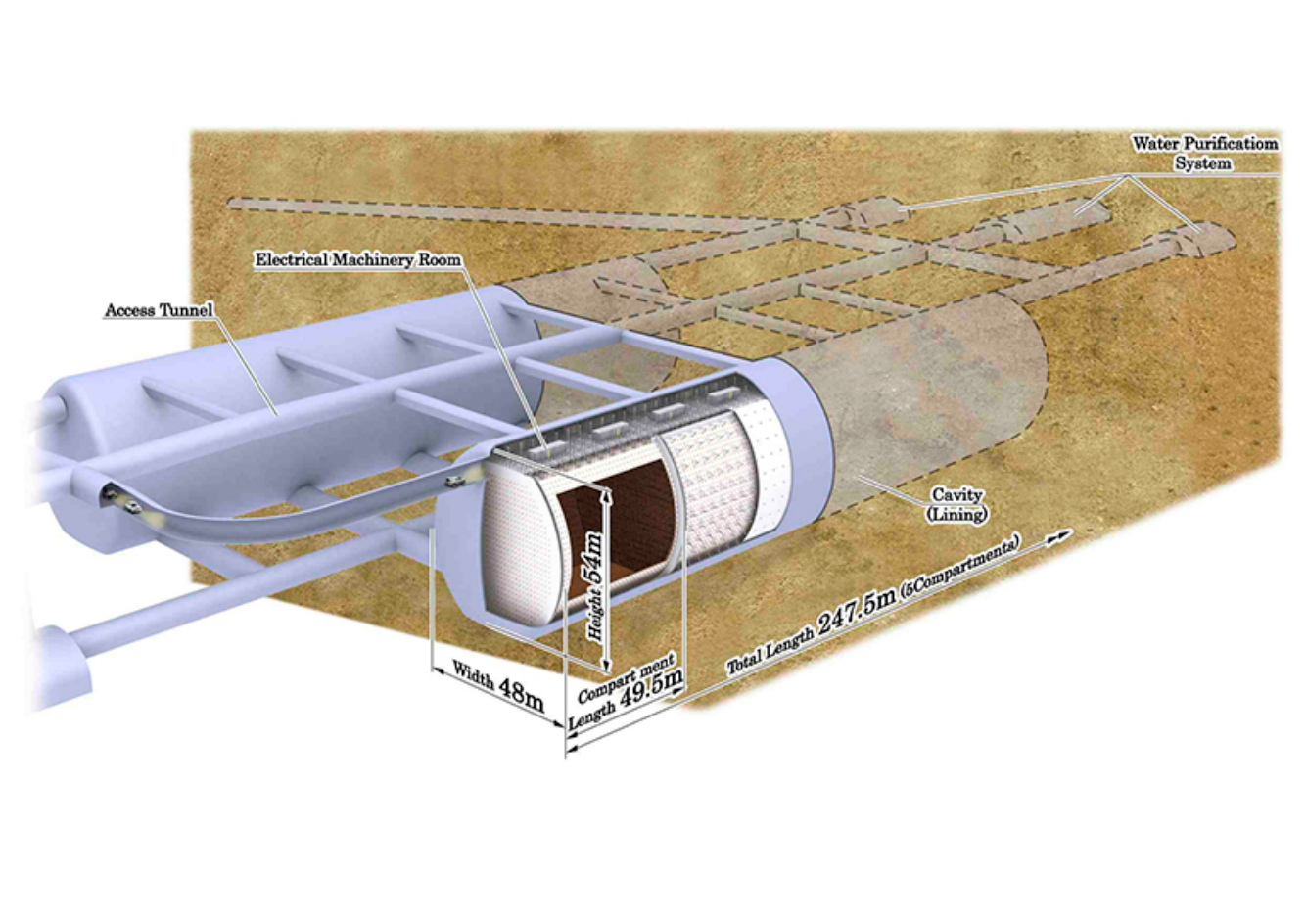}
  \caption{Schematic view of the Hyper-Kamiokande detector.}
  \label{fig:HK-schematic-view}
\end{figure}

The Hyper-K detector is designed as 
a next generation underground water Cherenkov detector that serves as a far
detector of a long baseline neutrino oscillation experiment for the J-PARC 
neutrino beam and as a detector capable of observing proton decays,
atmospheric and solar neutrinos, and neutrinos from other astrophysical
origins.
The baseline design of Hyper-K is based on the well-proven technologies 
employed and tested at Super-K.
Hyper-K consists of two cylindrical tanks lying side-by-side, 
the outer dimensions of each tank being 
$48 \ ({\rm W}) \times 54\ ({\rm H}) \times 250\ ({\rm L})\ {\rm m}^3.$
The total (fiducial) mass of the detector is 0.99 (0.56) million metric tons, 
which is about 20 (25) times larger than that of Super-K.
A proposed location for Hyper-K is about 8~km south of Super-K 
(and 295~km away from J-PARC) and 1,750 meters water equivalent 
(or 648~m of rock) deep.
The inner detector region is viewed by 99,000 20-inch PMTs, 
corresponding to the PMT density of 
$20\%$ photo-cathode coverage (the same as the second phase of Super-K).
The schematic view of the Hyper-K detector is illustrated in 
Fig.~\ref{fig:HK-schematic-view}. 

In addition to the long baseline neutrino oscillation experiment that is the main focus of this paper,
Hyper-K will provide a rich program in a wide range of science~\cite{Abe:2011ts}.
The scope of the project includes observation of atmospheric and solar neutrinos, proton decays, and neutrinos from other astrophysical origins.
The physics potential of Hyper-K is summarized in Table~\ref{tab:intro:phys}.

\begin{sidewaystable}[btp]
\caption{Physics targets and expected sensitivities of the Hyper-Kamiokande experiment,
based on the study shown in~\cite{Abe:2011ts} except for the long baseline experiment that is described in this paper.
Improvement is expected with further optimization of the detector design and development of reconstruction/analysis tools.
Also, only selected values are listed;
for example, other channels will be accessible for nucleon decays.
} 	
\label{tab:intro:phys}
\centering
\begin{tabular}{lll} \hline \hline
Physics Target & Sensitivity & Conditions \\
\hline \hline
Neutrino study w/ J-PARC $\nu$~~ && 7.5\,MW $\times$ $10^7$ sec\\
$-$ $CP$ phase precision & $<19^\circ$ & @ $\sin^22\theta_{13}=0.1$, mass hierarchy known \\
$-$ $CPV$ discovery coverage & 76\% (3\,$\sigma$), 58\% ($5\,\sigma$) & @ $\sin^22\theta_{13}=0.1$, mass hierarchy known \\
$-$ $\sin^2\theta_{23}$ & $\pm 0.015$ & 1$\sigma$ @ $\sin^2\theta_{23}=0.5$ \\
\hline
Atmospheric neutrino study && 10 years observation\\
$-$ MH determination & $> 3\,\sigma$ CL & @ $\sin^2\theta_{23}>0.4$ \\
$-$ $\theta_{23}$ octant determination & $> 3\,\sigma$ CL & @ $\sin^2\theta_{23}<0.46$ or $\sin^2\theta_{23}>0.56$ \\\hline
Nucleon Decay Searches && 10 years data \\
$-$ $p\rightarrow e^+ + \pi^0$ & $1.3 \times 10^{35}$ yrs (90\% CL UL) &\\
 & $5.7 \times 10^{34}$ yrs ($3\,\sigma$ discovery) &\\
$-$ $p\rightarrow \bar{\nu} + K^+$ & $3.2 \times 10^{34}$ yrs (90\% CL UL) &\\
 & $1.2 \times 10^{34}$ yrs ($3\,\sigma$ discovery) &\\ 
\hline
Astrophysical neutrino sources && \\
$-$ $^8$B $\nu$ from Sun & 200 $\nu$'s / day & 7.0\,MeV threshold (total
	 energy) w/ osc.\\
$-$ Supernova burst $\nu$ & 170,000$\sim$260,000 $\nu$'s & @ Galactic center (10 kpc)\\ 
 & 30$\sim$50 $\nu$'s & @ M31 (Andromeda galaxy) \\ 
$-$ Supernova relic $\nu$ & 830 $\nu$'s / 10 years & \\
$-$ WIMP annihilation at Sun & & 5 years observation\\
 ~~($\sigma_{SD}$: WIMP-proton spin & $\sigma_{SD}=10^{-39}$cm$^2$ & @ $M_{\rm WIMP}=10$\,GeV, $\chi\chi\rightarrow b\bar b$ dominant\\
 ~~~~dependent cross section)& $\sigma_{SD}=10^{-40}$cm$^2$ & @ $M_{\rm WIMP}=100$\,GeV, $\chi\chi\rightarrow W^+ W^-$ dominant\\
 \hline \hline
\end{tabular}
\end{sidewaystable}

\section{Neutrino Oscillations and CP Violation}
\subsection{Neutrino Oscillations in Three Flavor Framework}
Throughout this paper, unless stated otherwise, we consider 
the standard three flavor neutrino framework. 
The 3$\times3$ unitary matrix $U$ which describes the mixing of 
neutrinos~\cite{Maki:1962mu} (that is often referred to as the Maki-Nakagawa-Sakata-Pontecorvo (MNSP) or
Maki-Nakagawa-Sakata (MNS)~\cite{Pontecorvo:1967fh,Maki:1962mu} matrix) 
relates the flavor and mass eigenstates of neutrinos: 
\begin{eqnarray} 
\nu_\alpha = \sum_{i=1}^3 U_{\alpha i} \nu_i, 
\ \ (\alpha = e, \mu, \tau),
\end{eqnarray} 
where $\nu_\alpha  (\alpha = e, \mu, \tau)$ and 
$\nu_i  (i = 1,2,3)$ denote, respectively, flavor and 
mass eigenstates of neutrinos. 
Using the standard parametrization, which can be found, e.g. in 
Ref.~\cite{Agashe:2014kda}, $U$ can be expressed as,  
\begin{eqnarray}
U 
& = &
\left(
\begin{array}{ccc}
1 & 0   & 0 \\
0 & c_{23} & s_{23} \\
0 & -s_{23} & c_{23} \\
\end{array}
\right)
\left(
\begin{array}{ccc}
c_{13} & 0   & s_{13}e^{-i\deltacp} \\
0 & 1 & 0 \\
-s_{13}e^{i\deltacp} & 0 & c_{13} \\
\end{array}
\right)
\left(
\begin{array}{ccc}
c_{12} & s_{12}  & 0 \\
-s_{12} & c_{12} & 0 \\
0 & 0 & 1 \\
\end{array}
\right) \nonumber \\ 
& \times &
\left(
\begin{array}{ccc}
1 & 0 & 0 \\
0 & e^{i\frac{\alpha_{21}}{2}} & 0 \\
0 & 0 & e^{i\frac{\alpha_{31}}{2}} \\
\end{array}
\right)
\label{eq:mixing}
\end{eqnarray}
where $c_{ij} \equiv \cos\theta_{ij}$, $s_{ij} \equiv \sin\theta_{ij}$, 
and $\deltacp$ --- often called the Dirac $CP$ phase ---,  
is the Kobayashi-Maskawa type $CP$ phase~\cite{Kobayashi:1973fv} 
in the lepton sector. 
On the other hand, the two phases, $\alpha_{21}$ and $\alpha_{31}$, 
--- often called Majorana $CP$ phases --- 
exist only if neutrinos are of Majorana type~\cite{Schechter:1980gr, Bilenky:1980cx,Doi:1980yb}.
While the Majorana $CP$ phases can not be observed in neutrino 
oscillation,
they can be probed by lepton number violating processes such as 
neutrinoless double beta decay.

In vacuum,
the oscillation probability of $\nu_\alpha \to  \nu_\beta$ ($\alpha,\beta = e,\mu, \tau$)
for ultrarelativistic neutrinos is given by, 
\begin{eqnarray}
P(\nu_\alpha \to  \nu_\beta)  & = & 
\left|~\sum_{i=1}^3 U^*_{\alpha i} ~U_{\beta i} 
\text{e}^{-i\frac{m^2_i}{2E_\nu}L}~\right|^2 \nonumber \\
&= &
\delta_{\alpha \beta} 
-4 \sum_{i>j} {\Re}(U^*_{\alpha i} U_{\alpha j} U_{\beta i} U^*_{\beta j} )
\sin^2\left(  \frac{\Delta m^2_{ij}}{4E_\nu}L \right) \nonumber \\
&&
+2 \sum_{i>j} 
{\Im}(U^*_{\alpha i} U_{\alpha j} U_{\beta i} U^*_{\beta j} )
\sin \left( \frac{\Delta m^2_{ij}}{2E_\nu}L \right),
\label{eq:prob_vac}
\end{eqnarray}
where $E_\nu$ is the neutrino energy, $L$ is the baseline, 
$\Delta m^2_{ij} \equiv m_i^2-m_j^2$ $(i,j=1,2,3)$ are the mass squared
differences with $m_i$ and $m_j$ being the neutrino masses. 
For the $CP$ conjugate channel, $\bar{\nu}_\alpha \to  \bar{\nu}_\beta$, 
the same expression in Eq.~(\ref{eq:prob_vac}) holds, but the matrix $U$ 
is replaced by its complex conjugate (or equivalently $\deltacp \to -\deltacp$ in Eq.~(\ref{eq:mixing})), 
resulting in the third term in Equation~\ref{eq:prob_vac} switching sign.
For neutrinos traveling inside matter, coherent forward scattering induces
an asymmetry between the oscillation probabilities of neutrinos and antineutrinos supplementary to the intrinsic $CP$ violation.

The magnitude of the $CP$ violation in neutrino oscillation 
can be characterized by the difference of probabilities between neutrino and anti-neutrino channels, 
which, in vacuum, is given by~\cite{Barger:1980jm,Pakvasa:1980bz}, 
\begin{eqnarray}
\Delta P_{\alpha\beta} \equiv P(\nu_\alpha \to  \nu_\beta)
- P(\bar\nu_\alpha \to  \bar\nu_\beta)
= 16 J_{\alpha \beta} \sin \Delta_{21}
\sin \Delta_{32} \sin \Delta_{31},
\label{eq:DeltaP}
\end{eqnarray}
and 
\begin{eqnarray}
J_{\alpha \beta} \equiv 
{\Im}(U_{\alpha 1}U^*_{\alpha 2} U^*_{\beta 1} U_{\beta 2} )
= \pm J_{CP}, \ \ 
J_{CP} \equiv s_{12}c_{12}s_{23}c_{23}s_{13}c_{13}^2\sin\deltacp
\label{eq:Jarlskog} 
\end{eqnarray}
with positive (negative) sign for (anti-)cyclic permutation of the
flavor indices $e$, $\mu$ and $\tau$.  
The parameter $J_{CP}$ is the lepton analogue of 
the $CP$-invariant factor for quarks, 
the unique and phase-convention-independent measure 
for $CP$ violation~\cite{Jarlskog:1985ht}.
Using the current best fitted values of mixing parameters~\cite{Capozzi:2013csa}, we get 
$J_{CP} \simeq 0.034 \sin \deltacp$, or 
\begin{equation}
\Delta P_{\alpha\beta} 
\simeq  \pm 0.55  \sin \deltacp \sin \Delta_{21}
\sin \Delta_{32} \sin \Delta_{31}. 
\label{eq:DeltaP2}
\end{equation}
Thus, a large $CP$ violation effects are possible in the neutrino oscillation.

In general, it is considered that $CP$ violation in the neutrino sector which can be observed in the low energy regime, namely, in neutrino oscillation, does not directly imply the $CP$ violation required at high energy for the successful leptogenesis in the early universe. 
It has been discussed, however, that they could be related to each other and the $CP$ violating phase in the MNS matrix could be responsible also for the generation of the observed baryon asymmetry through leptogenesis in some scenarios. 
For example, in~\cite{Pascoli:2006ie,Pascoli:2006ci}, in the context of the seesaw mechanism, it has been pointed out that assuming the hierarchical mass spectrum for right handed Majorana neutrinos with the lightest mass to be $\lesssim 5\times 10^{12}$~GeV, observed baryon asymmetry could be generated through the leptogenesis if $|\sin\theta_{13}\sin\delta_{CP}|\gtrsim 0.1$, which is compatible with the current neutrino data. 
Hence, measurement of $CP$ asymmetry in neutrino oscillations may provide a clue for understanding the origin of matter-antimatter asymmetry of the Universe.

Since there are only three neutrinos, only two mass squared 
differences, $\Delta m^2_{21}$ and $\Delta m^2_{31}$, for example, 
are independent. 
Therefore, for a given energy and baseline, 
there are six independent parameters, namely, 
three mixing angles, one $CP$ phase, and two mass squared differences, 
in order to describe neutrino oscillations.
Among these six parameters, 
$\theta_{12}$ and $\Delta m^2_{21}$ have been measured
by solar~\cite{Ahmad:2002jz,Ahmad:2001an,Abe:2010hy} 
and reactor~\cite{Eguchi:2002dm,Araki:2004mb,Abe:2008aa}
neutrino experiments.
The parameters $\theta_{23}$ and $|\Delta m^2_{32}|$ 
(only its absolute value)
have been measured by 
atmospheric~\cite{Ashie:2005ik,Ashie:2004mr} and
accelerator~\cite{Ahn:2006zza,Adamson:2011ig,Abe:2012gx,Abe:2014ugx}
neutrino experiments.
Reactor experiment also starts to measure the atmospheric mass squared difference, 
$|\Delta m^2_{31}|$ though the uncertainty is still larger~\cite{An:2013zwz}.
Recently, \(\theta_{13}\) has also been measured by 
accelerator~\cite{Abe:2011sj,Adamson:2011qu,Abe:2013xua,Abe:2013hdq} and reactor 
experiments~\cite{Abe:2011fz,Ahn:2012nd,An:2012eh, An:2013zwz,Abe:2014lus}.
The relatively large value of $\theta_{13}$ opens the window to explore the $CP$ phase ($\deltacp$) and the mass hierarchy (the sign of $\Delta m^2_{31}$) using neutrino oscillation. 

\subsection{Physics Case with $\nu_\mu \to \nu_e$ Oscillation\label{sec:appearance}}
The oscillation probability from $\nu_\mu$ to $\nu_e$ in accelerator experiments
is expressed, to the first order of the matter effect, as follows~\cite{Arafune:1997hd}:
\begin{eqnarray}
P(\numu \to \nue) & = & 4 c_{13}^2s_{13}^2s_{23}^2 \cdot \sin^2\Delta_{31}  \nonumber \\
& & +8 c_{13}^2s_{12}s_{13}s_{23} (c_{12}c_{23}\cos\deltacp - s_{12}s_{13}s_{23})\cdot \cos\Delta_{32} \cdot \sin\Delta_{31}\cdot \sin\Delta_{21} \nonumber \\
& & -8 c_{13}^2c_{12}c_{23}s_{12}s_{13}s_{23}\sin\deltacp \cdot \sin\Delta_{32} \cdot \sin\Delta_{31}\cdot \sin\Delta_{21} \nonumber \\
& & +4s_{12}^2c_{13}^2(c_{12}^2c_{23}^2 + s_{12}^2s_{23}^2s_{13}^2-2c_{12}c_{23}s_{12}s_{23}s_{13}\cos\deltacp)\cdot \sin^2\Delta_{21} \nonumber \\
& & -8c_{13}^2s_{13}^2s_{23}^2\cdot \frac{aL}{4E_\nu} (1-2s_{13}^2)\cdot \cos\Delta_{32}\cdot \sin\Delta_{31} \nonumber \\
& & +8 c_{13}^2s_{13}^2s_{23}^2 \frac{a}{\Delta m^2_{31}}(1-2s_{13}^2)\cdot\sin^2\Delta_{31}, \label{Eq:cpv-oscprob}
\end{eqnarray}
\noindent where $\Delta_{ij}$ is $\Delta m^2_{ij}\, L/4E_\nu$, 
and $a =2\sqrt{2}G_Fn_eE_\nu= 7.56\times 10^{-5}\mathrm{[eV^2]} \times \rho \mathrm{[g/cm^3]} \times E_\nu[\mathrm{GeV}] $.
The corresponding probability for a $\numubar \to \nuebar$ transition is obtained by replacing $\deltacp \rightarrow -\deltacp$
and $a \rightarrow -a$.
The third term, containing $\sin\deltacp$, is the $CP$ violating term which flips sign between $\nu$ and $\bar{\nu}$ and thus introduces $CP$ asymmetry if $\sin\deltacp$ is non-zero.
The last two terms are due to the matter effect. 
Those terms which contain $a$ change their sign depending on the mass hierarchy.
As seen from the definition of $a$, the amount of asymmetry due to the matter effect is proportional to the neutrino energy at a fixed value of $L/E_\nu$.
A direct test of $CP$ violation, in a model independent way, is possible by measuring both neutrino and antineutrino appearance probabilities.
If the mass hierarchy is not known, the sensitivity of $CP$ violation is affected by the presence of the matter effect.
However, the mass hierarchy could be determined by the atmospheric neutrino measurement in Hyper-K and several measurements by other experiments.

Currently measured value of $\theta_{23}$ is consistent with maximal mixing, $\theta_{23} \approx \pi/4$~\cite{Abe:2014ugx, Adamson:2014vgd, Himmel:2013jva}.
It is of great interest to determine if $\sin^22\theta_{23}$ is maximal or not, and if not $\theta_{23}$ is less or greater than $\pi/4$, as 
it could constrain models of neutrino mass generation~\cite{King:2013eh,Albright:2010ap,Altarelli:2010gt,Ishimori:2010au,Albright:2006cw,Mohapatra:2006gs}.
When we measure $\theta_{23}$ with the survival probability $P(\numu \to \numu)$ which is proportional to $\sin^22\theta_{23}$ to first order, 
\begin{eqnarray}
P(\nu_\mu \rightarrow \nu_\mu) &\simeq& 1-4c^2_{13}s^2_{23} [1-c^2_{13}s^2_{23}]\sin^2(\Delta m^2_{32}\, L/4E_\nu) \\
&\simeq & 1-\sin^22\theta_{23}\sin^2(\Delta m^2_{32}\, L/4E_\nu), \hspace{2cm} \textrm{(for $c_{13}\simeq1$)}
\end{eqnarray}
there is an octant ambiguity:  either $\theta_{23} \le 45^\circ $  (in the first octant) or $\theta_{23} > 45^\circ $  (in the second octant).
By combining the measurement of $P(\numu \to \nue)$, the $\theta_{23}$ octant can be determined.

\subsection{Anticipated Neutrino Physics Landscape in the 2020s and Uniqueness of This Experiment}
Before Hyper-K commences data taking in $\sim 2025$, we expect a number of ongoing and planned neutrino experiments
as well as cosmological observations will advance our understanding of neutrino physics. 
In addition to accelerator and reactor experiments, Super-K will provide
precise measurements of neutrino oscillation parameters from atmospheric
neutrino observations, and will look for the mass hierarchy and the octant of $\theta_{23}$.
Cosmological observations will provide the information on neutrino masses. 
An observation of neutrino-less double $\beta$ decay in the next 10 years would be evidence that the neutrino is a Majorana particle with the inverted mass hierarchy.
Following this progress, we definitely need a new experiment to discover $CP$ violation in neutrinos, and to unambiguously establish the mass hierarchy and $\theta_{23}$ octant. 
For these purposes, we propose the Hyper-K experiment with the J-PARC neutrino beam.

The Hyper-K experiment will have several unique advantages.
\begin{itemize}
\item The experiment will have high statistics of neutrino events thanks to the large fiducial mass and the high power J-PARC neutrino beam. 
\item The relatively short baseline among the proposed long baseline experiments results in a small ambiguity from the matter effect.
\item The experiment will operate in the same beam line as T2K with the same off-axis configuration. The features of the neutrino beam and the operation of the high power beam are well understood.
\item The systematc errors are already well understood based on Super-K and T2K, allowing reliable extrapolations.
\end{itemize}
With these features, Hyper-K will be one of the most sensitive experiments to probe neutrino $CP$ violation, as we present in this paper.

\section{Experimental setup}

\subsection{J-PARC accelerator and neutrino beamline}
An intense and high quality neutrino beam is a key for the success of a long baseline neutrino oscillation experiment.
J-PARC (Japan Proton Accelerator Research Complex) is one of world leading facilities in neutrino physics, currently providing a beam for the T2K experiment.
We will utilize the full potential of this existing facility with future increase of the beam power to the design value of 750~kW and beyond.

The J-PARC accelerator cascade~\cite{JPARCTDR} 
consists of a normal-conducting LINAC as an injection 
system, a Rapid Cycling Synchrotron (RCS), and a Main Ring synchrotron (MR). 
In the fast extraction mode operation, MR has achieved 
1.24$\times$10$^{14}$ protons per pulse (ppp) beam intensity, 
which is a world record for extracted ppp for any 
synchrotron.
The corresponding beam power is 240\,kW. 
The upgrade scenario of J-PARC accelerator~\cite{jparc-midterm1}
is being implemented to reach the design power of 750\,kW in forthcoming years, 
with a typical planned parameter set as listed in Table~\ref{jparc:MRFXpara}.
This will double the current repetition rate by 
(i) replacing the magnet power supplies, 
(ii) replacing the RF system, and 
(iii) upgrading injection/extraction devices.
The design power of 750\,kW will be achieved well before Hyper-K will start data taking.
Furthermore, conceptual studies on how to realize 1$\sim$2\,MW 
beam powers and even beyond are now underway~\cite{jparc-longterm}, 
such as by raising the RCS top energy, enlarging the MR aperture, or 
inserting an ``emittance-damping'' ring between the RCS and MR. 

\begin{table}[tbp]
  \begin{center}
    \caption{Planned parameters of the J-PARC Main Ring for fast extraction. 
             Numbers in parentheses are those achieved up until May 2013.}
    \begin{tabular}{lcc}
      \hline \hline
      Parameter &  \multicolumn{2}{c}{Value} \\
      \hline
      Circumference (m)         & \multicolumn{2}{c}{1567.5} \\
      Kinetic energy (GeV)   & \multicolumn{2}{c}{30   } \\
      Beam intensity   (ppp)     & $2.0\times 10^{14}$
       & ($1.24\times 10^{14}$) \\
      ~~~~~~~~~~~~~~~~~~~~~(ppb)  & $2.5\times 10^{13}$
      & ($1.57\times 10^{13}$) \\
      Harmonic number        & \multicolumn{2}{c}{9} \\
      Number of bunches per spill     & \multicolumn{2}{c}{8} \\
      Spill width  ($\mu$s)          & \multicolumn{2}{c}{$\sim$~5} \\
      Bunch full width at extraction (ns)     & \multicolumn{2}{c}{$\sim$~50}  \\
      Maximum RF voltage  (kV)           & 560   & (280)\   \\
      Repetition period  (sec)    & 1.28 & (2.48) \\
      Beam power  (kW)           & 750 &  (240)   \\
      \hline \hline
    \end{tabular}
    \label{jparc:MRFXpara}
  \end{center}
\end{table}

\begin {figure}[tbp]
  \begin{center}
    \includegraphics[width=0.7\textwidth]{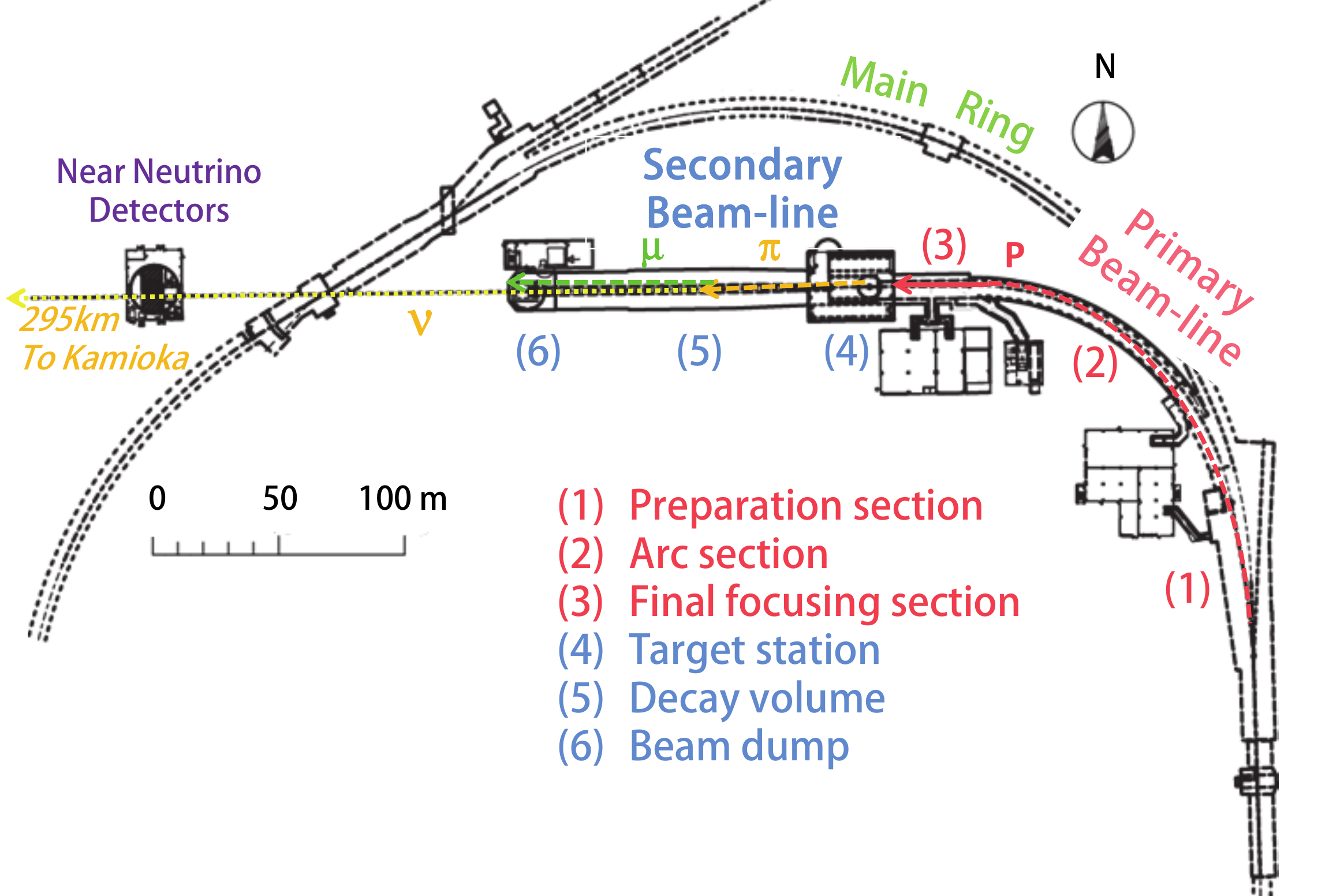}
    \caption{The neutrino experimental facility 
             (neutrino beamline) at J-PARC.}
    \label{fig:beamline}
  \end{center}
\end {figure}

Figure~\ref{fig:beamline} shows an overview of the neutrino experimental 
facility~\cite{Abe:2011ks,Sekiguchi:2012xma}. The primary beamline guides the extracted 
proton beam to a production target/pion-focusing horn system
in a target station. The pions decay into muons and neutrinos during 
their flight in a 110 m-long decay volume. A graphite beam dump 
is installed at the end of the decay volume, and muon monitors 
downstream of the beam dump monitor the muon profile. 
A neutrino near detector complex is situated 280 m 
downstream of the target to monitor neutrinos at production.
To generate a narrow band neutrino beam, 
the beamline utilizes an off-axis beam configuration~\cite{OffAxisBeam},
with the capability to vary the off-axis angle 
in the range from 2.0$^\circ$ to 2.5$^\circ$. 
The latter value has been used for the T2K experiment and 
is assumed also for the proposed project. 
The centerline of the beamline extends 295 km to the west, 
passing midway between Tochibora (Hyper-K candidate site) and Mozumi (where Super-K is located), so that both 
sites have identical off-axis angles.  

\begin{table}[btp]
  \begin{center}
    \caption{Acceptable beam power and achievable parameters 
             for each beamline component~\cite{Ishida:2014uxa}.
             Limitations as of May 2013 are also given in parentheses.}
    \begin{tabular}{lcc}
      \hline \hline
      Component   & \multicolumn{2}{c}{Beam power/parameter} \\ 
      \hline
      Target      & \multicolumn{2}{c}{3.3$\times$10$^{14}$ ppp } \\
      Beam window & \multicolumn{2}{c}{3.3$\times$10$^{14}$ ppp } \\
      Horn        &   ~  & ~ \\
      \multicolumn{1}{c}{Cooling for conductors} & 
      \multicolumn{2}{c}{2\,MW       } \\
      \multicolumn{1}{c}{Stripline cooling}   
      & 1$\sim$2\,MW & ( 400\,kW )\\
      \multicolumn{1}{c}{Hydrogen production} 
      & 1$\sim$2\,MW & ( 300\,kW ) \\ 
      \multicolumn{1}{c}{Horn current} & 320 kA & ( 250 kA ) \\
      \multicolumn{1}{c}{Power supply repetition} &  1 Hz  &  ( 0.4 Hz ) \\
      Decay volume   &  \multicolumn{2}{c}{4\,MW} \\
      Hadron absorber/beam dump  &  \multicolumn{2}{c}{3\,MW} \\
      \multicolumn{1}{c}{Water cooling facilities}  
       & $\sim$2\,MW & ( 750\,kW ) \\                  
      Radiation shielding   & 4\,MW & ( 750\,kW )\\
      Radioactive air leakage to the target station ground floor 
      & $\sim$2\,MW &  ( 500\,kW ) \\ 
      Radioactive cooling water treatment 
      & $\sim$2\,MW &  ( 600\,kW ) \\
      \hline \hline
    \end{tabular}
    \label{jparc:BLupgrade}
  \end{center}
\end{table}

Based on the considerable experience gained on the path to achieving 240\,kW beam power operation, 
improvement plans to realize 750\,kW operation,
such as improving the activated air confinement in the target station
and expanding the facilities for the treatment of activated water,
are being implemented and/or proposed.
Table~\ref{jparc:BLupgrade} gives a summary of acceptable beam power and/or 
achievable parameters for each 
beamline component~\cite{Ishida:2014uxa},
after the proposed improvements in forthcoming years. 

\subsection{Near detectors\label{sec:ND}}
The accelerator neutrino event rate observed at Hyper-K depends on the oscillation probability, neutrino flux, neutrino
interaction cross-section, detection efficiency, and the detector fiducial mass of Hyper-K.  To 
extract estimates of the oscillation parameters from data, one must model the neutrino flux, cross-section and 
detection efficiency with sufficient precision.  In the case of the neutrino cross-section, the model must describe the exclusive differential
cross-section that includes the dependence on the incident neutrino energy, $E_{\nu}$, the kinematics of the outgoing
lepton, $p_{l}$ and $\theta_{l}$, and the kinematics of final state hadrons and photons.  
In our case, the neutrino energy is inferred from the lepton kinematics, while the modeling of reconstruction efficiencies depends on the hadronic final state as well.

The neutrino flux and cross-section models can be constrained by data collected at near detectors, situated close enough
to the neutrino production point so that oscillation effects are negligible.  
Our approach to using near detector data will build on the experience of T2K while considering new near detectors that may address important uncertainties in the neutrino flux or cross-section modeling.

The conceptual design of the near detectors is being developed based on the physics sensitivity studies described in Section~\ref{sec:physics_sensitivities}.
In this section, we present basic considerations on the near detector requirements and conceptual designs.
More concrete requirements and detector design will be presented in future.

We assume to use T2K near detectors~\cite{Abe:2011ks}, INGRID and ND280, possibly with an upgrade.
The INGRID detector~\cite{Otani:2010zza} consists of 16 iron-scintillator modules configured in a cross pattern centered on the beam axis 280 m downstream
from the T2K target.
The rate of interactions in each module is measured and a profile is constructed to 
constrain the neutrino beam direction.  The ND280 off-axis detector is located 280 m downstream from the T2K target as well, but
at an angle of 2.5 degrees away from the beam direction.
The P0D $\pi^{0}$ detector~\cite{Assylbekov201248}, time projection chambers (TPCs)~\cite{Abgrall:2010hi},
fine grain scintillator bar detectors (FGDs)~\cite{Amaudruz:2012pe} and surrounding
electromagnetic calorimeters (ECALs)~\cite{Allan:2013ofa}.  The detectors are immersed in a 0.2 T magnetic field and the magnetic yoke is instrumented with
plastic scintillator panels for muon range detection~\cite{Aoki:2012mf}.  The magnetic field allows for momentum measurement and 
sign selection of charged particles.  The magnetization of ND280 is particularly important for operation in antineutrino mode where
the neutrino background is large.  In that case, ND280 is able to separate the ``right-sign" $\mu^{+}$ from the ``wrong-sign" $\mu^{-}$. The 
P0D and FGDs act as the neutrino targets, while the TPCs provide measurements of momentum and ionizing energy loss for particle
identification.  The P0D and one of the FGDs include passive water layers that allow for neutrino interaction rate measurements on 
the same target as Super-K.
ND280 has been employed to measure the rates of charged current $\nu_{\mu}$ and $\nu_{e}$ interactions, as well as
NC$\pi^0$ interactions.

The T2K collaboration is in the process of discussing various upgrade possibilities at the ND280 site~\cite{T2KPACreport}. These include the deployment of heavy water ($\mbox{D}_2\mbox{O}$) within the passive water targets in FGD2 that would allow the extraction of neutrino interaction properties on the quasi-free neutron in deuterium via a subtraction with data taken with light water $\mbox{H}_2\mbox{O}$. The use of a water-based liquid scintillator (WbLS) developed at BNL~\cite{Yeh:2011zz} is  being explored in the context of a tracking detector with comparable or finer granularity than the FGD to allow the detailed reconstruction of hadronic system emerging from the neutrino interactions or a larger detector with coarser segmentation that would allow high statistics studies. Either would significantly enhance the study of neutrino interactions on water by reducing the reliance on subtraction and enhancing the reconstruction capabilities relative to the currently deployed passive targets. Finally, a high pressure TPC that can contain various noble gases (He, Ne, Ar) to serve both as the target and tracking medium is being studied. Such a detector would allow the ultimate resolution of the particles emitted from the target nucleus while allowing a study of the $A$-dependence of the cross-sections and final state interactions to rigorously test models employed in neutrino event generators.

Since many of the uncertainties on the modeling of neutrino interactions arise from uncertainties on nuclear effects, the ideal 
near detector should include the same nuclear targets as the far detector.  In T2K near detectors, the P0D~\cite{Assylbekov201248} and FGD~\cite{Amaudruz:2012pe} detectors include passive water 
layers, however extracting water only cross sections requires complicated analyses that subtract out the interactions on other 
materials in the detectors.  An alternative approach is to build a water Cherenkov (WC) near detector to measure the cross section 
on H$_2$O directly and with no need for a subtraction analysis.  This approach was taken by K2K~\cite{Ahn:2006zza} and was proposed 
for T2K~\cite{t2k2km}.   The MiniBooNE experiment has also employed a mineral oil Cherenkov detector at a short baseline to great 
success~\cite{AguilarArevalo:2008qa}.  A WC near detector design is largely guided by two requirements:
\begin{enumerate}
\item The detector should be large enough to contain muons up to the momentum of interest for measurements at the far detector, and to provide sufficient radiation length for detection of gamma rays.
\item The detector should be far enough from the neutrino production point so that there is minimal pile-up of interactions in the same beam timing bunch.
\end{enumerate}
These requirements lead to designs for kiloton size detectors located at intermediate distances, 1--2 km from the target, for the J-PARC neutrino beam.  

The main disadvantage of the WC detector is the inability to separate positively and negatively charged leptons, and hence 
antineutrino and neutrino interactions. 
This ability is especially important for a $CP$ asymmetry measurement where the wrong 
sign contribution to the neutrino flux should be well understood.  Hence, the WC detector will most likely be used in conjunction with a magnetized tracking detector such as ND280.  
Recent developments in the addition of Gadolinium (Gd)~\cite{Beacom:2003nk} and Water-based Liquid Scintillator (WbLS) compounds~\cite{Yeh:2011zz} to water do raise the possibility to separate neutrino and antineutrino interactions by detecting the presence of neutrons or protons in the final state.

Two conceptual designs for possible intermediate WC detectors have been studied.
Unoscillated Spectrum (TITUS) is a 2 kiloton WC detector located about 2 km from the target at the same off-axis angle as the far detector.  
At this baseline the detector sees fluxes for the neutral current and $\nu_e$ backgrounds that are nearly identical to the Hyper-K fluxes. 
The detector geometry and the presence of a muon range detector are optimized to detect the high momentum tail of the muon spectrum.  
The use of Gd in TITUS to separate neutrino and antineutrino interactions is being studied.  The $\nu$PRISM detector is located 1 km from the target and is 
50 m tall, covering a range of off-axis angles from 1-4 degrees. 
The $\nu$PRISM detector sees a range of neutrino spectra, peaked at energies 
from 0.4 to 1.0\,GeV depending on the off-axis angle.  
The purpose of $\nu$PRISM is to use these spectra to better probe the relationship between the incident neutrino energy and final state lepton kinematics, a part of the interaction model with larger uncertainties arising from nuclear effects.

\subsection{Hyper-Kamiokande\label{sec:HK-detector}}
Hyper-Kamiokande
is to be the third generation water Cherenkov detector in Kamioka, 
designed for a wide variety of neutrino studies and nucleon decay searches.
Its total (fiducial) water mass of one (0.56) million tons would be approximately 20 (25) times larger than
that of Super-Kamiokande.
Table~\ref{tab:detector-parameters} 
summarizes the baseline design parameters of the Hyper-K detector.

\begin{table}[tbp]
\caption{Parameters of the Hyper-Kamiokande baseline design.}
\begin{center}
\begin{tabular}{lll} \hline \hline
Detector type & & Ring-imaging \\
& & water Cherenkov detector \\ \hline 
Candidate site & Address & Tochibora mine \\
& & Kamioka town, Gifu, JAPAN \\
& Lat. & $36^\circ21'20.105''$N $^{\dagger}$ \\
& Long. & $137^\circ18'49.137''$E $^{\dagger}$\\
& Alt. & 508 m \\
& Overburden & 648 m rock \\
& &(1,750 m water equivalent)  \\
& Cosmic Ray Muon flux & $\sim$ 8 $\times$ 10$^{-7}$ sec$^{-1}$cm$^{-2}$  \\
& Off-axis angle for the J-PARC $\nu$ & $2.5^\circ$ (same as Super-K)  \\
& Distance from the J-PARC & 295 km (same as Super-K)  \\ \hline 
Detector geometry & Total Water Mass & 0.99 Megaton  \\
 & Inner Detector (Fiducial) Mass & 0.74 (0.56) Megaton  \\
 & Outer Detector Mass & 0.2 Megaton  \\ \hline 
Photo-sensors & Inner detector & 99,000 20-inch $\phi$ PMTs \\
& & 20\% photo-coverage \\ 
& Outer detector & 25,000 8-inch $\phi$ PMTs \\ \hline 
Water quality & light attenuation length & $>100$ m @ 400 nm  \\
 & Rn concentration & $<1$ mBq/m$^3$ \\ \hline 
\hline 
\multicolumn{3}{r}{$^{\dagger}$ World geographical coordination system}\\
\end{tabular}
\end{center}
\label{tab:detector-parameters} 
\end{table}

\begin{figure}[tbp]
  \centering
  \includegraphics{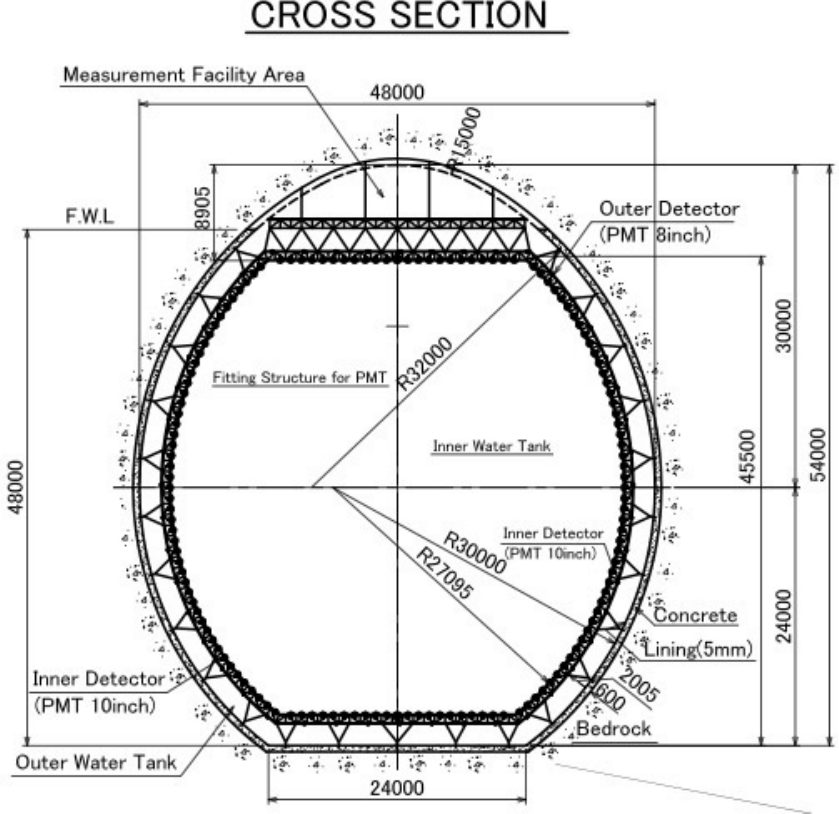}
  \caption{Cross section view of the Hyper-Kamiokande detector.\label{fig:crosssection}}
\end{figure}

\begin{figure}[htbp]
  \centering
  \includegraphics[width=\textwidth]{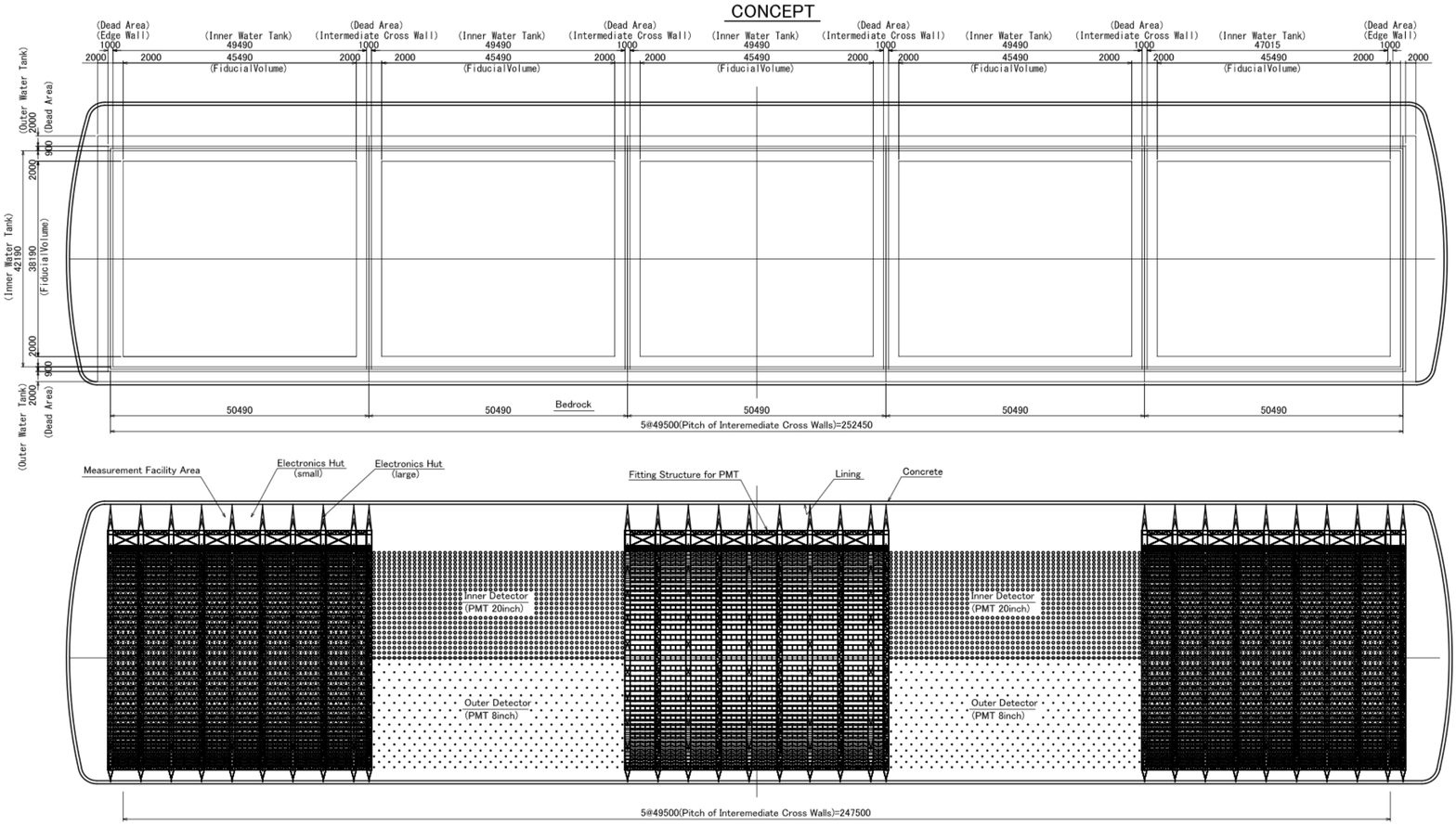}
  \caption{
    Profile of the Hyper-K detector.
    Top: the detector segmentation.
    Bottom: PMT arrays and the support structure for the inner
    and outer detectors.
    Each quasi-cylindrical tank
    lying horizontally is segmented by intermediate walls
    into five compartments.\label{fig:profile}}
\end{figure}

In the baseline design, the Hyper-K detector is composed of two separated caverns
as shown in Fig.~\ref{fig:HK-schematic-view}, each having an egg-shape cross section
48\,meters wide, 54\,meters tall, and 250\,meters long as shown
in Fig.~\ref{fig:crosssection} and \ref{fig:profile}.
The welded polyethylene tanks are filled up to a depth of 48\,m with
ultra-pure water: the total water mass equals 0.99 million tons.

Each tank will be optically separated by segmentation walls located every 49.5\,m to form 5 (in total 10) compartments as shown in Fig.~\ref{fig:profile}, such that event triggering and event reconstruction can be performed in each compartment separately and independently.
Because the compartment dimension of 50 m is comparable with that of Super-K (36 m) and is shorter than the typical light attenuation length in water achieved by the Super-K water filtration system ($>100$\,m @ 400~nm), we expect that the detector performance of Hyper-K for beam and atmospheric neutrinos will be effectively the same as that of Super-K.

The water in each compartment is further optically
separated into three regions.
The inner region has a barrel shape of 42 m in height and width,
and 48.5 m in length,
and is viewed by an inward-facing array of 20-inch diameter photomultiplier tubes (PMTs).
The entire array consists of 99,000 Hamamatsu R3600  PMTs,
uniformly surrounding the region and giving a photocathode coverage of 20\%.
The PMT type, size, and number density are subject to optimization.
We have been also developing new photosensors as possible alternative options to the R3600,
such as a PMT with a box-and-line dynode and a hybrid photo-detector (HPD), both with a high quantum efficiency photocathode.
An outer region completely surrounds the 5 (in total 10) inner regions
and is equipped with 25,000 8-inch diameter PMTs.
This region is 2 m thick at the top, bottom,
and barrel sides, except at both ends of each cavern, where
the outer region is larger than 2 m due to rock engineering considerations.
A primary function of the outer detector is to reject
entering cosmic-ray muon backgrounds and to help in identifying
nucleon decays and neutrino interactions occurring in the inner detector.
The middle region or dead space is an uninstrumented, 0.9\,m thick
shell between the inner and outer detector volumes
where the stainless steel PMT support structure is located.
Borders of both inner and outer regions are lined with opaque sheets.
This dead space, along with the outer region, acts as a shield against
radioactivity from the surrounding rock.
The total water mass of the inner region is 0.74 million tons
and the total fiducial mass is 10 times 0.056 = 0.56 million tons.
The fiducial volume is defined as the region formed by
a virtual boundary located 2\,m away
from the inner PMT plane. 

The estimated cosmic-ray muon rate around the Hyper-K detector
candidate site is $\sim$ 8 $\times$ 10$^{-7}$ sec$^{-1}$cm$^{-2}$ which is
roughly 5 times larger than the flux at Super-K's location
($\sim$ 1.5 $\times$ 10$^{-7}$ sec$^{-1}$cm$^{-2}$).
The expected deadtime due to these muons is less than 1\% and
negligible for long baseline experiments, as well as nucleon decay searches and atmospheric neutrino studies.

Water is the target material and signal-sensitive medium of the detector, and thus its quality directly affects the physics sensitivity.
In Super-Kamiokande the water purification system has been continually modified and improved over the course of two decades.
As a result, the transparency is now kept above 100~m and is very stable, and the radon concentration in the tank is held below 1~mBq/m$^3$.  Following this success, the Hyper-Kamiokande water system has been designed based on the current Super-Kamiokande water system with scaling up the process speeds to 1200~m$^3$/hour for water circulation and 400~m$^3$/hour for radon free air generation.
With these systems, the water quality in Hyper-Kamiokande is expected to be same as that in Super-Kamiokande.
Adding dissolved gadolinium sulfate for efficient tagging of neutrons has been studied as an option to enhance Hyper-K physics capability.
The feasibility of adding Gd to Super-K~\cite{Beacom:2003nk} is now under study with EGADS (Evaluating Gadolinium's Action on Detector Systems) project in Kamioka.
We have been careful to keep the possibility of gadolinium loading in mind when designing the overall Hyper-Kamiokande water system.  

We have evaluated the expected performance of the Hyper-K detector using 
the MC simulation and reconstruction tools under development.
We have been developing a detector simulation dedicated to Hyper-K based on
``WCSim,''~\cite{WCsim} which is an open-source water
Cherenkov detector simulator based on the GEANT4 library~\cite{Agostinelli:2002hh,Allison:2006ve}.
A new reconstruction algorithm developed for Super-K/T2K~\cite{Abe:2013hdq}, named ``fiTQun,'' has been adopted for the Hyper-K analysis.
It uses a maximum likelihood fit with charge and time probability density functions constructed for every PMT hit
assuming several sets of physics variables (such as vertex, direction, momentum, and particle type)~\cite{Patterson:2009ki, Abe:2013hdq}.

\begin{figure}[tb]
\centering
  \hspace*{-0.5cm}\includegraphics[width=8cm,clip]{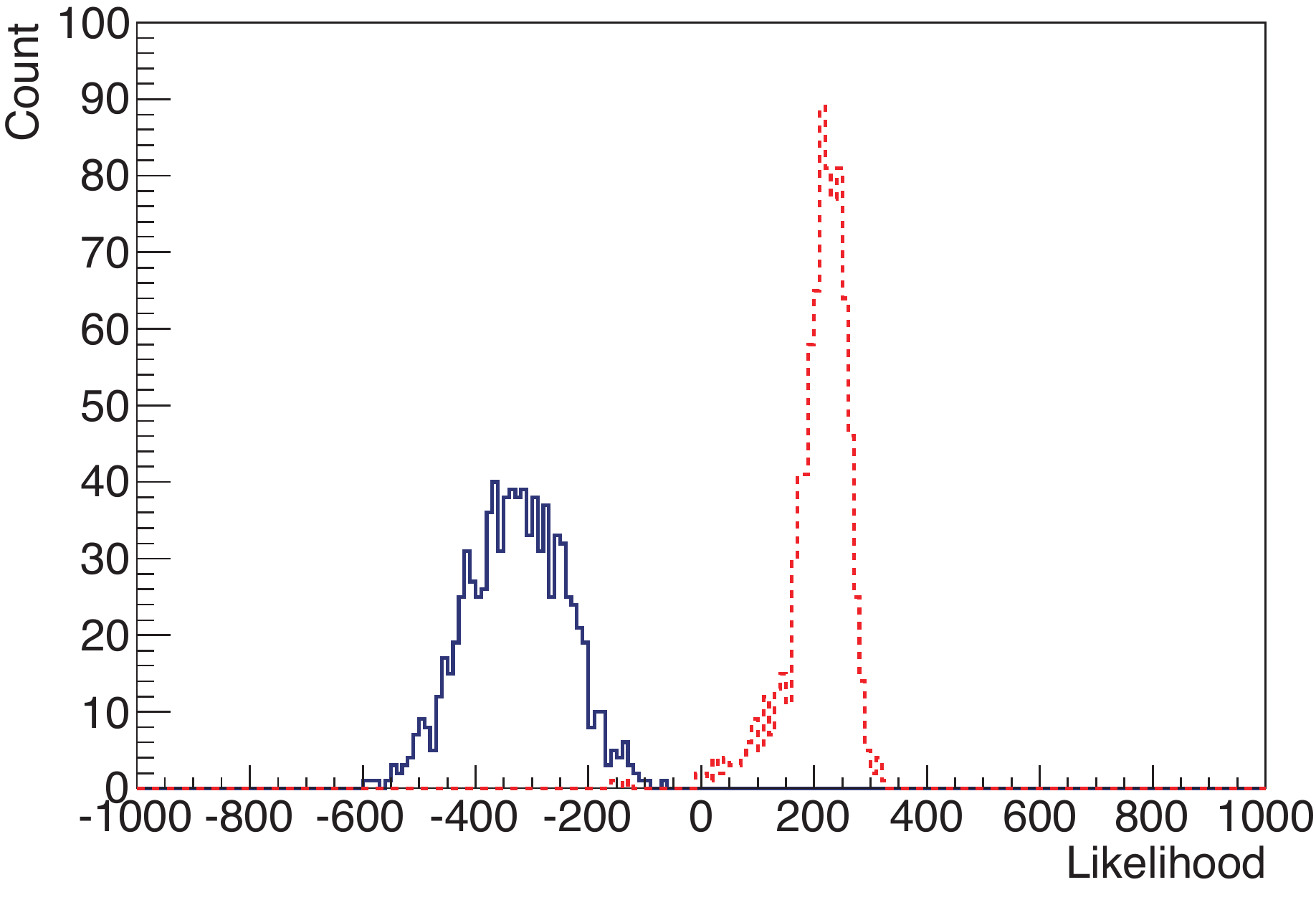}
  \caption
  { PID likelihood functions for electron (blue solid histogram) and $\mu$ (red dashed histogram) with 500\,MeV/$c$ momentum. A negative (positive) value indicates electron-like ($\mu$-like) particle.
    }
  \label{pid}
\end{figure}

As an example of the evaluation, electrons and muons with 500\,MeV/$c$ are generated with a fixed vertex (at the center of
the tank) and direction (toward the barrel of the tank) 
in the Hyper-K detector simulation.
Figure~\ref{pid} shows the likelihood function for the particle identification.
A negative (positive) value indicates electron-like ($\mu$-like) particle.
It demonstrates a clear separation of electrons and muons.
The obtained performance of Hyper-Kamiokande is compared with the performance of 
SK-II (20\% photo coverage, old electronics) and SK-IV (40\% photo coverage, new electronics) in Table~\ref{tab:performance}.
The vertex resolution for muon events will be improved to the same level as
Super-K with an update of the reconstruction program.
From the preliminary studies, 
the performance of Hyper-K is similar to or possibly better than
SK-II or SK-IV with the new algorithm.
In the physics sensitivity study described in Section~\ref{sec:physics_sensitivities},
a Super-K full MC simulation with the SK-IV configuration is used because 
it includes the simulation of new electronics and is tuned with the real data,
while giving similar performance with Hyper-K as demonstrated above.

\begin{table}
  \begin{center}
   \caption{
Comparison of performance of SK-II (20\% photo-coverage), SK-IV (40\% photo-coverage), 
and the expected performance of Hyper-Kamiokande baseline design (20\% photo-coverage) with preliminary Hyper-K simulation and reconstruction.
}
  \begin{tabular}{l|cc|cc|cc}
      \hline \hline
      & \multicolumn{2}{c|}{SK-II}  & \multicolumn{2}{c|}{SK-IV}  & \multicolumn{2}{c}{Hyper-K}\\
      \hline
      Particle type ($p=$500\,MeV/$c$)  & $e$     & $\mu$   & $e$     & $\mu$   & $e$       & $\mu$\\
      \hline 
      Vertex resolution                &  28~cm  &  23~cm  &  25~cm  &  17~cm  & 27~cm     &  30~cm \\
      Particle identification          &  98.5\% &  99.0\% &  98.8\% &  99.5\% & $>$99.9\% &  99.2\%  \\
      Momentum resolution              &  5.6\%  &  3.6\%  &  4.4\%  &  2.3\%  & 4.0\%     &  2.6\%   \\
      \hline \hline
    \end{tabular}
  \label{tab:performance}
  \end{center}
\end{table}

\section{Physics Sensitivities}
\label{sec:physics_sensitivities}

\subsection{Overview}
As discussed in Sec.~\ref{sec:appearance}, a comparison of muon-type to electron-type transition probabilities between neutrinos and anti-neutrinos is 
one of the most promising methods to observe the lepton $CP$ asymmetry.
Recent observation of a nonzero, rather large value of $\theta_{13}$~\cite{Abe:2011sj, Abe:2011fz,Ahn:2012nd,An:2012eh} makes 
this exciting possibility more realistic.

\begin{figure}[tbp]
\centering
\includegraphics[width=0.48\textwidth]{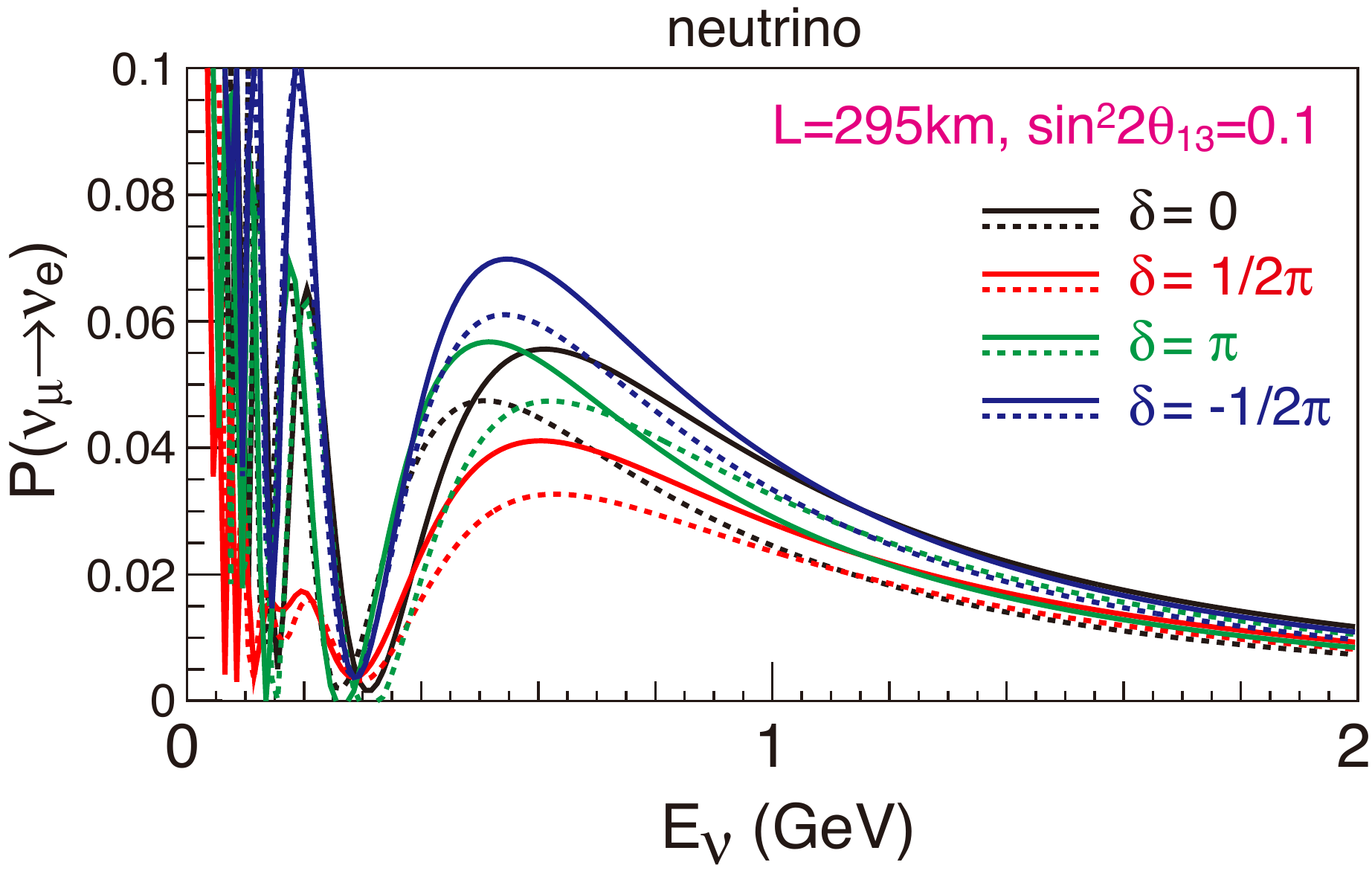}
\includegraphics[width=0.48\textwidth]{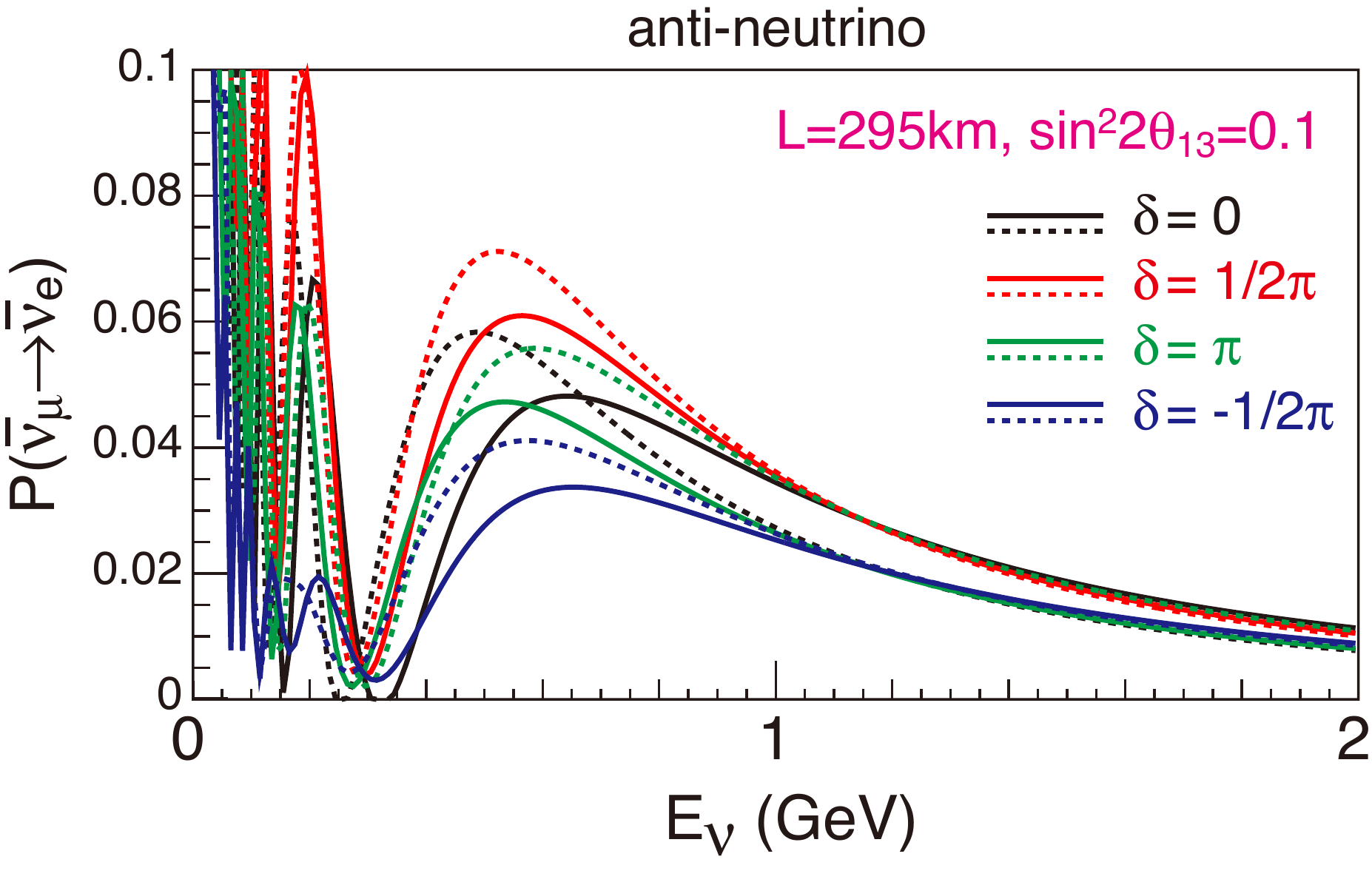}
\caption{Oscillation probabilities as a function of the neutrino energy for $\numu \to \nue$ (left) and $\numubar \to \nuebar$ (right) transitions with L=295~km and $\sin^22\theta_{13}=0.1$. 
Black, red, green, and blue lines correspond to $\deltacp = 0, \frac{1}{2}\pi, \pi$, and $-\frac{1}{2}\pi$, respectively.
Solid (dashed) line represents the case for a normal (inverted) mass hierarchy.
\label{fig:cp-oscpob}}
\end{figure}

Figure~\ref{fig:cp-oscpob} shows the $\numu \to \nue$ and $\numubar \to \nuebar$ oscillation probabilities as a function of the true neutrino energy for a baseline of 295~km.
The Earth matter density of 2.6\,$g$/cm$^3$ is used in this analysis.
The cases for $\deltacp = 0, \frac{1}{2}\pi, \pi$, and $-\frac{1}{2}\pi$, are overlaid. 
Also shown are the case of normal mass hierarchy ($\Delta m^2_{32}>0$) with solid lines and inverted mass hierarchy ($\Delta m^2_{32}<0$) with dashed lines.
The oscillation probabilities depend on the value of $\deltacp$, and by comparing the neutrinos and anti-neutrinos,
one can see the effect of $CP$ violation.

\begin{figure}[tbp]
\centering
\includegraphics[width=0.48\textwidth]{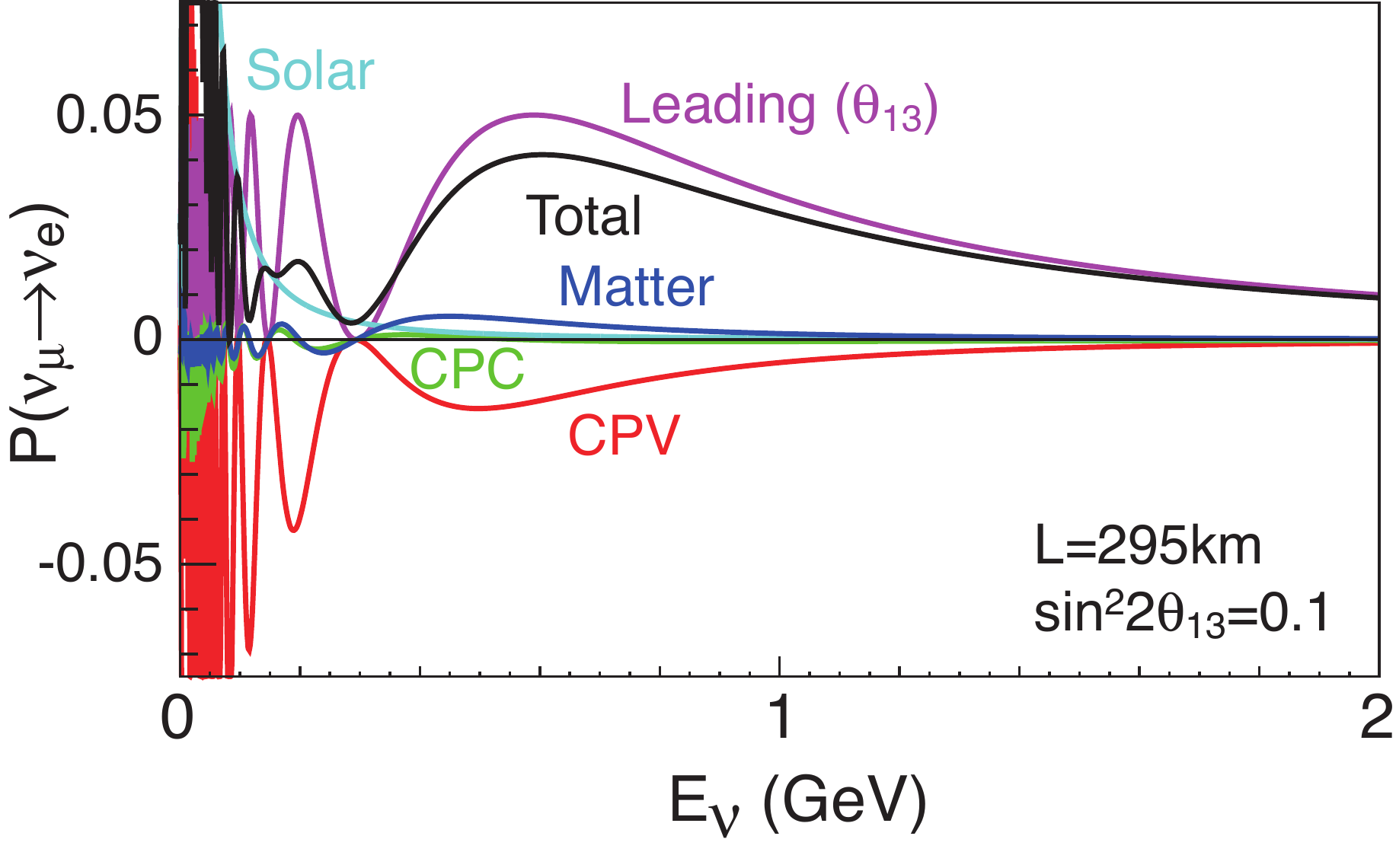}
\caption{Oscillation probability of $\numu \to \nue$  as a function of the neutrino energy with a baseline of 295~km.  $\sin^22\theta_{13}=0.1$,
$\deltacp = \frac{1}{2}\pi$, and normal hierarchy are assumed.
Contribution from each term of the oscillation probability formula is shown separately.
\label{fig:cp-oscpob-bd}}
\end{figure}

There are sets of different mass hierarchy and values of $\deltacp$ which give similar oscillation probabilities.
This is known as the degeneracy due to unknown mass hierarchy and may introduce an ambiguity 
if we do not know the true mass hierarchy.
Because there are a number of experiments planned to determine mass hierarchy
in the near future, it is expected that the mass hierarchy will be determined by the time Hyper-K
starts to take data.
If not, Hyper-K itself has a sensitivity to the mass hierarchy by the atmospheric neutrino measurements as shown in Table~\ref{tab:intro:phys}.
Furthermore, a combined analysis of the accelerator and atmospheric neutrino data in Hyper-K will enhance the sensitivity as shown in Sec.~\ref{sec:lbl-atm}.
Thus, the mass hierarchy is assumed to be known in this analysis, unless otherwise stated.

Figure~\ref{fig:cp-oscpob-bd} shows the contribution from each term of the $\numu \to \nue$ oscillation probability formula, Eq.(\ref{Eq:cpv-oscprob}),
for $L=295$\,km, $\sin^22\theta_{13}=0.1$, $\sin^22\theta_{23}=1.0$, $\deltacp = \pi/2$, and normal mass hierarchy.
For $E_\nu\simeq 0.6$\,GeV which gives $\sin\Delta_{32} \simeq \sin\Delta_{31} \simeq 1$,
\begin{eqnarray}
\frac{P(\nu_\mu \to  \nu_e) - P(\bar\nu_\mu \to  \bar\nu_e)}{P(\nu_\mu \to  \nu_e) + P(\bar\nu_\mu \to  \bar\nu_e)} &\simeq& 
\frac{-16J_{CP}\sin \Delta_{21} + 16 c_{13}^2s_{13}^2s_{23}^2 \frac{a}{\Delta m^2_{31}}(1-2s_{13}^2) }{8c_{13}^2s_{13}^2s_{23}^2}\\
&=& -0.28 \sin\delta + 0.07.
\end{eqnarray}
The effect of $CP$ violating term can be as large as 28\%, while the matter effect is much smaller.

The uncertainty of Earth density between Tokai and Kamioka is estimated to be at most 6\%\cite{Hagiwara:2011kw}. Because the matter effect contribution to the total appearance probability is less than 10\% for 295km baseline, the uncertainty from matter density is estimated to be less than 0.6\% and neglected in this analysis.

Due to the relatively short baseline and thus lower neutrino energy at the oscillation maximum, 
the contribution of the matter effect is smaller for the J-PARC to Hyper-Kamiokande experiment
compared to other proposed experiments like LBNE in the United States~\cite{Adams:2013qkq} or LBNO in Europe~\cite{Agarwalla:2014tca}.
Thus the $CP$ asymmetry measurement with J-PARC to Hyper-K long baseline experiment has less uncertainty related to the matter effect, while other experiments with $>1000$~km baseline have much better sensitivity to the mass hierarchy with accelerator neutrino beams~\footnote{Note that Hyper-K has sensitivity to the mass hierarchy using atmospheric neutrinos as shown in Table~\ref{tab:intro:phys}.}.
The sensitivities for $CP$ violation and mass hierarchy can be further enhanced by combining measurements with different baseline.

The analysis method is based on a framework developed for the sensitivity study by T2K reported in~\cite{Abe:2014tzr}.
A binned likelihood analysis based on the reconstructed neutrino energy distribution is performed using both \nue\ (\nuebar) appearance and \numu\ (\numubar) disappearance samples simultaneously.
In addition to $\sin^22\theta_{13}$ and $\deltacp$, $\sin^2\theta_{23}$ and $\Delta m^2_{32}$ are also included as free parameters in the fit.
Table~\ref{Tab:oscparam} shows the nominal oscillation parameters used in the study presented in this paper, and the treatment during the fitting.
Systematic uncertainties are estimated based on the experience and prospects of the T2K experiment, and implemented as a covariance matrix which takes into account the correlation of uncertainties.

An integrated beam power of 7.5~MW$\times$10$^7$~sec is assumed in this study. It corresponds to $1.56\times10^{22}$ protons on target with 30\,GeV J-PARC beam.
We have studied the sensitivity to $CP$ violation with various assumptions of neutrino mode and anti-neutrino mode beam running time ratio for both normal and inverted mass hierarchy cases.
The dependence of the sensitivity on the $\nu$:$\overline{\nu}$ ratio is found to be not significant between $\nu$:$\overline{\nu}$=1:1 to 1:5.
In this paper, $\nu$:$\overline{\nu}$ ratio is set to be 1:3 so that the expected number of events are approximately the same for neutrino and anti-neutrino modes.

\begin{table}[tbp]
\caption{Oscillation parameters used for the sensitivity analysis and treatment in the fitting. The \textit{nominal} values are used for figures and numbers in this section, unless otherwise stated.}
\centering
\begin{tabular}{ccc} \hline \hline
Parameter & Nominal value & Treatment \\ \hline
$\sin^22\theta_{13}$ & 0.10 & Fitted \\
$\deltacp$ & 0 & Fitted \\
$\sin^2\theta_{23}$ & 0.50 & Fitted \\
$\Delta m^2_{32}$ & $2.4\times10^{-3}~\mathrm{eV}^2$ & Fitted \\
Mass hierarchy & Normal or Inverted & Fixed \\
$\sin^22\theta_{12}$ & $0.8704$ & Fixed \\
$\Delta m^2_{21}$ & $7.6\times10^{-5}~\mathrm{eV}^2$ & Fixed \\ \hline \hline
\end{tabular}
\label{Tab:oscparam}
\end{table}%

\subsection{Neutrino flux}
The neutrino flux is estimated by T2K collaboration~\cite{Abe:2012av} by simulating the J-PARC neutrino beam line while tuning the modeling of hadronic interactions using data from NA61/SHINE~\cite{Abgrall:2011ae,Abgrall:2011ts} and other experiments measuring hadronic interactions on nuclei. 
To date, NA61/SHINE has provided measurements of pion and kaon production multiplicities for proton interactions on a 0.04 interaction length graphite target, as well as the inelastic cross section for protons on carbon.
Since ``thin" target data are used, the secondary interactions of hadrons inside and outside of the target are modeled using other data or scaling the NA61/SHINE data to different center of mass energies or target nuclei.
NA61/SHINE also took data with a replica of 90\,cm-long T2K target, which will reduce the uncertainties related to the secondary interactions inside of the target.

\begin {figure}[tbp]
  \centering
    \includegraphics[width=0.48\textwidth]{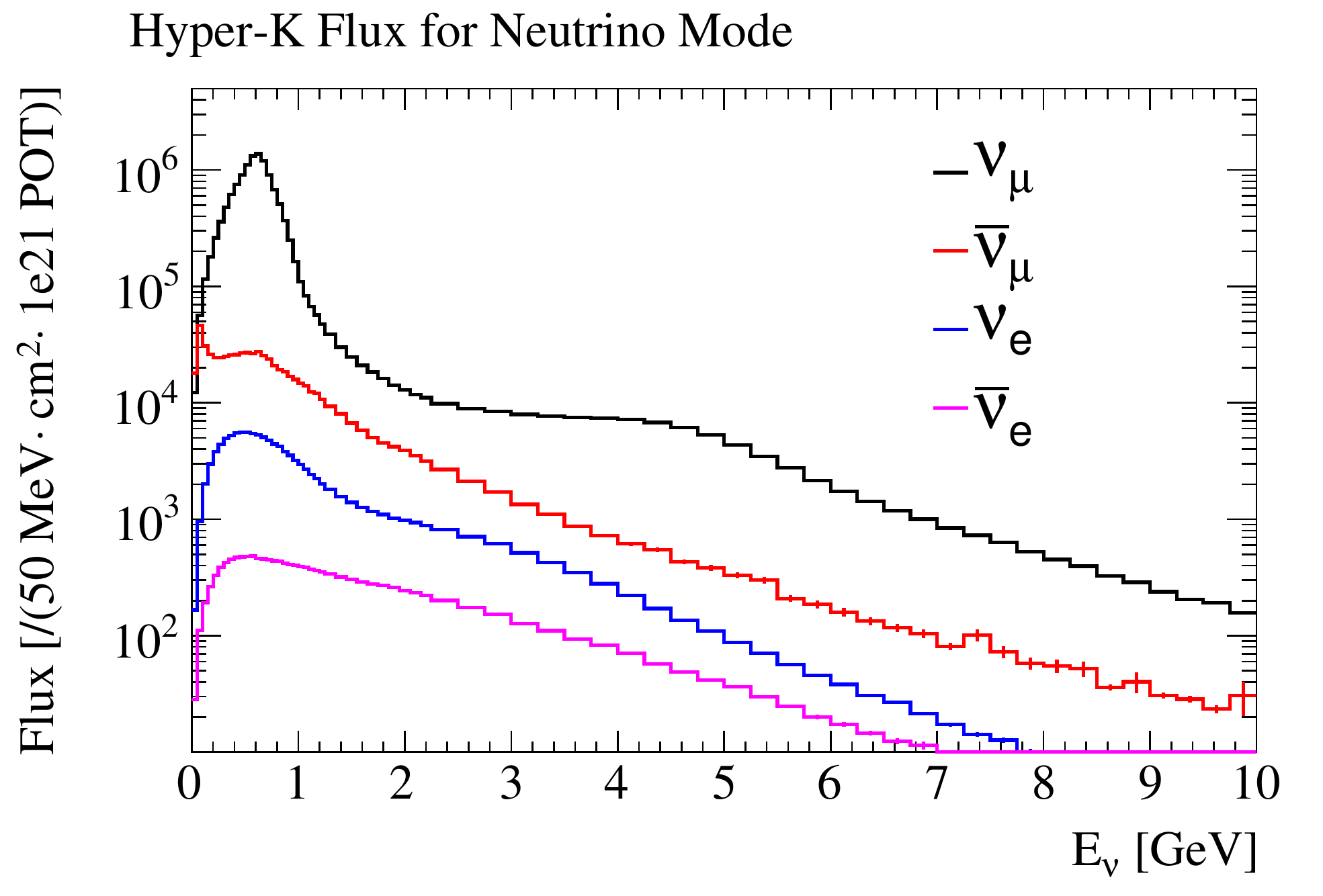}
    \includegraphics[width=0.48\textwidth]{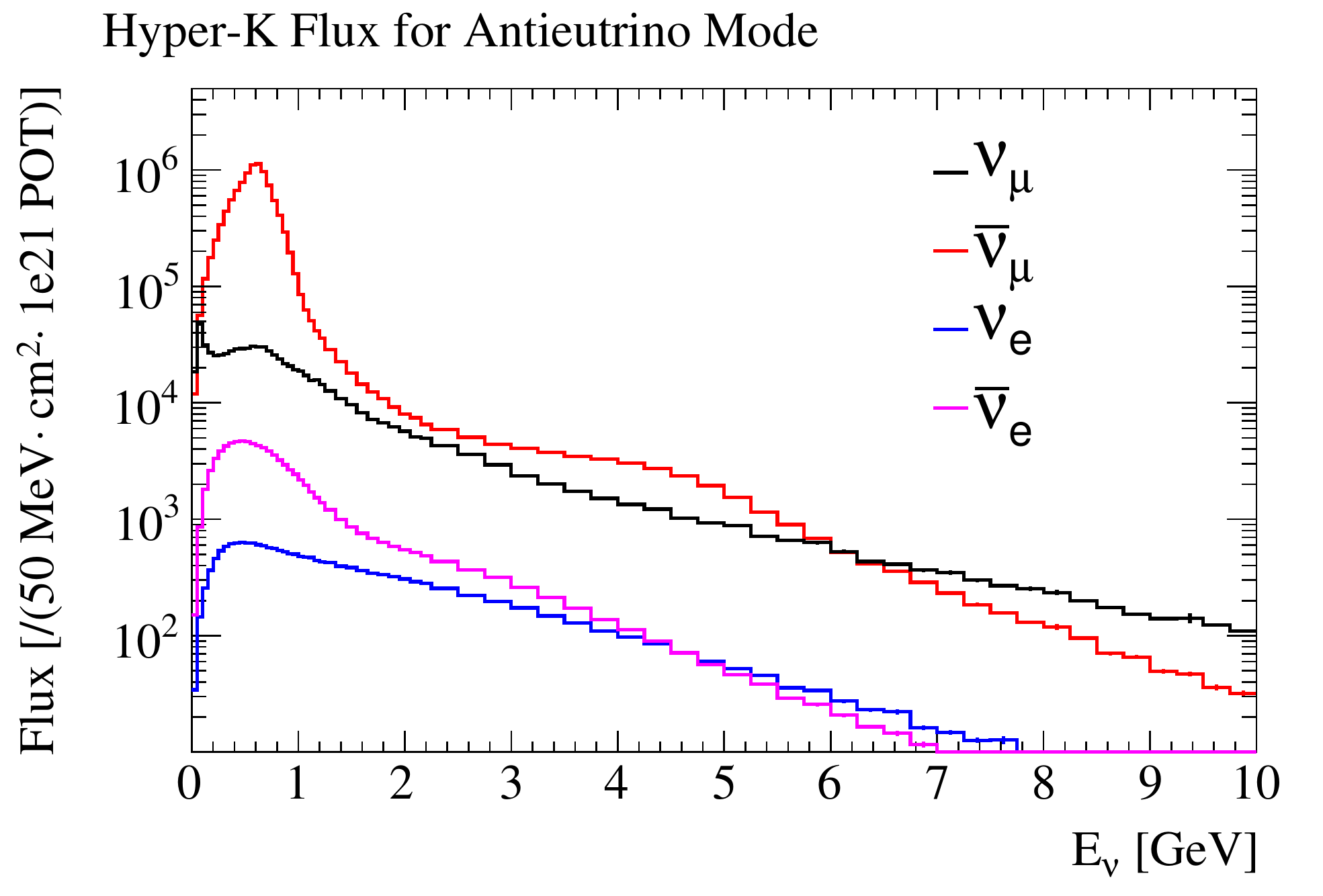}
    \caption{The predicted Hyper-K neutrino fluxes from the J-PARC beam without oscillations. The
neutrino enhanced beam is shown on the left and the antineutrino enhanced beam is shown on the right.}
    \label{fig:flux_pred}
\end {figure}

 For the studies presented in this
document, the T2K flux simulation has been used with the horn currents raised from 250 kA to 320 kA. 
The flux is estimated for both polarities of the horn fields, corresponding to neutrino enhanced and antineutrino enhanced fluxes. 
The calculated fluxes at Hyper-K, without oscillations, are shown in Fig.~\ref{fig:flux_pred}.  

The sources of uncertainty in the T2K flux calculation include:
\begin{itemize}
\item{Uncertainties on the primary production of pions and kaons in proton on carbon collisions.}
\item{Uncertainties on the secondary hadronic interactions of particles in the target or 
beam line materials after the initial hadronic scatter.} 
\item{Uncertainties on the properties of the proton beam incident on the target, including the absolute
current and the beam profile.}
\item{Uncertainties on the alignment of beam line components, including the target and magnetic horns.}
\item{Uncertainties on the modeling of the horn fields, including the absolute field strength and 
asymmetries in the field.}
\end{itemize}

\begin {figure}[tbp]
  \centering
    \includegraphics[width=0.48\textwidth]{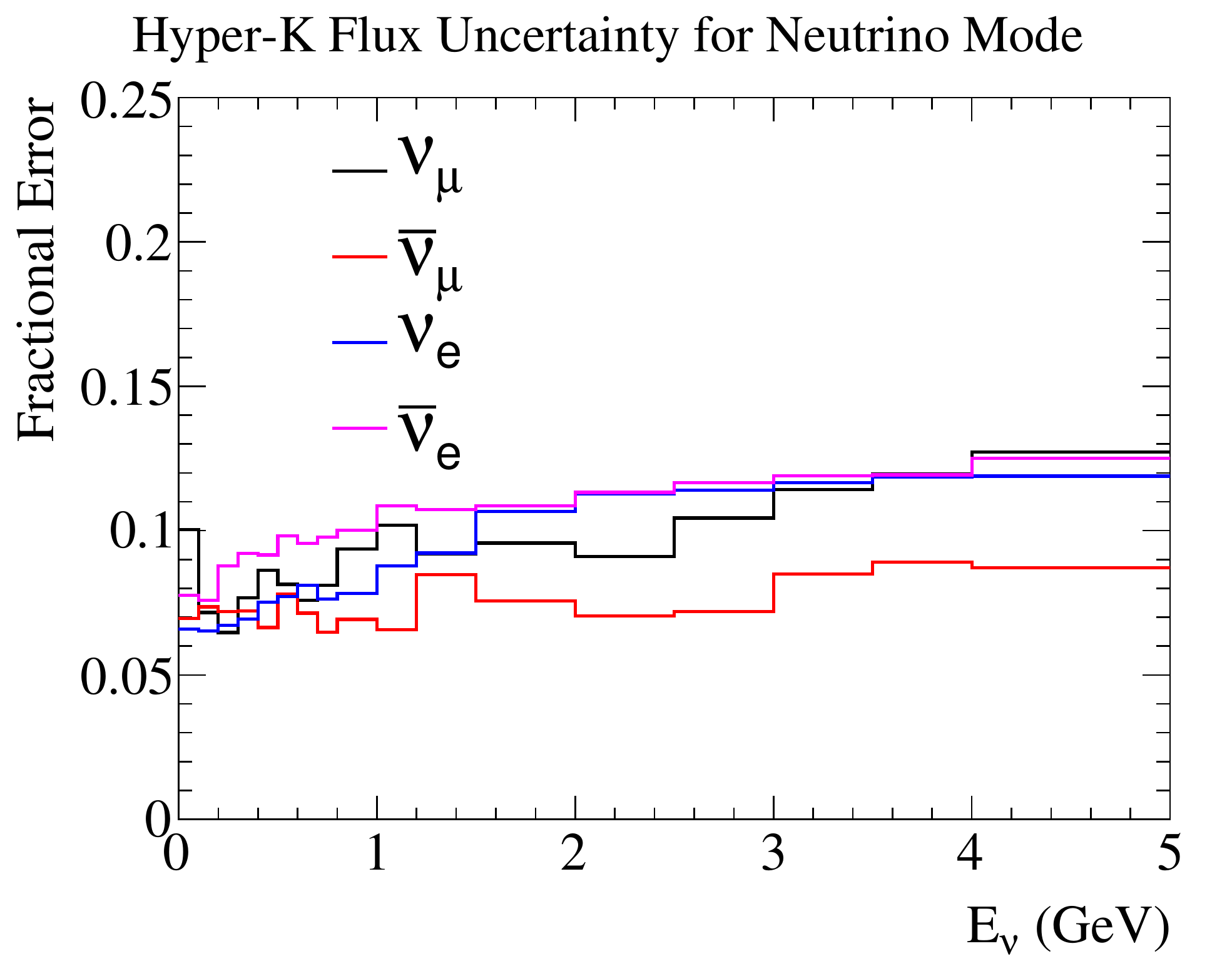}
    \includegraphics[width=0.48\textwidth]{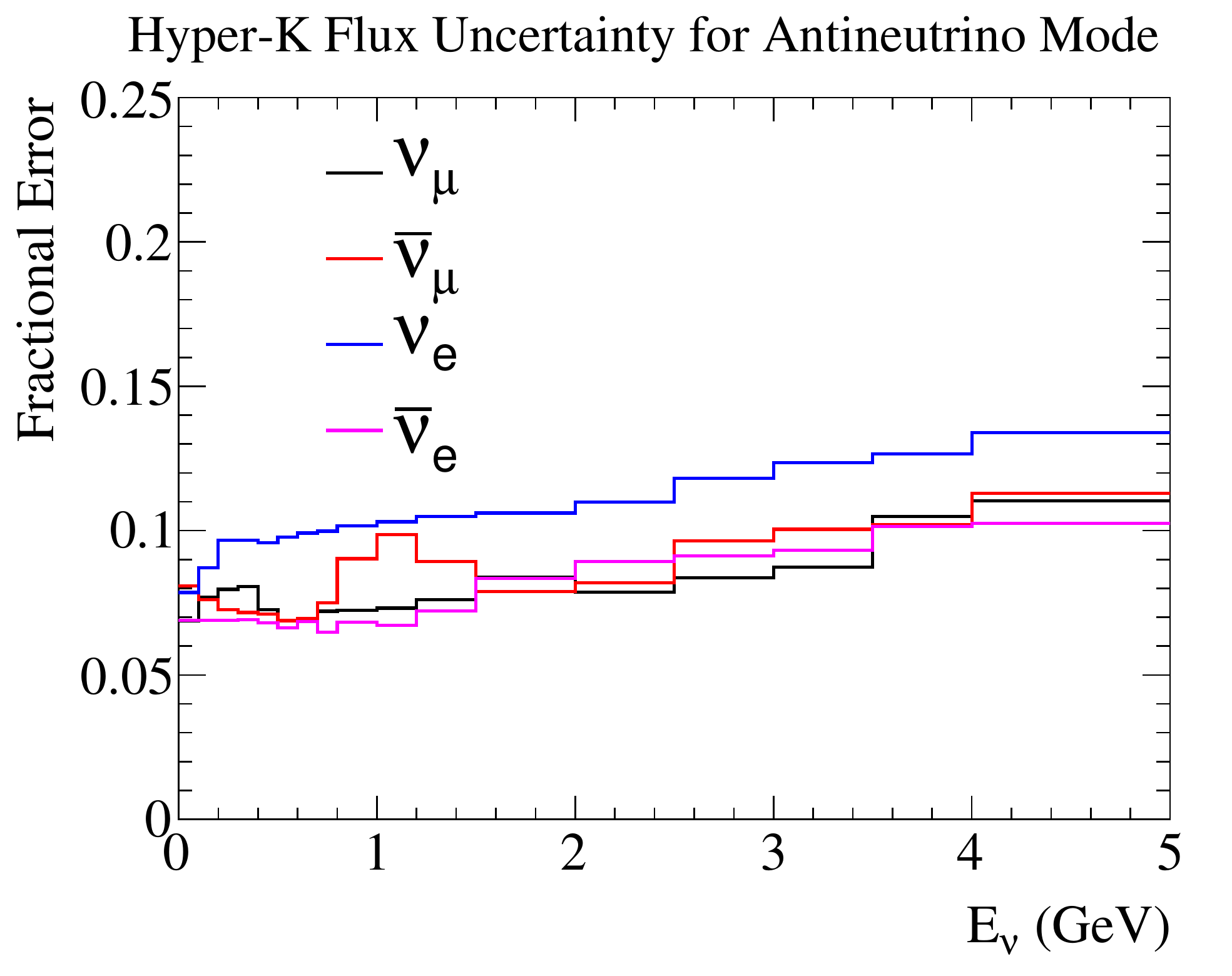}
    \caption{The predicted uncertainty on the neutrino flux calculation assuming replica target hadron production data
are available.}
    \label{fig:flux_unc}
\end {figure}

\begin {figure}[tbp]
  \centering
    \includegraphics[width=0.48\textwidth]{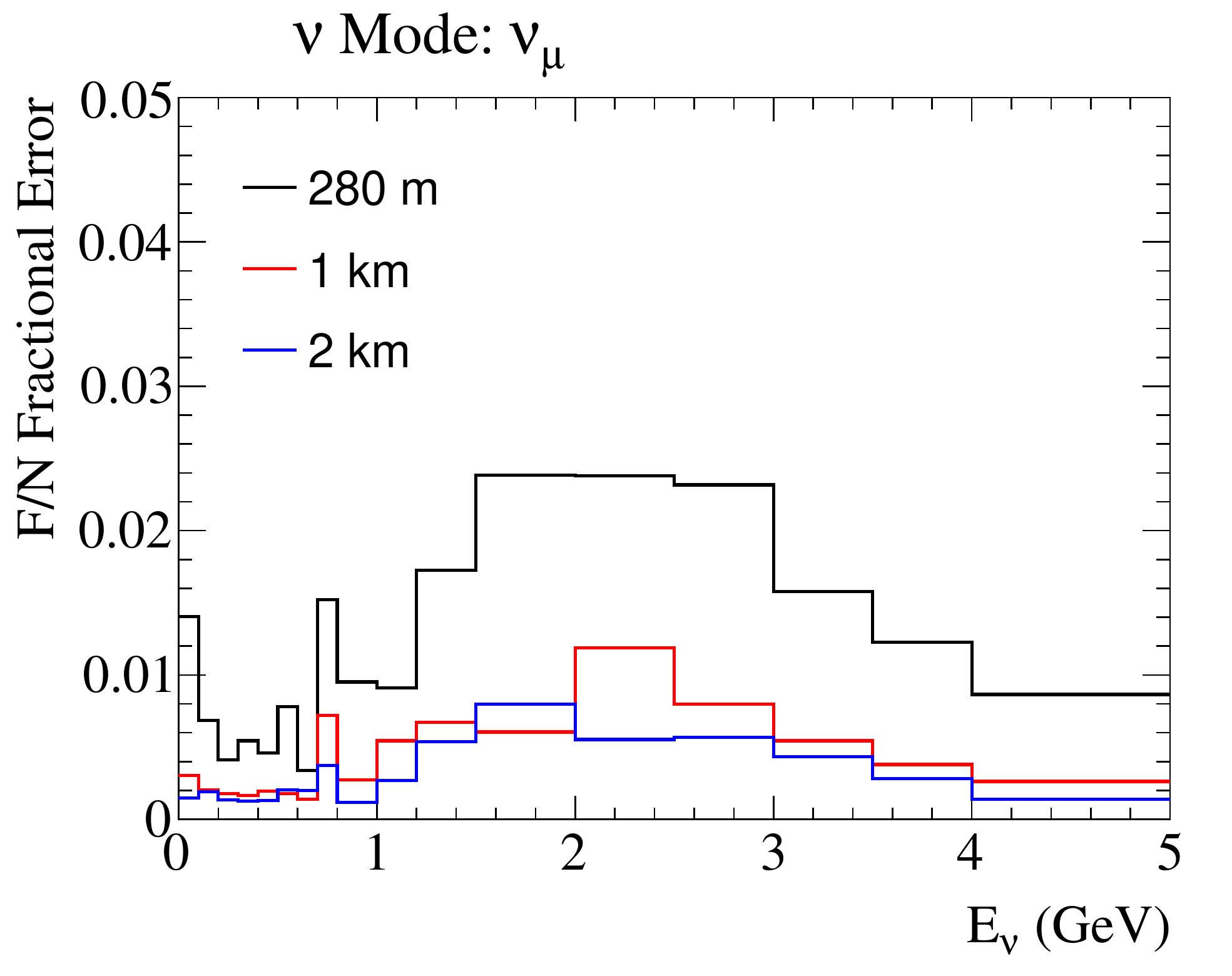}
    \includegraphics[width=0.48\textwidth]{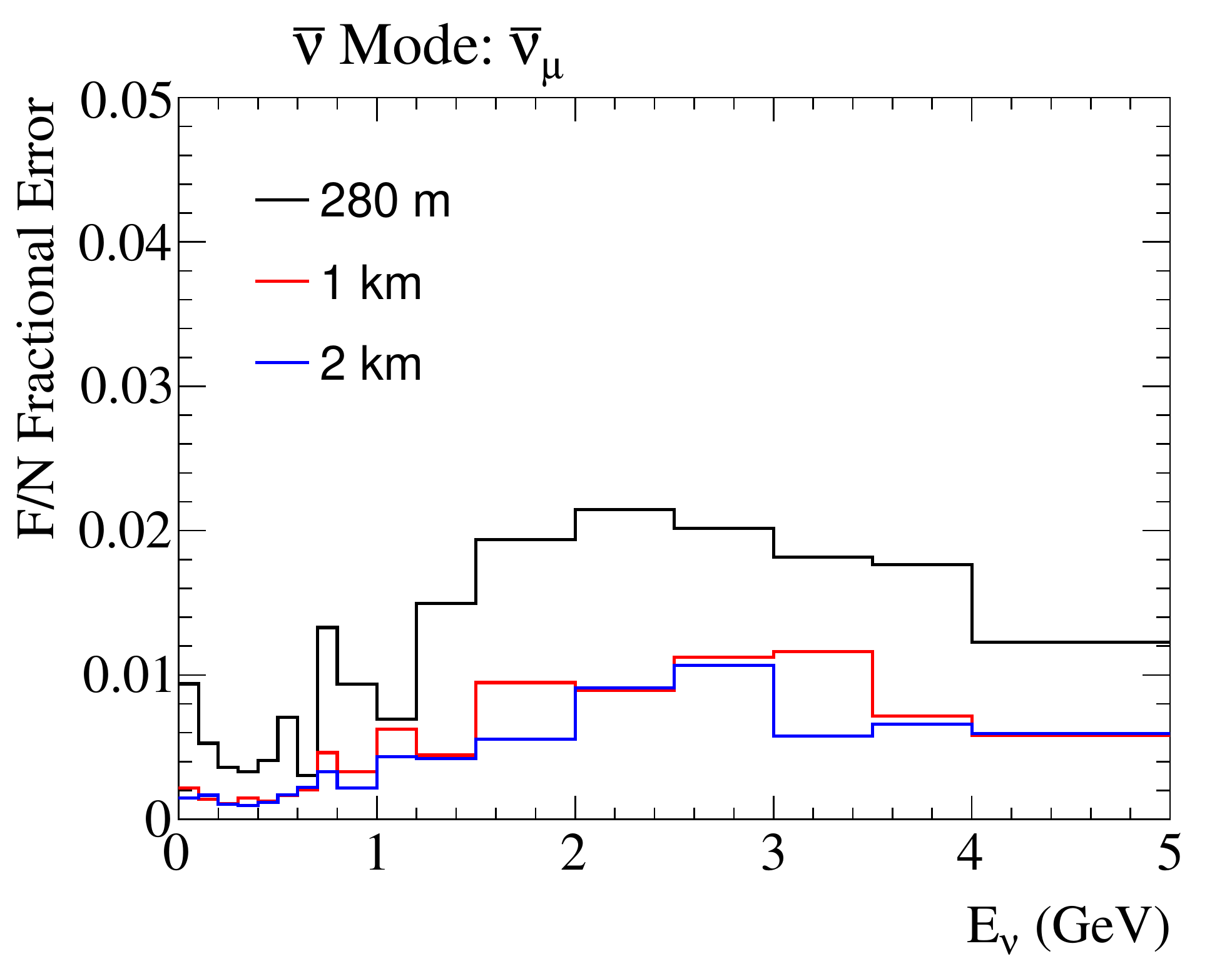}
    \includegraphics[width=0.48\textwidth]{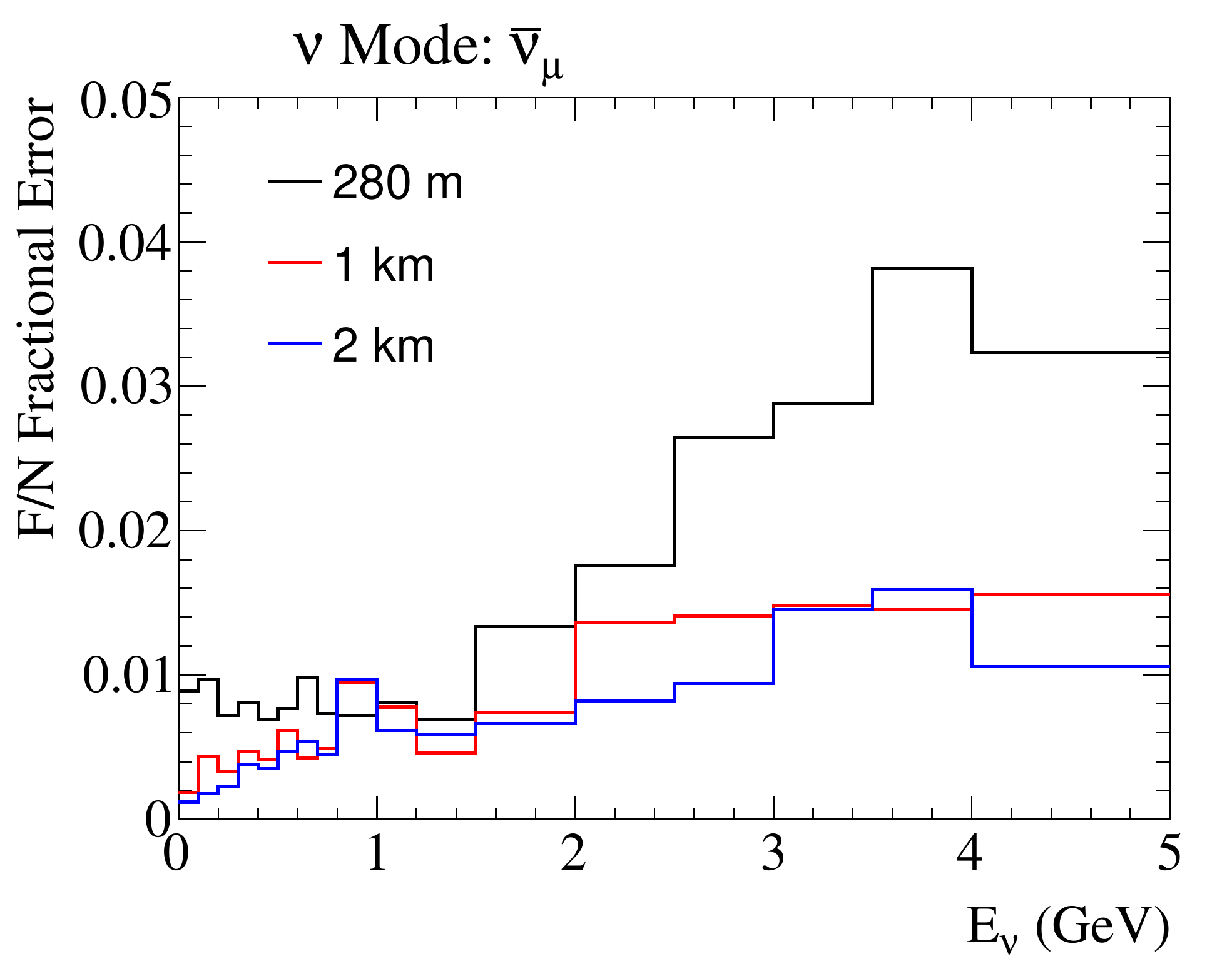}
    \includegraphics[width=0.48\textwidth]{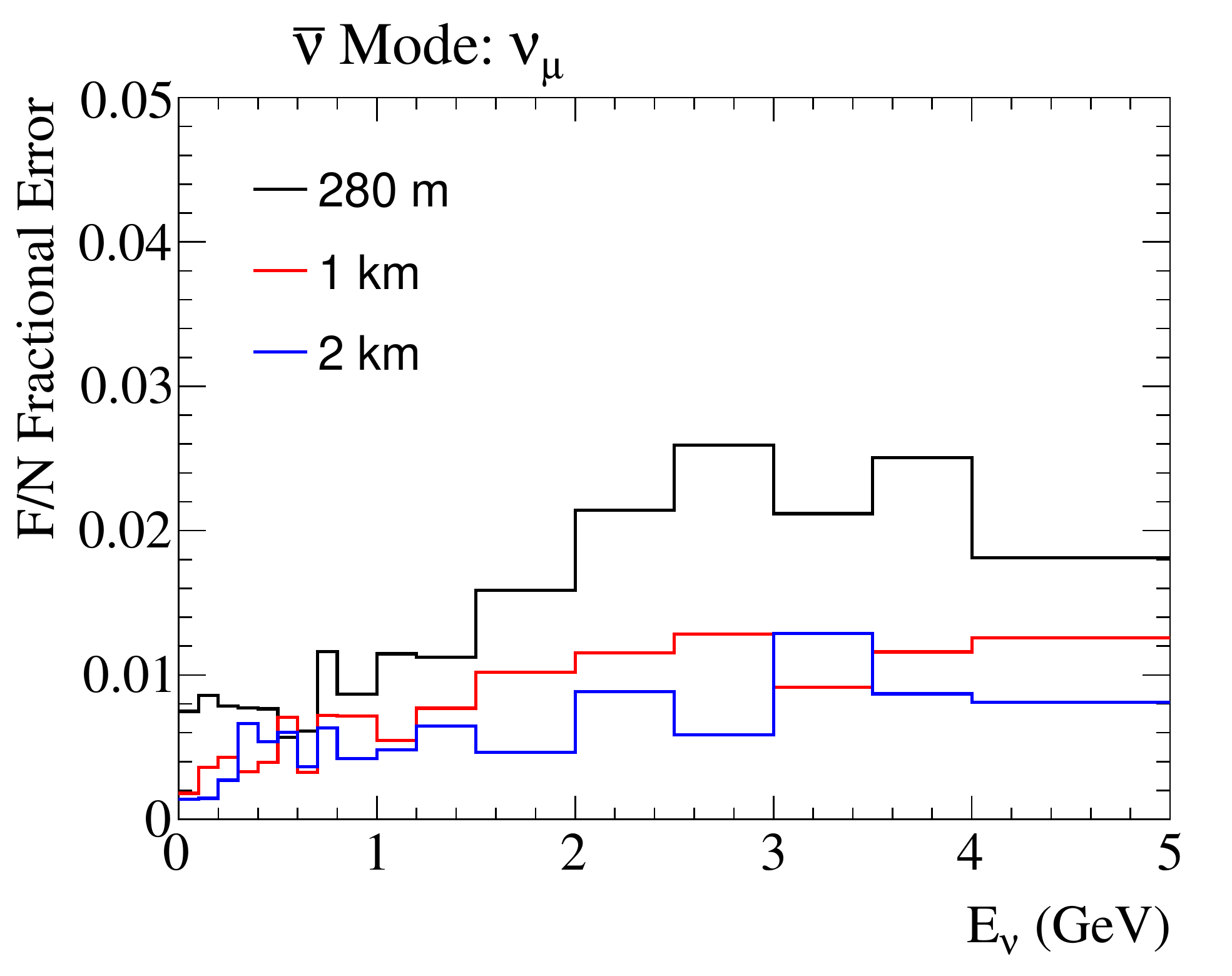}
    \caption{The uncertainty on the far-to-near flux ratio for near detectors at 280 m, 1 km and 2 km.  Left: neutrino 
enhanced beam.  Right: antineutrino enhanced beam.  Top: the focused component of the beam.  Bottom: the defocused component of
the beam.}
    \label{fig:fn_unc}
\end {figure}

The uncertainties on the hadronic interaction modeling are the largest contribution to the flux
uncertainty and may be reduced by using the hadron production data with a replica of T2K target.
A preliminary analysis using a subset of the replica target data from NA61/SHINE
has shown that it can be used 
to predict the T2K flux~\cite{Abgrall:2012pp}.  Since it is expected that replica target data will
be available for future long baseline neutrino experiments, the Hyper-K flux uncertainty is 
estimated assuming the expected uncertainties on the measurement of particle multiplicities from the replica
target.  Hence, uncertainties related to the modeling of hadronic interactions inside the target are 
no longer relevant, however, uncertainties for interactions outside of the target are considered.  
The uncertainties on the measured replica target multiplicities are estimated by applying the same uncertainties
that NA61/SHINE has reported for the thin target multiplicity measurements.

The total uncertainties on the flux as function of the neutrino energy are shown in Fig.~\ref{fig:flux_unc}.
In oscillation measurements, the predicted flux is used in combination with measurements of the neutrino
interaction rate from near detectors.  Hence, it is useful to consider the uncertainty on the ratio of the flux
at the far and near detectors: 
\begin{equation}
\delta_\mathrm{F/N}(E_{\nu}) = \delta \left (\frac{\phi_\mathrm{HK}(E_{\nu})}{\phi_\mathrm{ND}(E_{\nu})} \right )
\end{equation}
Here $\phi_\mathrm{HK}(E_{\nu})$ and $\phi_\mathrm{ND}(E_{\nu})$ are the predicted fluxes at Hyper-K and the near detector
respectively.  T2K uses the ND280 off-axis detector located 280 m from the T2K target.  At that distance,
the beam-line appears as a line source of neutrinos, compared to a point source seen by Hyper-K, and the 
far-to-near ratio is not flat.  For near detectors placed further away, at 1 or 2 km for example, the 
far-to-near flux ratio becomes more flat and there is better cancellation of the flux uncertainties 
between the near and far detectors.  Fig.~\ref{fig:fn_unc} shows how the uncertainty on the far-to-near
ratio evolves for baselines of 280 m, 1 km and 2 km.  While this extrapolation uncertainty is reduced for 
near detectors further from the production point, even the 280 m to Hyper-K uncertainty is less than $1\%$ near
the flux peak energy of 600\,MeV.

\subsection{Expected observables at Hyper-K}\label{sec:sens-events}
Interactions of neutrinos in the Hyper-K detector are simulated with the NEUT program library~\cite{hayato:neut,Mitsuka:2007zz,Mitsuka:2008zz}, which is used in both Super-K and T2K.
The response of the detector is simulated using the Super-K full Monte Carlo simulation 
based on the GEANT3 package~\cite{Brun:1994zzo}.
The simulation is based on the SK-IV configuration with the upgraded electronics and DAQ system.
Events are reconstructed with the Super-K reconstruction software.
As described in Sec.~\ref{sec:HK-detector}, the performance of Hyper-K detector for neutrinos with J-PARC beam energy is expected to be similar to that of Super-K.
Thus, the Super-K full simulation gives a realistic estimate of the Hyper-K performance.

The criteria to select \nue\ and \numu\ candidate events are based on those developed for and established with the Super-K and T2K experiments.
Fully contained (FC) events with a reconstructed vertex inside the fiducial volume (FV) and visible energy ($E_\mathrm{vis}$) greater than 30\,MeV are selected as FCFV neutrino event candidates.
In order to enhance charged current quasielastic (CCQE, $\nu_l + n \rightarrow l^- + p$ or $\overline{\nu}_l + p \rightarrow l^+ + n$) interaction, a single Cherenkov ring is required.

Assuming a CCQE interaction, the neutrino energy  ($E_\nu ^{\rm rec}$) is reconstructed from the energy of the final state charged lepton ($E_\ell$) and the angle between the neutrino beam and the charged lepton directions ($\theta_\ell$) as
\begin{eqnarray}
E_\nu ^{\rm rec}=\frac {2(m_n-V) E_\ell +m_p^2 - (m_n-V)^2 - m_\ell^2} {2(m_n-V-E_\ell+p_\ell\cos\theta_\ell)},
\label{eq:Enurec}
\end{eqnarray}
where $m_n, m_p, m_\ell$ are the mass of neutron, proton, and charged lepton, respectively, $p_\ell$ is the charged lepton momentum, and $V$ is the nuclear potential energy (27\,MeV).

\begin{figure}[tbp]%
\includegraphics[width=0.48\textwidth]{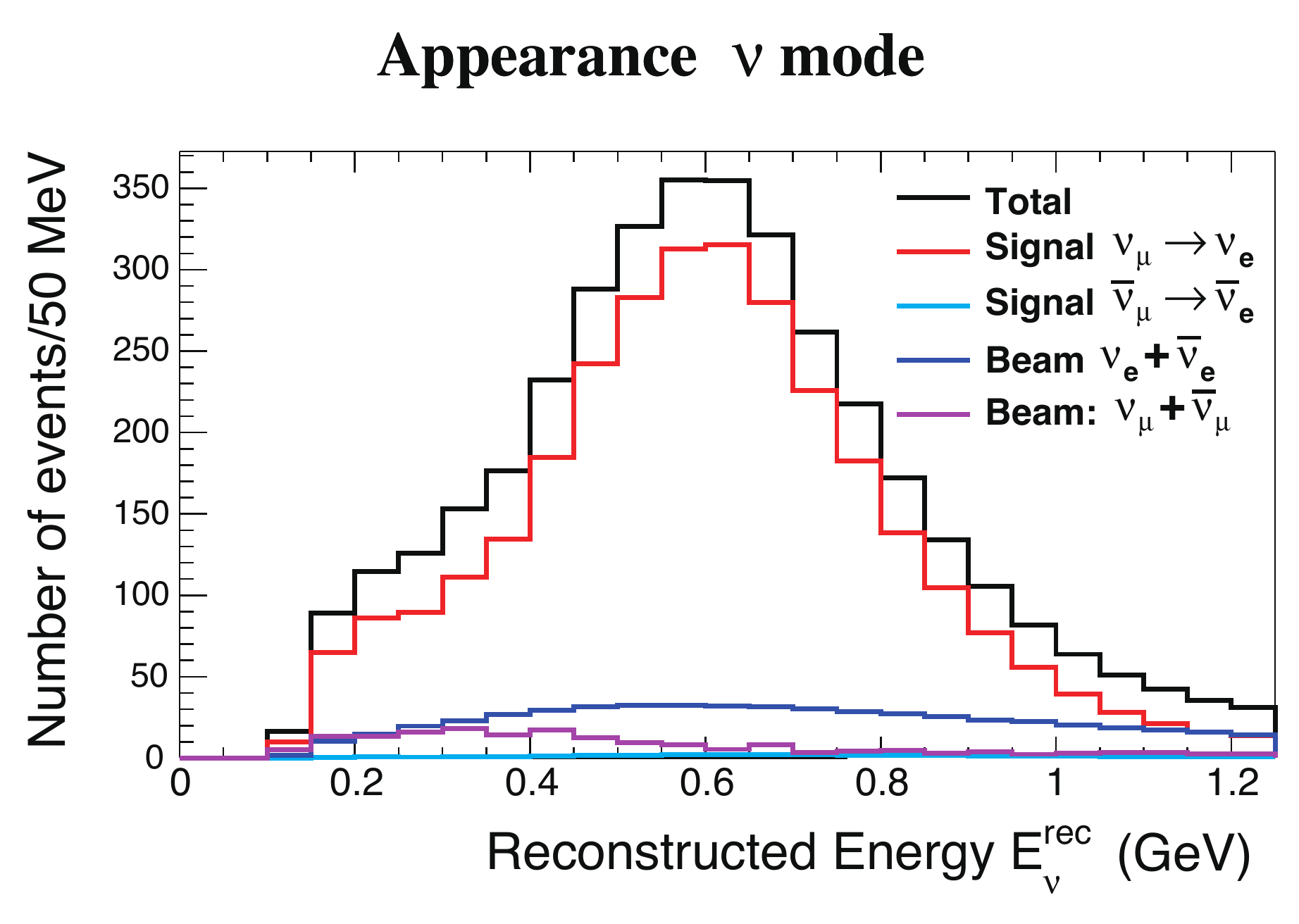}
\includegraphics[width=0.48\textwidth]{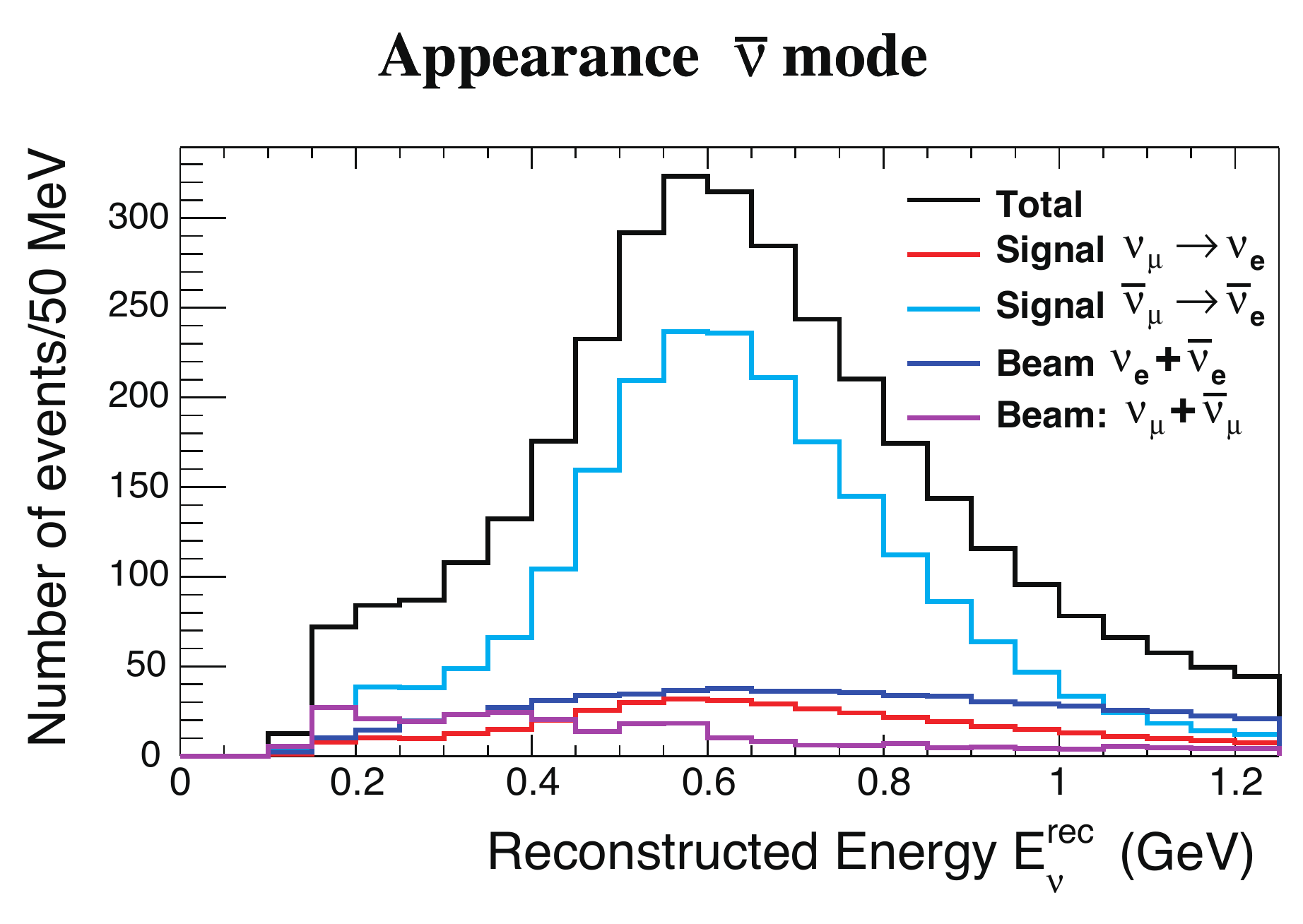}\\
\caption{
Reconstructed neutrino energy distribution of the $\nue$ candidate events. Normal mass hierarchy with $\sin^22\theta_{13}=0.1$ and $\deltacp=0$ is assumed.
\label{Fig:sens-enurec-nue}
}
\end{figure}

Then, to select \nue/\nuebar\ candidate events the following criteria are applied;
\begin{itemize}
\item The reconstructed ring is identified as electron-like ($e$-like).
\item The visible energy ($E_\mathrm{vis}$) is greater than 100\,MeV.
\item There is no decay electron associated to the event.
\item The reconstructed energy ($E_\nu^\mathrm{rec}$) is less than 1.25\,GeV.
\item In order to reduce the background from mis-reconstructed $\pi^0$ events, additional criteria using a reconstruction algorithm recently developed for T2K (fiTQun, see Sec.~\ref{sec:HK-detector}) is applied.
With a selection based on the reconstructed $\pi^0$ mass and the ratio of the best-fit likelihoods of the $\pi^0$ and electron fits as used in T2K~\cite{Abe:2013hdq}, the remaining $\pi^0$ background is reduced to about 30\% compared to the previous study~\cite{Abe:2011ts}.
\end{itemize}

\begin{table}[tbp]%
\caption{\label{Tab:sens-selection-nue}%
The expected number of \nue\ candidate events. Normal mass hierarchy with $\sin^22\theta_{13}=0.1$ and $\deltacp=0$ are assumed.
Background is categorized by the flavor before oscillation. }
\begin{center}%
\begin{tabular}{c|cc|ccccc|c|c} \hline \hline
				& \multicolumn{2}{c|}{signal} & \multicolumn{6}{c|}{BG} & \multirow{2}{*}{Total} \\ 
				&$\numu \to \nue$	& $\numubar \to \nuebar$ 	&$\numu$ CC	&$\numubar$ CC	&$\nue$  CC& $\nuebar$ CC & NC & BG Total	&  \\ \hline  
$\nu$ mode		& 3016				&	28						& 11		& 0				& 503	& 20		& 172		&	706 & 3750 \\ 
$\bar{\nu}$ mode	& 396				&	2110					& 4			& 5				& 222	& 396		& 265		&	891 &3397 \\ \hline \hline
\end{tabular}%
\end{center}
\end{table}%

Figure~\ref{Fig:sens-enurec-nue} shows the reconstructed neutrino energy distributions of $\nue$ events after all the selections.
The expected number of \nue\ candidate events is shown in Table~\ref{Tab:sens-selection-nue} for each signal and background component.
In the neutrino mode, the dominant background component is intrinsic $\nue$ contamination in the beam.
The mis-identified neutral current $\pi^0$ production events
are suppressed thanks to the improved $\pi^0$ rejection. 
In the anti-neutrino mode, in addition to $\nuebar$ and $\numubar$, $\nue$ and $\numu$ components have non-negligible contributions due to larger fluxes and cross-sections compared to their counterparts in the neutrino mode.

\begin{figure}[tbp]%
\includegraphics[width=0.48\textwidth]{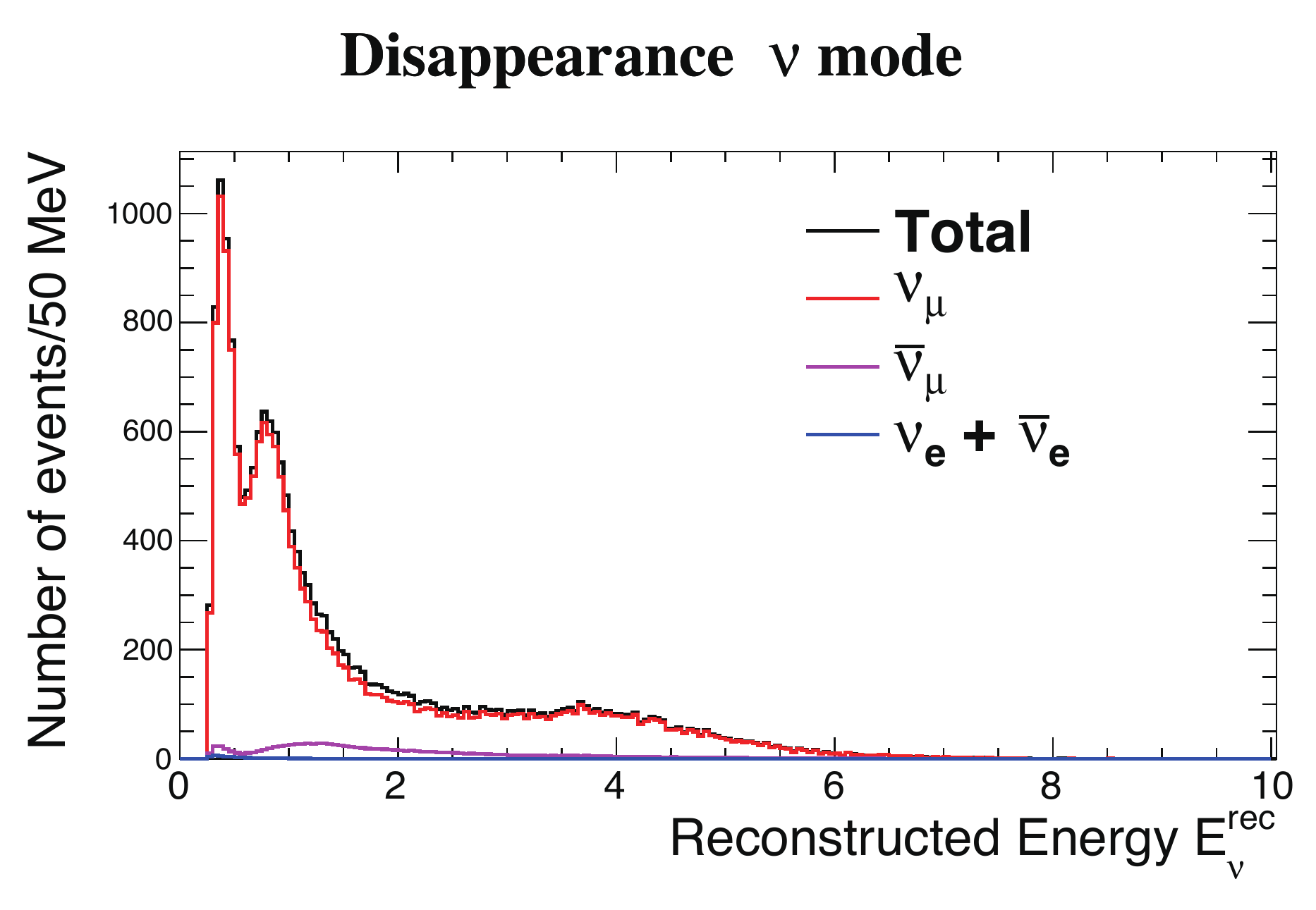}
\includegraphics[width=0.48\textwidth]{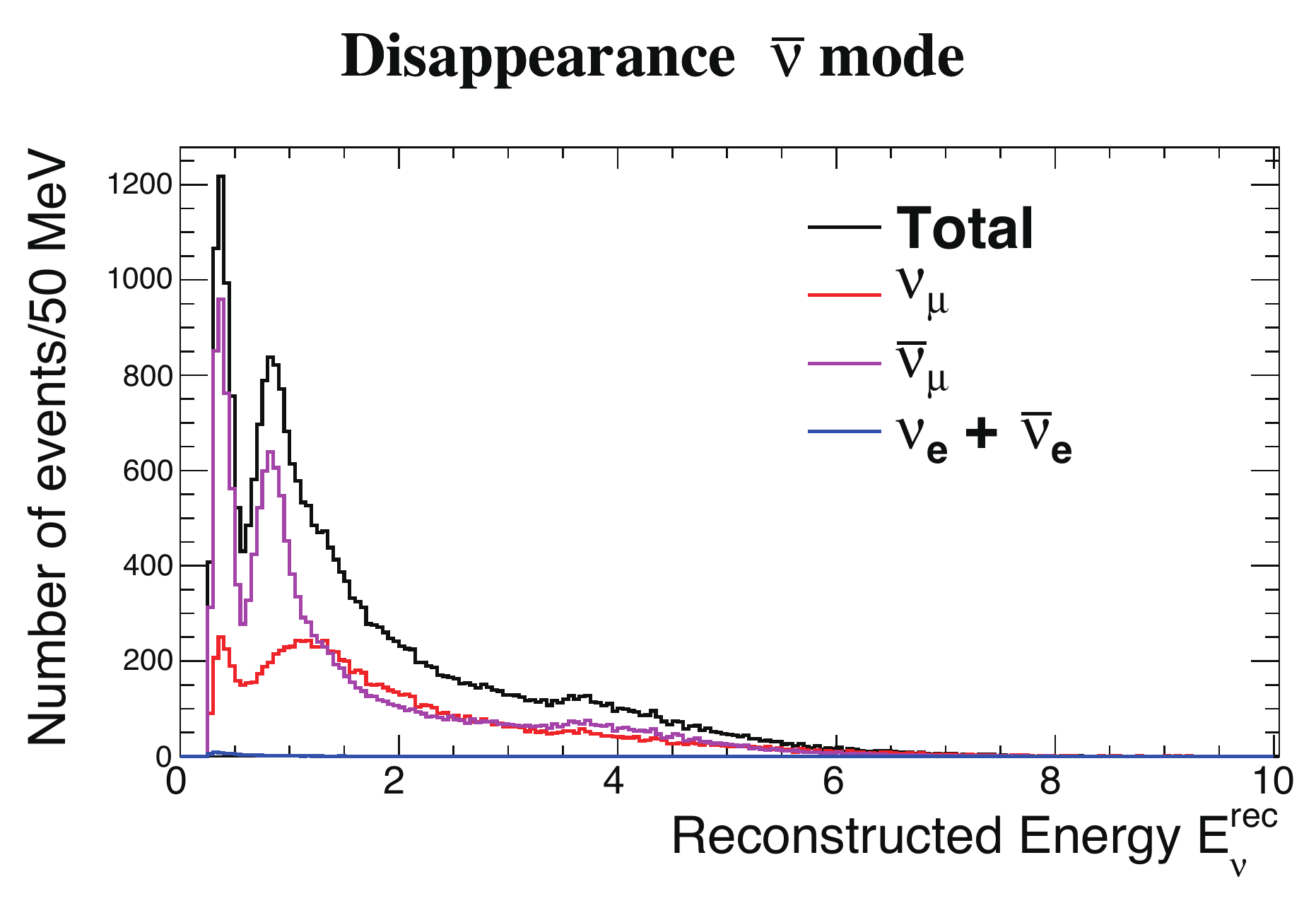}
\caption{%
Reconstructed neutrino energy distribution of the $\numu$ candidate events.
\label{Fig:sens-enurec-numu}
}
\end{figure}

\begin{table}[tbp]%
\caption{\label{Tab:sens-selection-numu}%
The expected number of $\numu$ candidate events.}
\begin{center}%
\begin{tabular}{cccccccc} \hline \hline
				&~$\numu$ CC~	& ~$\numubar$ CC~	&~$\nue$ CC~ & ~$\nuebar$ CC~ 	&~~NC~~ 	& ~$\numu \to \nue$~		& ~~total~~ 		\\ \hline 
$\nu$ mode~~		& 17225		&	1088			& 11			& 1				& 999 		& 49			& 19372		 \\ 
$\bar{\nu}$ mode~~	& 10066		&	15597			& 7				& 7				& 1281		& 6  			& 26964		 \\ \hline \hline
\end{tabular}%
\end{center}
\end{table}%

For the \numu/\numubar\ candidate events the following criteria are applied;
\begin{itemize}
\item The reconstructed ring is identified as muon-like ($\mu$-like).
\item The reconstructed muon momentum is greater than 200\,MeV/c.
\item There is at most one decay electron associated to the event.
\end{itemize}
Figure~\ref{Fig:sens-enurec-numu} shows the reconstructed neutrino energy distributions of the selected $\numu$/$\numubar$ events.
Table~\ref{Tab:sens-selection-numu} shows the number of $\numu$ candidate events for each signal and background component.
For the neutrino mode, most of the events are due to $\numu$, 
while in the anti-neutrino mode the contribution from wrong-sign $\numu$
components is significant.

\begin{figure}[tbp]
\centering
\includegraphics[width=0.48\textwidth]{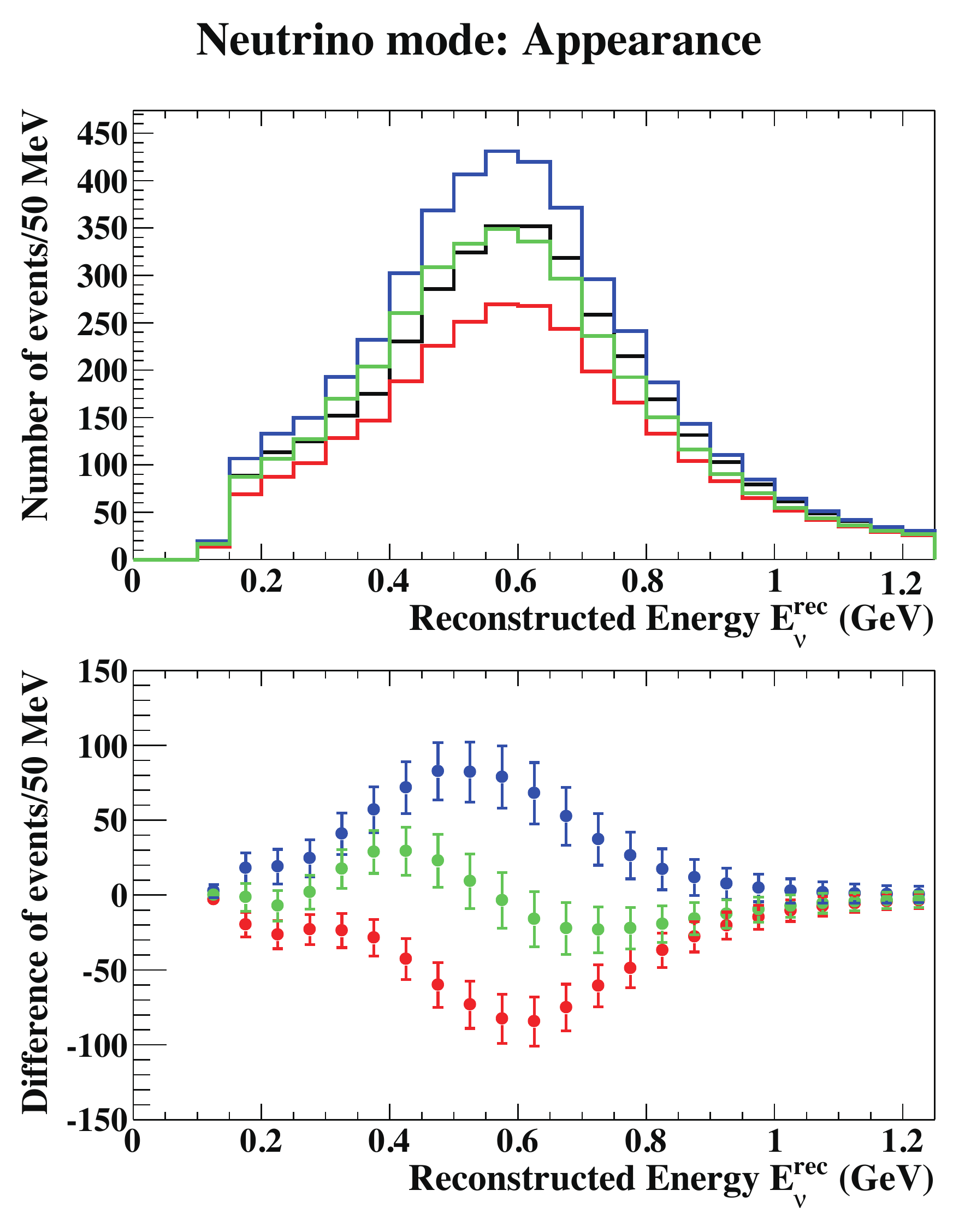}
\includegraphics[width=0.48\textwidth]{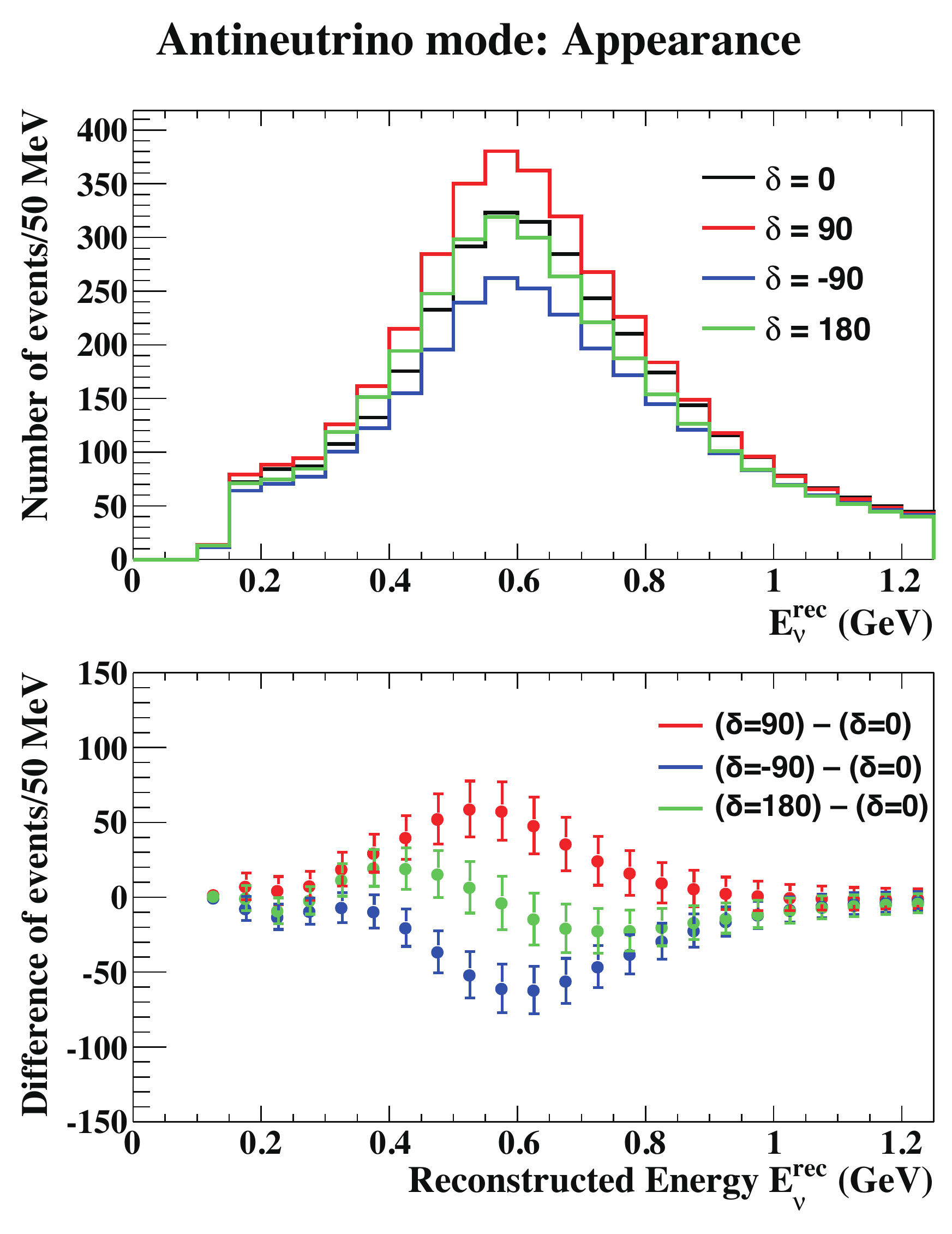}
\caption{
Top: Reconstructed neutrino energy distribution for several values of $\deltacp$.  
$\sin^22\theta_{13}=0.1$ and normal hierarchy is assumed. 
Bottom: Difference of the reconstructed neutrino energy distribution from the case with $\deltacp=0^\circ$.
The error bars represent the statistical uncertainties of each bin.
}
\label{enurecdiff-nue}
\end{figure}

\begin{figure}[tbp]
\centering
\includegraphics[width=0.48\textwidth]{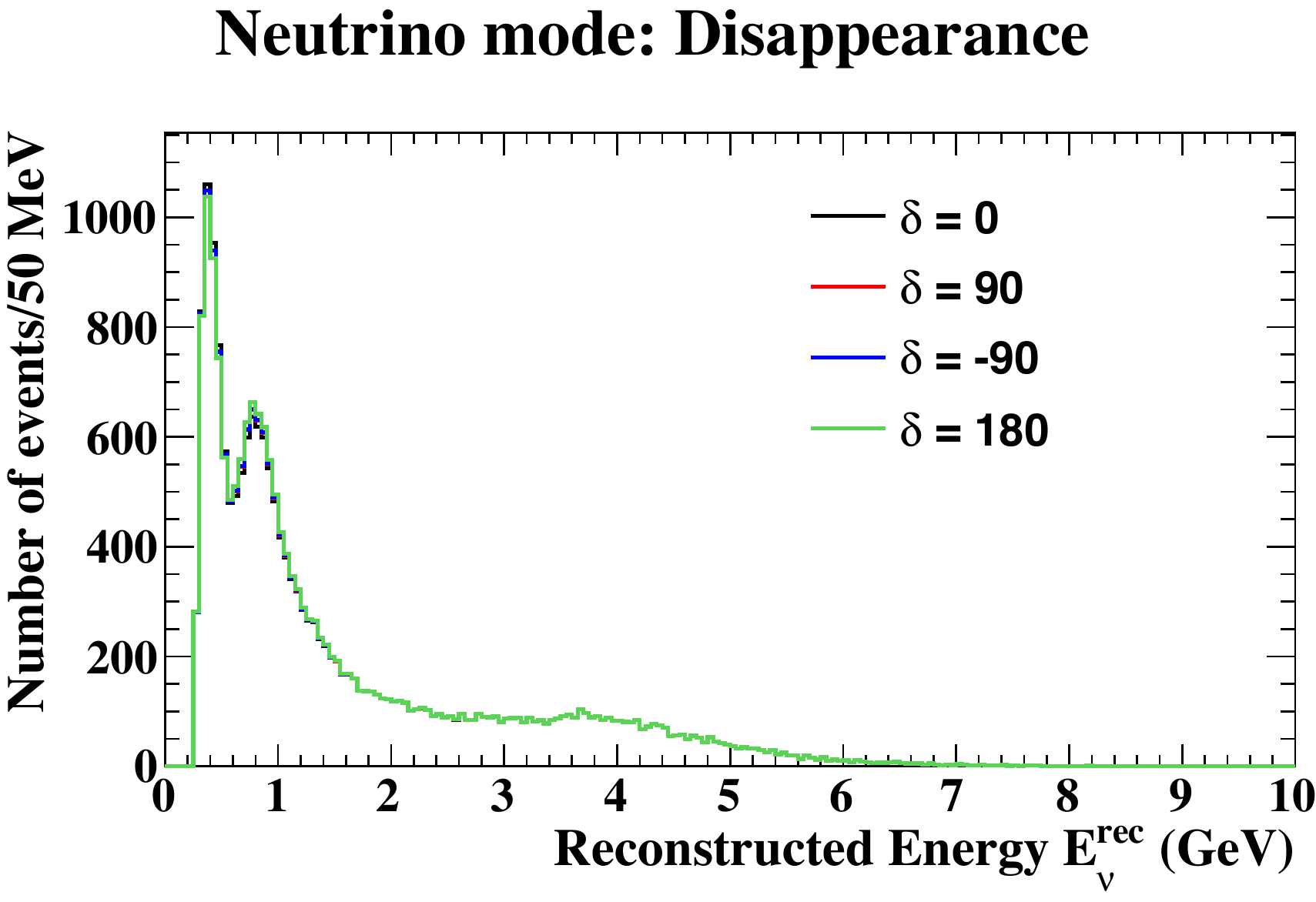}
\includegraphics[width=0.48\textwidth]{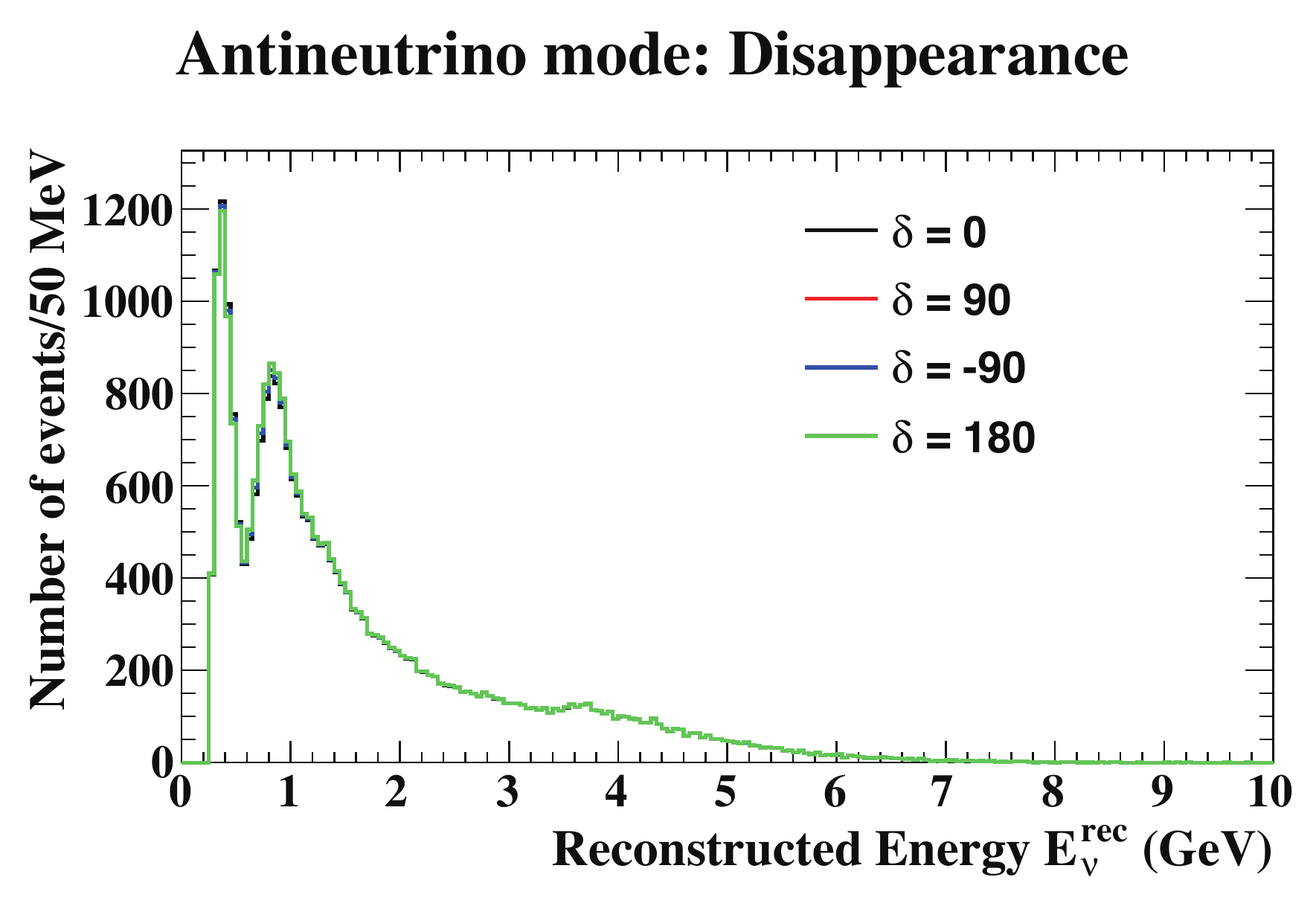}
\caption{
Reconstructed neutrino energy distribution of $\numu$ candidates for several values of $\deltacp$.
}
\label{enurecdiff-numu}
\end{figure}

The reconstructed neutrino energy distributions of $\nue$ events for several values of $\deltacp$
are shown in the top plots of Fig.~\ref{enurecdiff-nue}.
The effect of $\deltacp$ is clearly seen using the reconstructed neutrino energy.
The bottom plots show the difference of reconstructed energy spectrum from $\deltacp=0^\circ$ for the cases $\delta = 90^\circ, -90^\circ$ and $180^\circ$.
The error bars correspond to the statistical uncertainty.
By using not only the total number of events but also the reconstructed energy distribution,
the sensitivity to $\deltacp$ can be improved, 
and one can discriminate all the values of $\deltacp$, including the difference between $\deltacp = 0$ and $\pi$.
Figure~\ref{enurecdiff-numu} shows the reconstructed neutrino energy
distributions of the $\numu$ sample for several values of $\deltacp$.
As expected the difference is very small for $\numu$ events.

\subsection{Analysis method}
The sensitivity of a long baseline experiment using Hyper-K and J-PARC neutrino beam is studied using a binned likelihood analysis based on the reconstructed neutrino energy distribution.
Both \nue\ appearance and \numu\ disappearance samples, in both neutrino and
antineutrino runs, are simultaneously fitted.

The $\chi^2$ used in this study is defined as 
\begin{equation} \label{eq:sens:chi2}
\chi^2 =  -2 \ln \mathcal{L}  + P,
\end{equation}
where $\ln \mathcal{L}$ is the log likelihood for a Poisson distribution,
\begin{equation}
-2\ln \mathcal{L} = \sum_k \left\{ -{N_k^\mathrm{test}(1+f_i)} + N_k^\mathrm{true} \ln \left[ N_k^\mathrm{test}(1+f_i) \right] \right\}.
\end{equation}
Here, $N_k^\mathrm{true}$ ($N_k^\mathrm{test}$) is the number of events in $k$-th reconstructed energy bin for the true (test) oscillation parameters.
The index $k$ runs over all reconstructed energy bins for
muon and electron neutrino samples and for neutrino and anti-neutrino mode running.
The parameters $f_i$ represent systematic uncertainties.
For anti-neutrino mode samples, an additional overall normalization parameter
with 6\% prior uncertainty is introduced to account for a possible uncertainty in the anti-neutrino interaction, which is less known experimentally in this energy region.
This additional uncertainty is expected to decrease as we accumulate and analyze more anti-neutrino data in T2K, but we conservatively assign the current estimate for this study.
A normalization weight $(1+f^{\overline{\nu}}_\mathrm{norm})$ is multiplied to $N_k^\mathrm{test}$ in the anti-neutrino mode samples.

The penalty term $P$ in Eq.~\ref{eq:sens:chi2} constrains the systematic parameters $f_i$ with the normalized covariance matrix $C$,
\begin{equation}
P = \sum_{i,j} f_i (C^{-1})_{i,j} f_j.
\end{equation}
In order to reduce the number of the systematic parameters, 
several reconstructed energy bins that have similar covariance values are merged for $f_i$.

The size of systematic uncertainty is evaluated based on the experience and prospects of the T2K experiment, as it provides the most realistic estimate as the baseline.
We estimate the systematic uncertainties assuming the T2K neutrino beamline and
near detectors, taking into account improvements expected with future T2K
running and analysis improvements.
For Hyper-K a further reduction of systematic uncertainties will be possible
with upgrade of beamline and near detectors, improvements in detector calibration and analysis techniques, and improved understanding of neutrino interaction with more measurements.
In particular, as described in Sec.~\ref{sec:ND}, studies of near detectors are ongoing with a goal of further reducing systematic uncertainties.
The sensitivity update is expected in the near future as the near detector design studies advance.

There are three main categories of systematic uncertainties. We assume improvement from the current T2K uncertainties for each category as follows.
\begin{description}
\item[i) Flux and cross section uncertainties constrained by the fit to current near detector data: ] 
These arise from systematics of the near detectors. 
The understanding of the detector will improve in the future, but this category of
uncertainties is conservatively assumed to stay at the same level as currently estimated.
\item[ii) Cross section uncertainties that are not constrained by the fit to current near detector data: ]
Errors in this category will be reduced as more categories of samples are added to the near detector data fit, which constrains the cross section models.
We assume the uncertainties arising from different target nucleus between the near and the far detectors will become negligible by including the measurement with the water target in the near detector.
\item[iii) Uncertainties on the far detector efficiency and reconstruction modeling: ]
Because most of them are estimated by using atmospheric neutrinos as a control sample, errors in this category are expected to decrease with more than an order of magnitude larger statistics available with Hyper-K than currently used for T2K.
Uncertainties arising from the energy scale is kept the same because it is not estimated by the atmospheric neutrino sample.
\end{description}
The flux and cross section uncertainties are assumed to be uncorrelated between the neutrino and anti-neutrino running, except for the uncertainty of \nue/\numu\ cross section ratio which is treated to be anti-correlated considering the theoretical uncertainties studied in~\cite{Day:2012gb}.
Because some of the uncertainties, such as those from the cross section
modeling or near detector systematics, are expected to be correlated and give
more of a constraint, this is a conservative assumption.
The far detector uncertainty is treated to be fully correlated between the neutrino and anti-neutrino running.

\begin{figure}[tbp]
\centering
\includegraphics[width=0.48\textwidth]{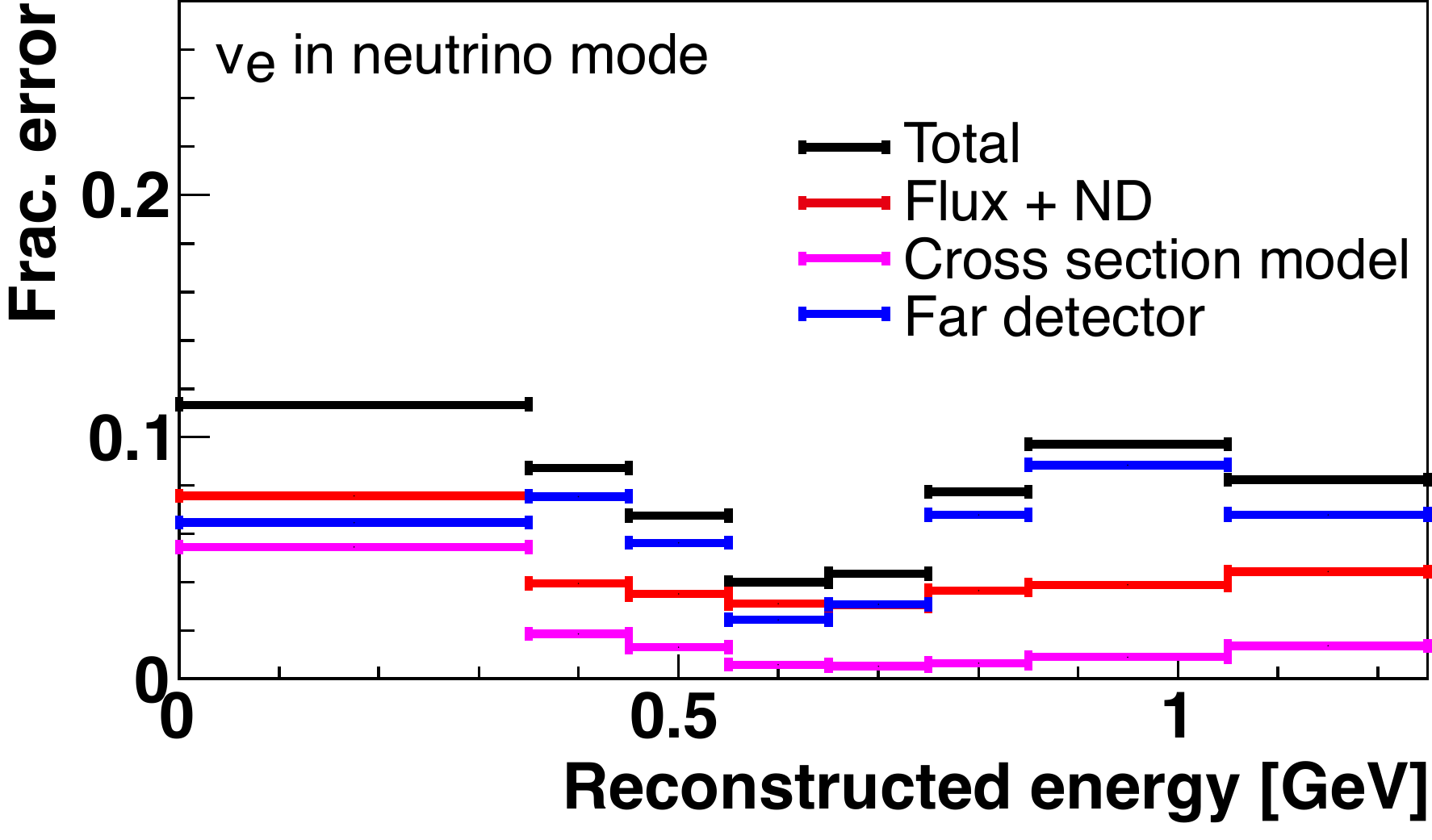}
\includegraphics[width=0.48\textwidth]{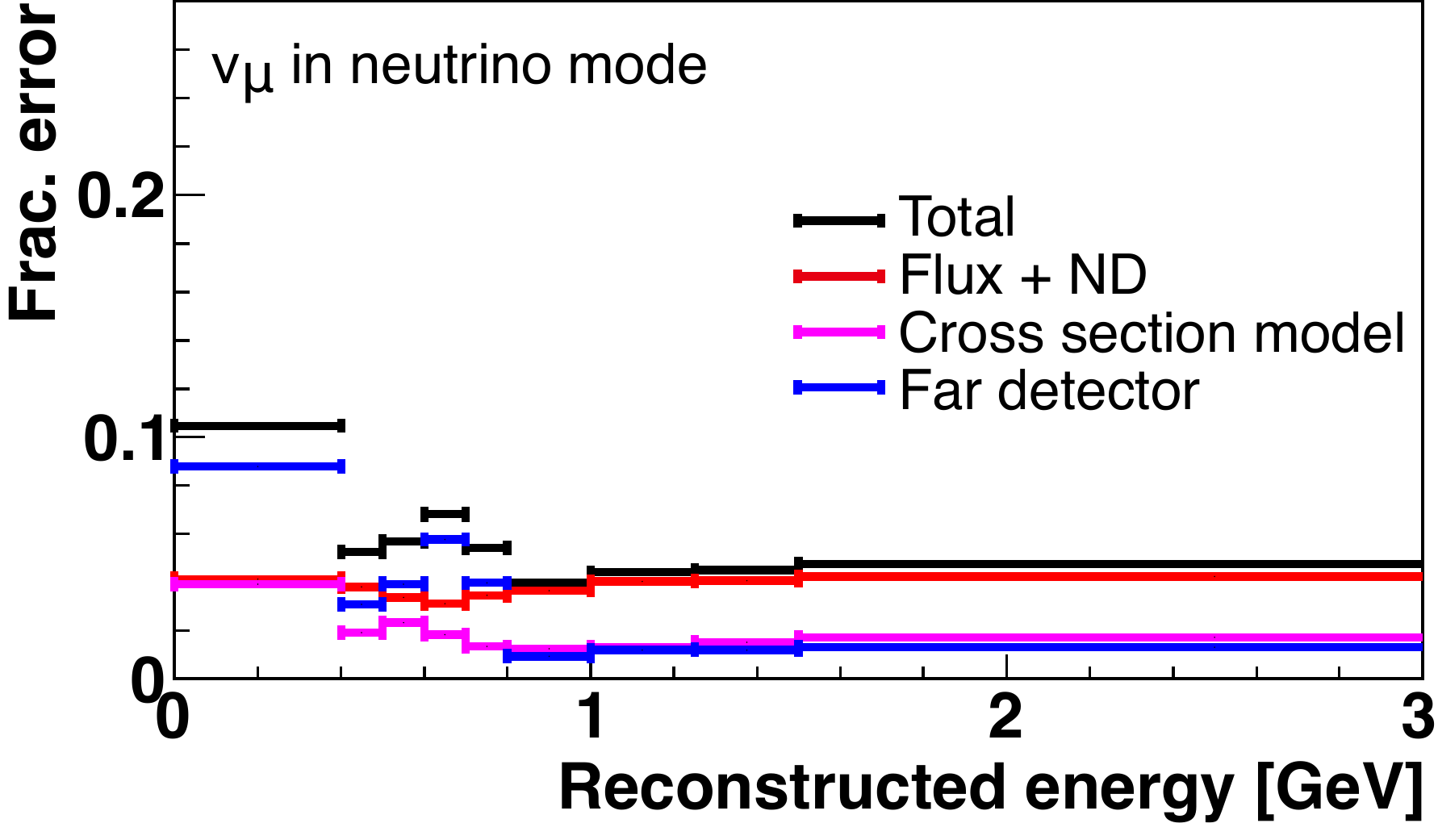}
\caption{
Fractional error size for the appearance (left) and the disappearance (right)  samples in the neutrino mode.
Black: total uncertainty, red: the flux and cross-section constrained by the near detector, 
magenta: the near detector non-constrained cross section,
blue: the far detector error.
\label{Fig:systerror}
}
\end{figure}

\begin{figure}[tbp]
\centering
\includegraphics[width=0.48\textwidth]{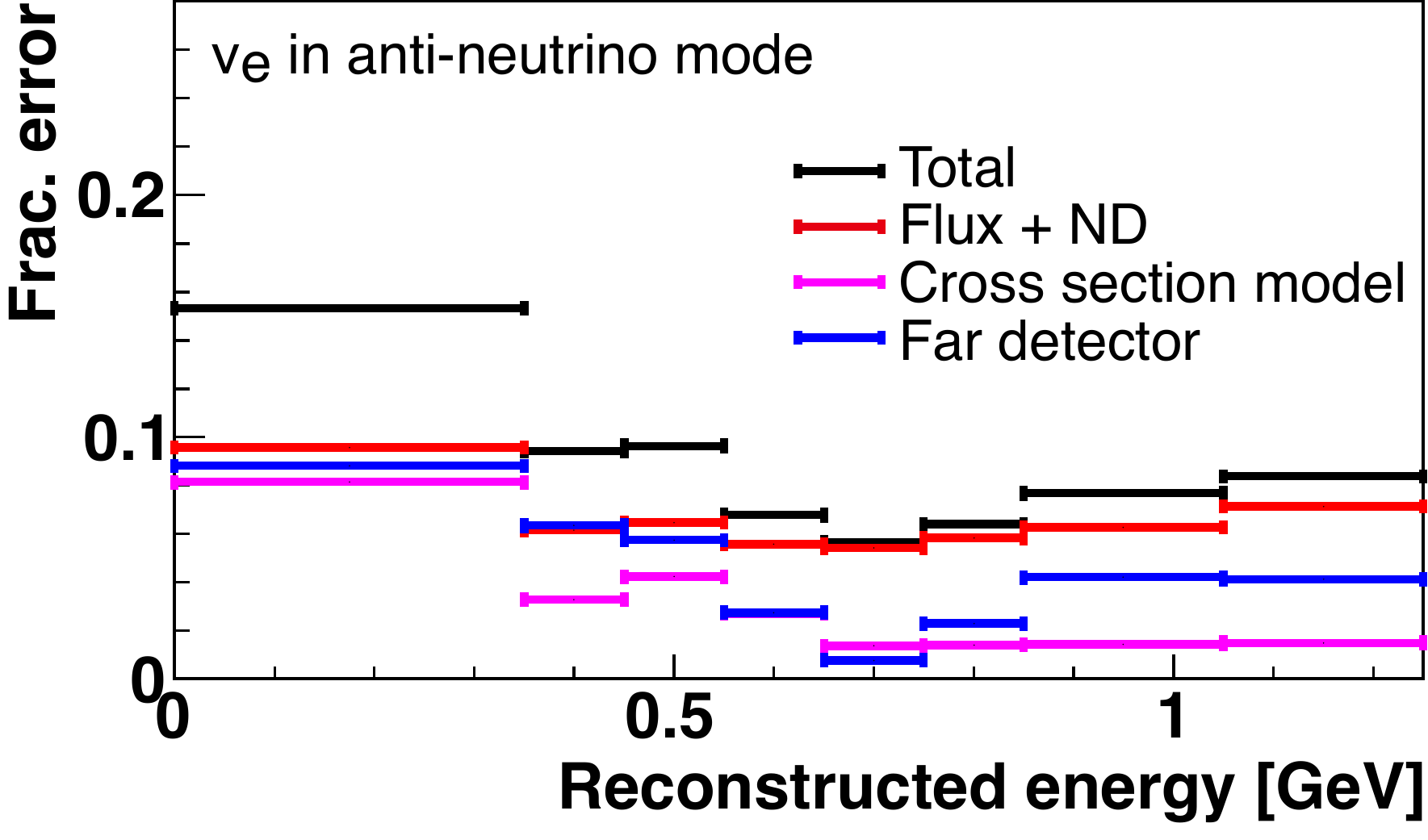}
\includegraphics[width=0.48\textwidth]{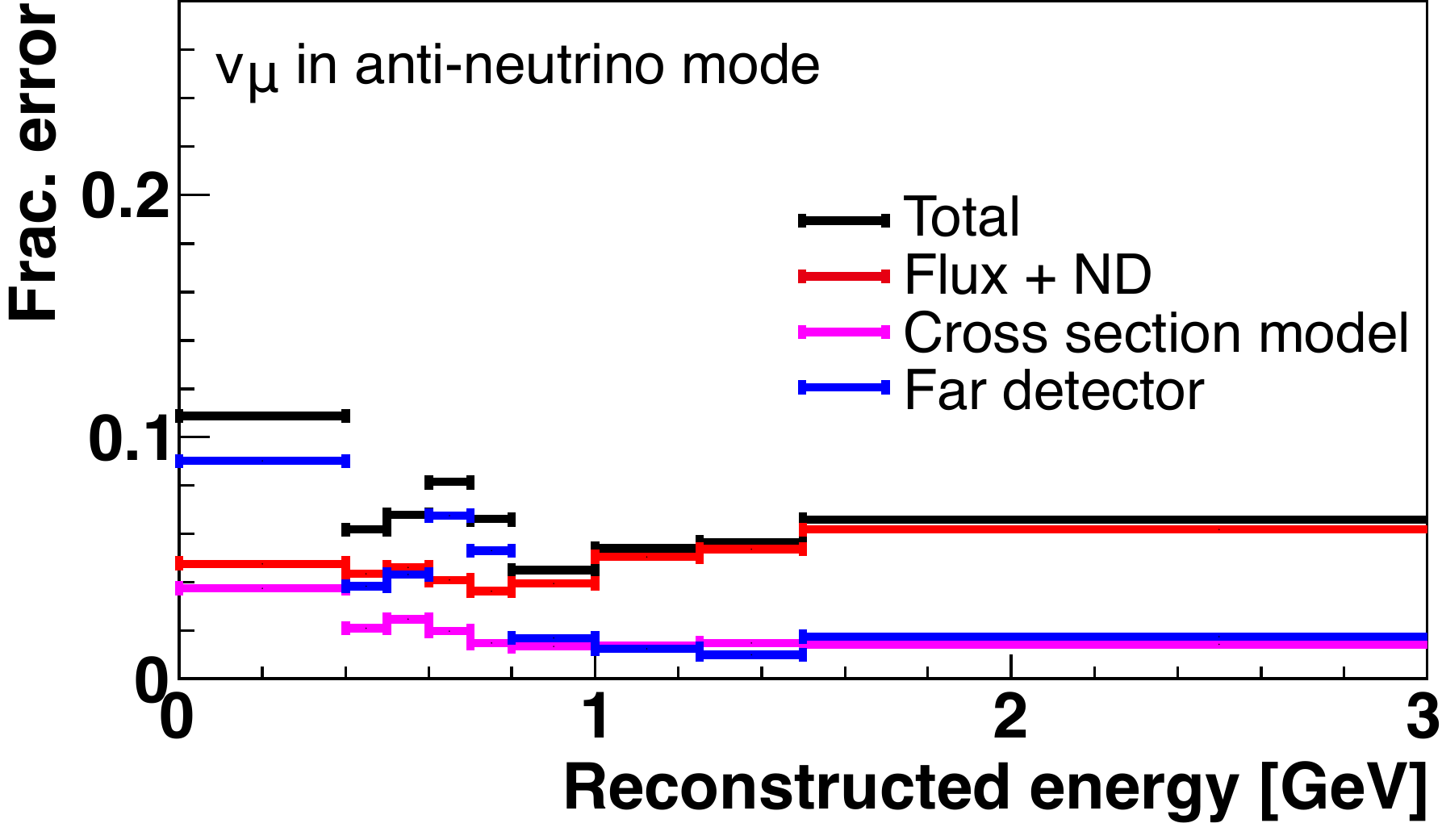}
\caption{
Fractional error size for the appearance (left) and the disappearance (right)  samples in the anti-neutrino mode.
Black: total uncertainty, red: the flux and cross-section constrained by the near detector, 
magenta: the near detector non-constrained cross section,
blue: the far detector error.
\label{Fig:systerror-anti}
}
\end{figure}

\begin{figure}[tbp]
\centering
\includegraphics[width=0.75\textwidth]{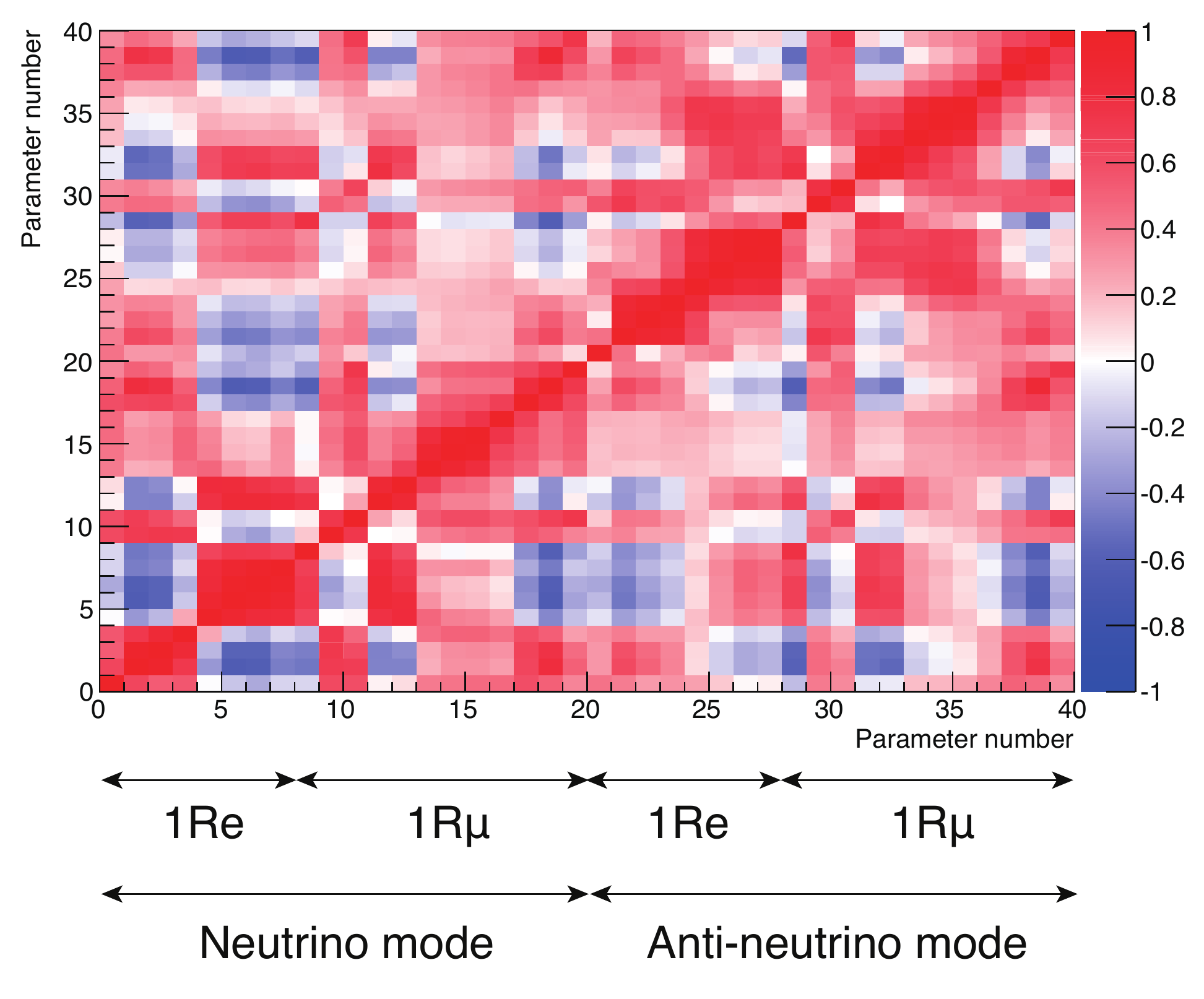}
\caption{
Correlation matrix between reconstructed energy bins of the four samples due to the systematic uncertainties.
Bins 1--8, 9--20, 21--28, and 29--40 correspond to 
the neutrino mode single ring $e$-like,
the neutrino mode single ring $\mu$-like,
the anti-neutrino mode single ring $e$-like, and
the anti-neutrino mode single ring $\mu$-like samples, respectively.
\label{Fig:correlationmatrix}
}
\end{figure}

Figures~\ref{Fig:systerror} and \ref{Fig:systerror-anti} show the fractional systematic uncertainties for the appearance and disappearance reconstructed energy spectra in neutrino and anti-neutrino mode, respectively.
Black lines represent the prior uncertainties and bin widths of the systematic parameters $f_i$,
while colored lines show the contribution from each uncertainty source.
Figure~\ref{Fig:correlationmatrix} shows the correlation matrix of the systematic uncertainties between the reconstructed neutrino energy bins of the four samples.
The systematic uncertainties (in \%) of the number of expected events at the far detector are summarized in Table~\ref{tab:sens:systsummary}.

\begin{table}[htdp]
\caption{Uncertainties (in \%) for the expected number of events at Hyper-K from the systematic uncertainties assumed in this study.}
\centering
\begin{tabular}{cccccc}  \hline \hline
&  & Flux \& ND-constrained & ND-independent  & \multirow{2}{*}{Far detector}  & \multirow{2}{*}{Total} \\
 & &   cross section &  cross section \\ \hline
\multirow{2}{*}{$\nu$ mode} & Appearance & 3.0 & 1.2 & 0.7 & 3.3 \\
 & Disappearance & 2.8 & 1.5 & 1.0 & 3.3 \\ \hline
\multirow{2}{*}{$\overline{\nu}$ mode} & Appearance & 5.6 & 2.0 & 1.7 & 6.2 \\
 & Disappearance & 4.2 & 1.4 & 1.1 & 4.5 \\
\hline \hline
\end{tabular}

\label{tab:sens:systsummary}
\end{table}%

\subsection{Expected sensitivity to CP violation}

\begin{figure}[tbp]
\centering
\includegraphics[width=1.0\textwidth]{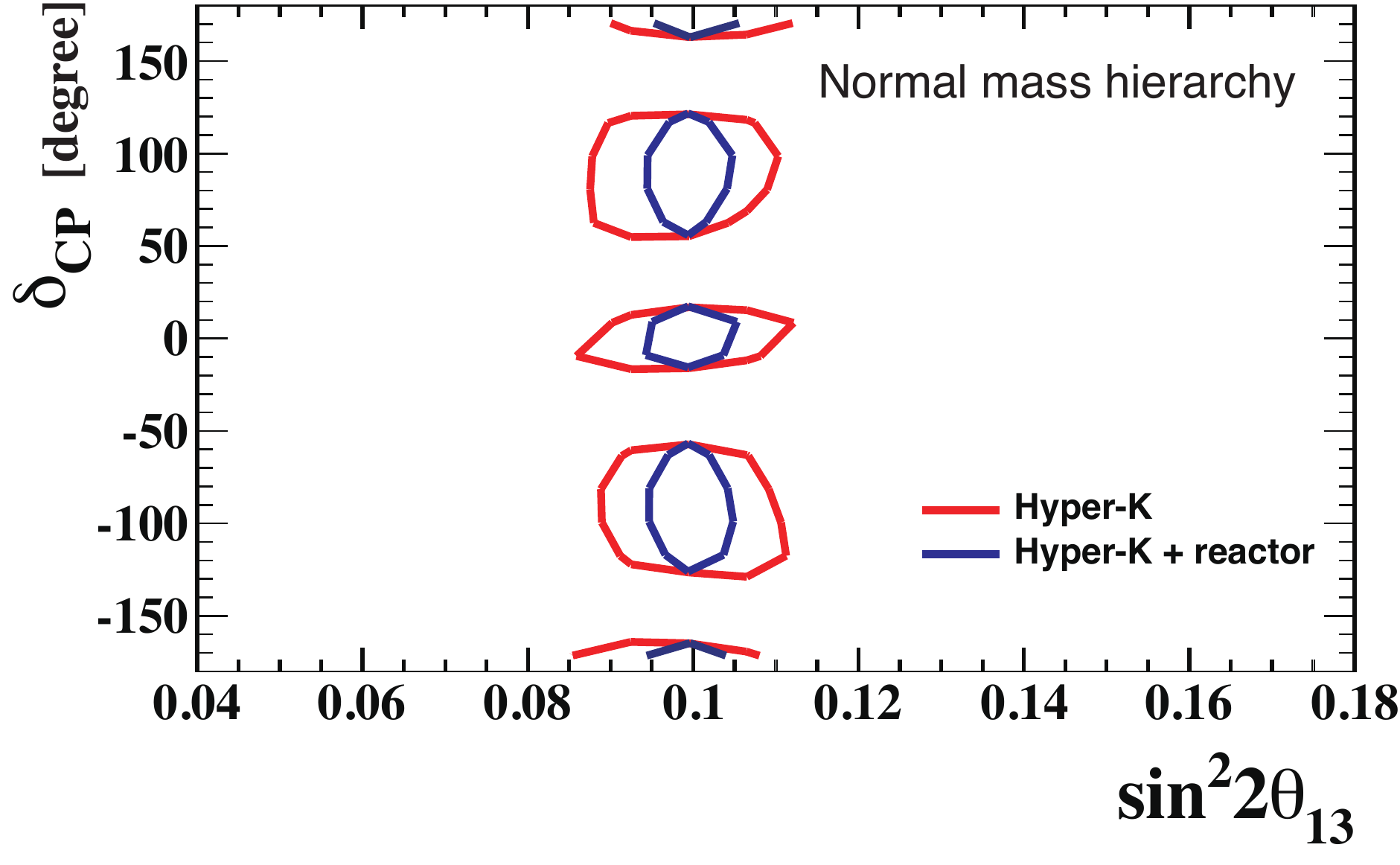}
\includegraphics[width=1.0\textwidth]{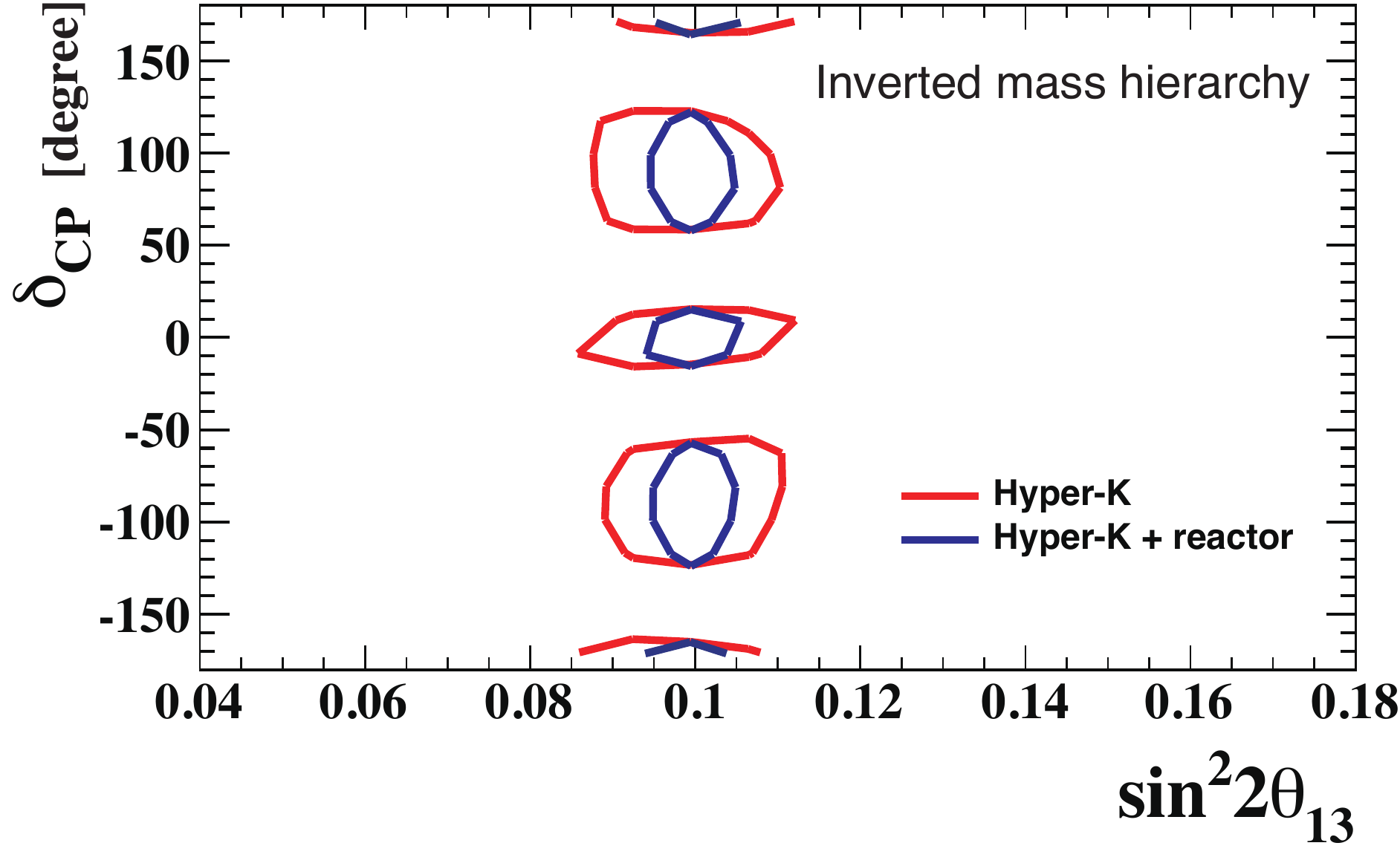}
\caption{The 90\% CL allowed regions in the $\sin^22\theta_{13}$-$\deltacp$ plane.
The results for the true values of $\deltacp = (-90^\circ, 0, 90^\circ, 180^\circ)$ are overlaid.
Top: normal hierarchy case. Bottom: inverted hierarchy case.
Red (blue) lines show the result with Hyper-K only (with $\sin^22\theta_{13}$ constraint from reactor experiments).
\label{fig:CP-contour}}
\end{figure}

\begin{figure}[tbp]
\centering
\includegraphics[width=1.0\textwidth]{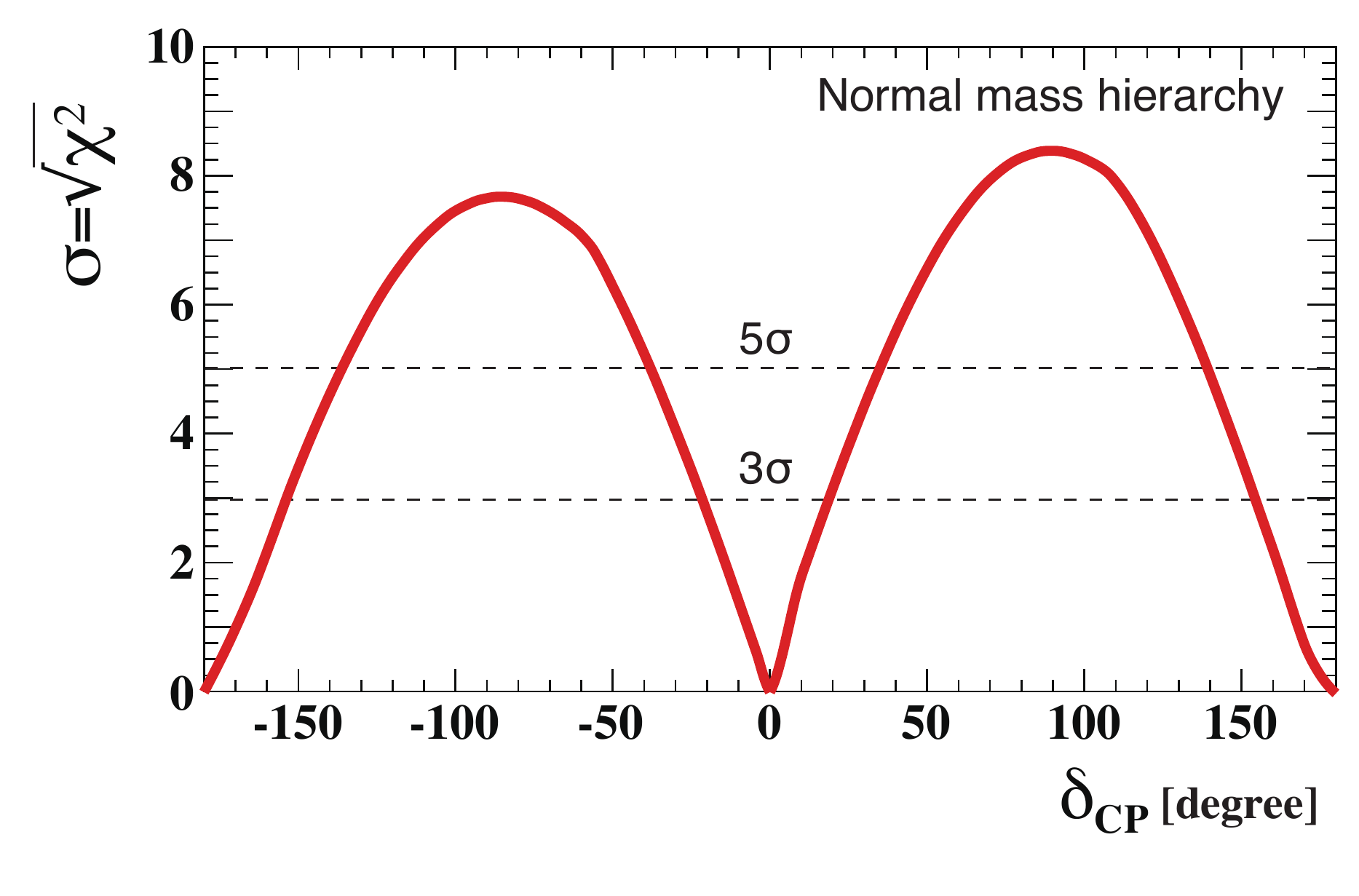}
\includegraphics[width=1.0\textwidth]{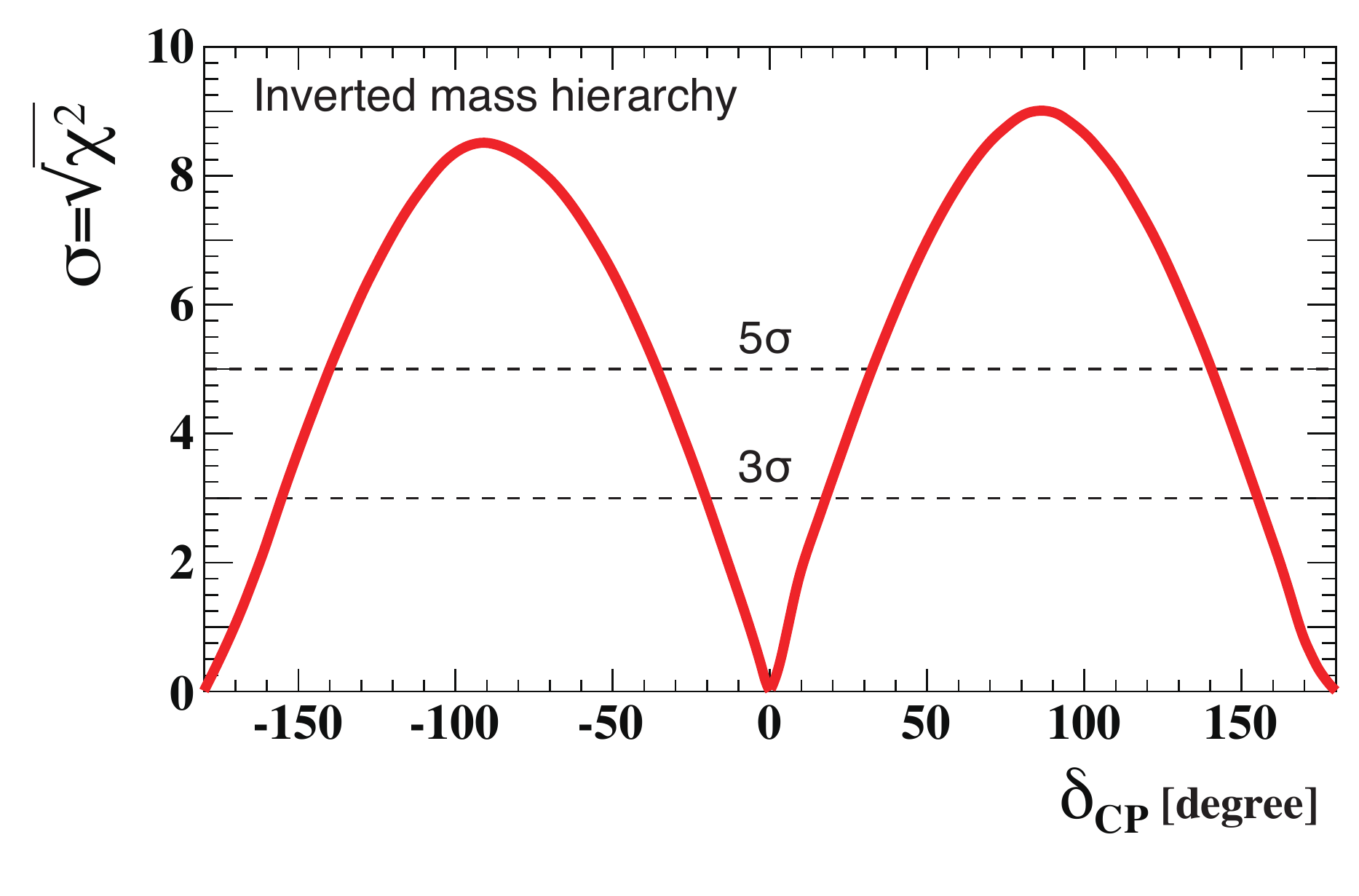}
\caption{Expected significance to exclude $\sin\deltacp = 0$.
Top: normal hierarchy case. Bottom: inverted hierarchy case.
\label{fig:CP-chi2}}
\end{figure}

\begin{figure}[tbp]
\centering
\includegraphics[width=0.9\textwidth]{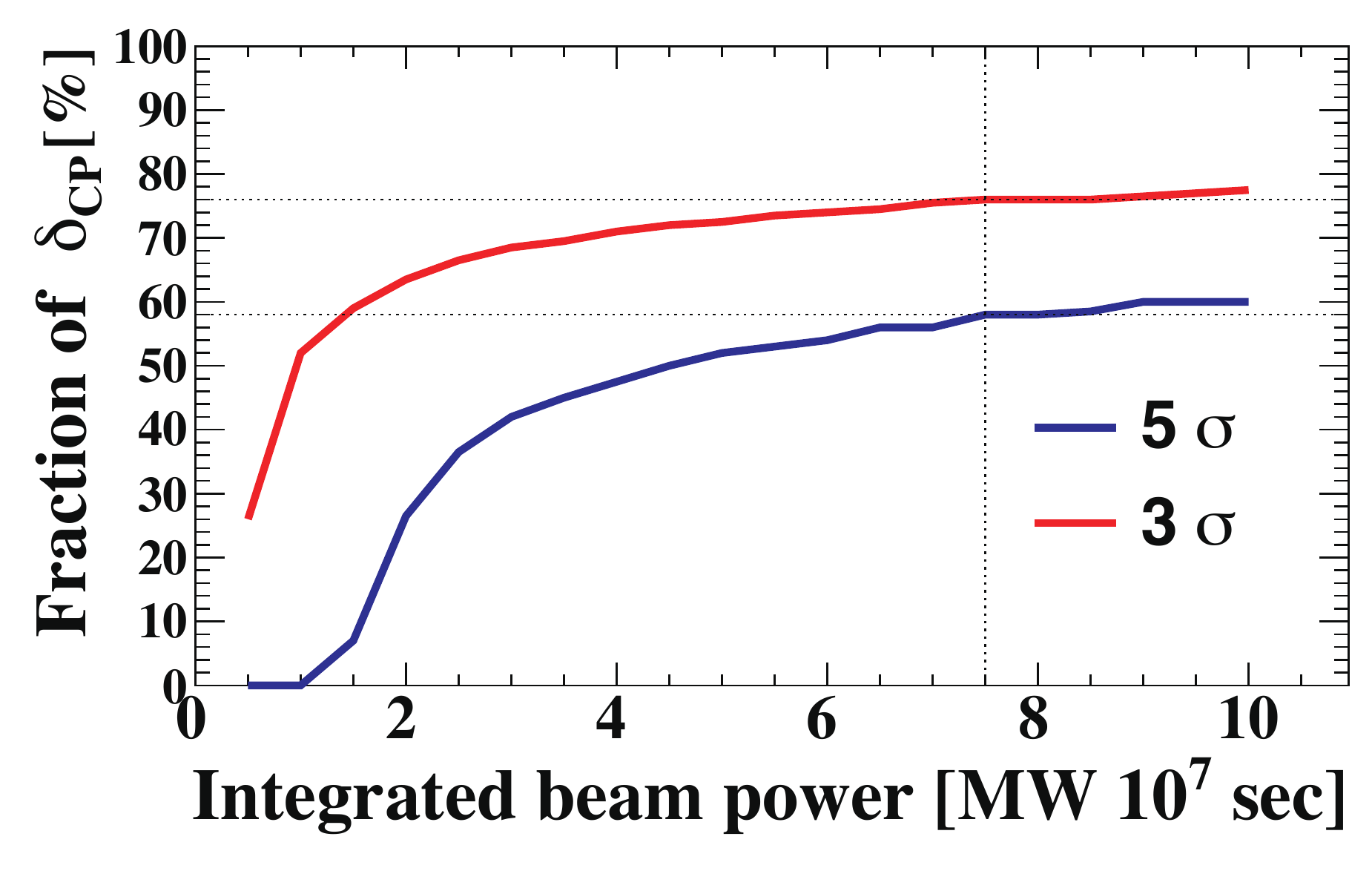}
\caption{Fraction of $\deltacp$ for which $\sin\deltacp= 0$ can be excluded with  more than 3\,$\sigma$ (red) and 5\,$\sigma$ (blue) significance as a function of the integrated beam power. For the normal hierarchy case.
The ratio of neutrino and anti-neutrino mode is fixed to 1:3. 
\label{fig:delta-sens-time}}
\end{figure}

Figure~\ref{fig:CP-contour} shows the 90\% CL allowed regions on the $\sin^22\theta_{13}$-$\deltacp$ plane.
The results for the true values of $\deltacp = (-90^\circ, 0, 90^\circ, 180^\circ)$ are overlaid.
The top (bottom) plot shows the case for the normal (inverted) mass hierarchy.
Also shown are the allowed regions when we include a constraint from the reactor experiments, $\sin^22\theta_{13}=0.100 \pm0.005$.
With reactor constraints, although the contour becomes narrower in the direction of $\sin^22\theta_{13}$, the sensitivity to $\deltacp$ does not significantly change because $\delta_{CP}$ is constrained by the comparison of neutrino and anti-neutrino oscillation probabilities by Hyper-K and not limited by the uncertainty of $\theta_{13}$.

Figure~\ref{fig:CP-chi2} shows the expected significance to exclude $\sin\deltacp = 0$ (the $CP$ conserved case).
The significance is calculated as $\sqrt{\Delta \chi^2}$, where $\Delta \chi^2$ is the difference of $\chi^2$ for the \textit{trial} value of \deltacp\ and for $\deltacp = 0^\circ$ or 180$^\circ$ (the smaller value of difference is taken).
We have also studied the case with a reactor constraint, but the result changes only slightly.
Figure~\ref{fig:delta-sens-time} shows the fraction of $\deltacp$ for which $\sin\deltacp= 0$ is excluded with more than 3\,$\sigma$ and 5\,$\sigma$ of significance as a function of the integrated beam power.
The ratio of integrated beam power for the neutrino and anti-neutrino mode is fixed to 1:3.
The normal mass hierarchy is assumed.
The results for the inverted hierarchy is almost the same.
$CP$ violation in the lepton sector can be observed with more than 3(5)\,$\sigma$ significance for 76(58)\% of the possible values of $\deltacp$.

\begin{figure}[tbp]
\centering
\includegraphics[width=0.9\textwidth]{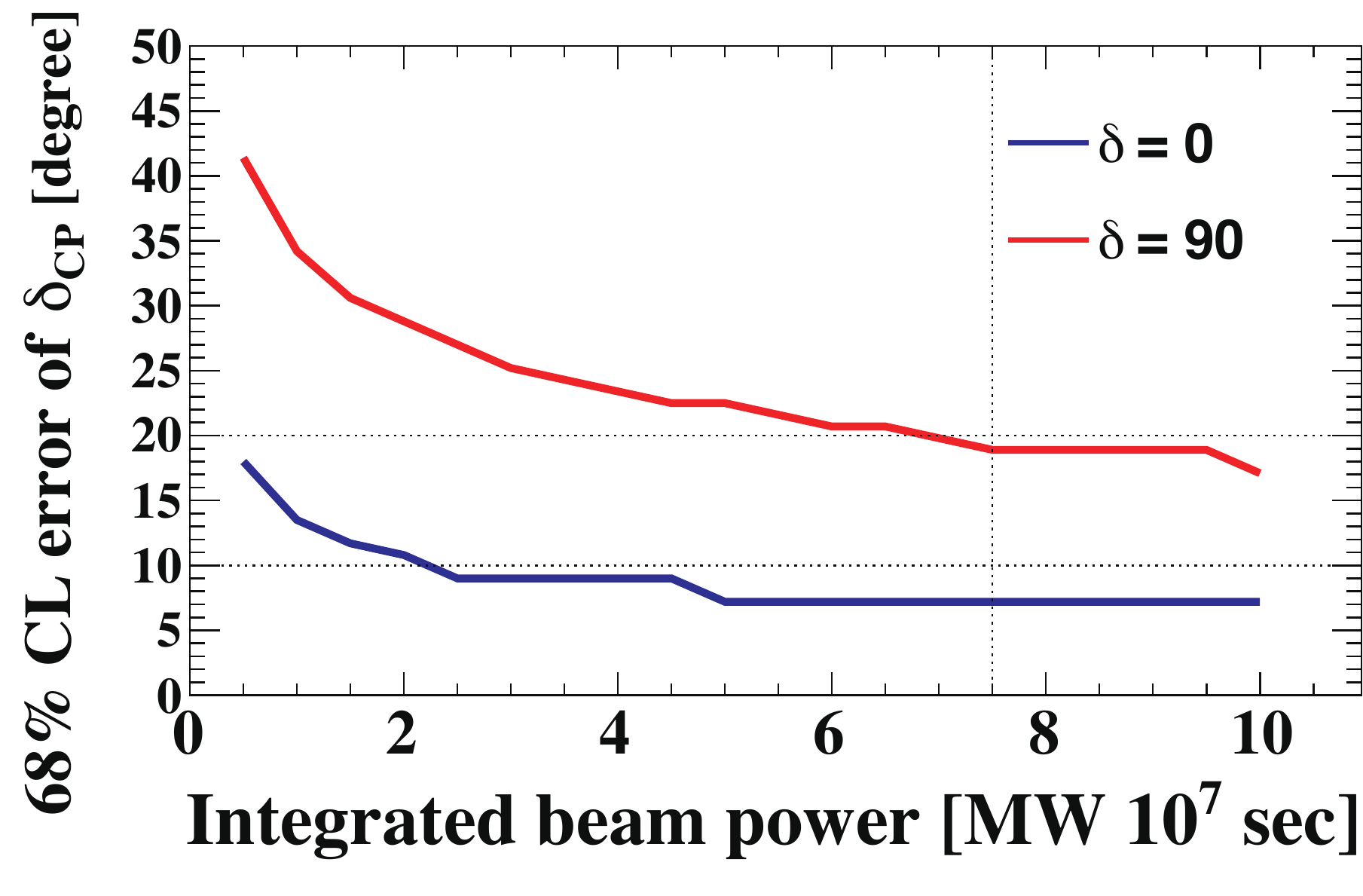}
\caption{Expected 68\% CL uncertainty of $\deltacp$ as a function of integrated beam power. 
\label{fig:delta-error-time}}
\end{figure}

Figure~\ref{fig:delta-error-time} shows the 68\% CL uncertainty of $\deltacp$ as a function of the integrated beam power.
With 7.5~MW$\times$10$^7$sec of exposure (1.56$\times$10$^{22}$ protons on target), 
the value of $\deltacp$ can be determined to better than 19$^\circ$ for all values of $\deltacp$.

\begin{figure}[tbp]
\centering
\includegraphics[width=0.9\textwidth]{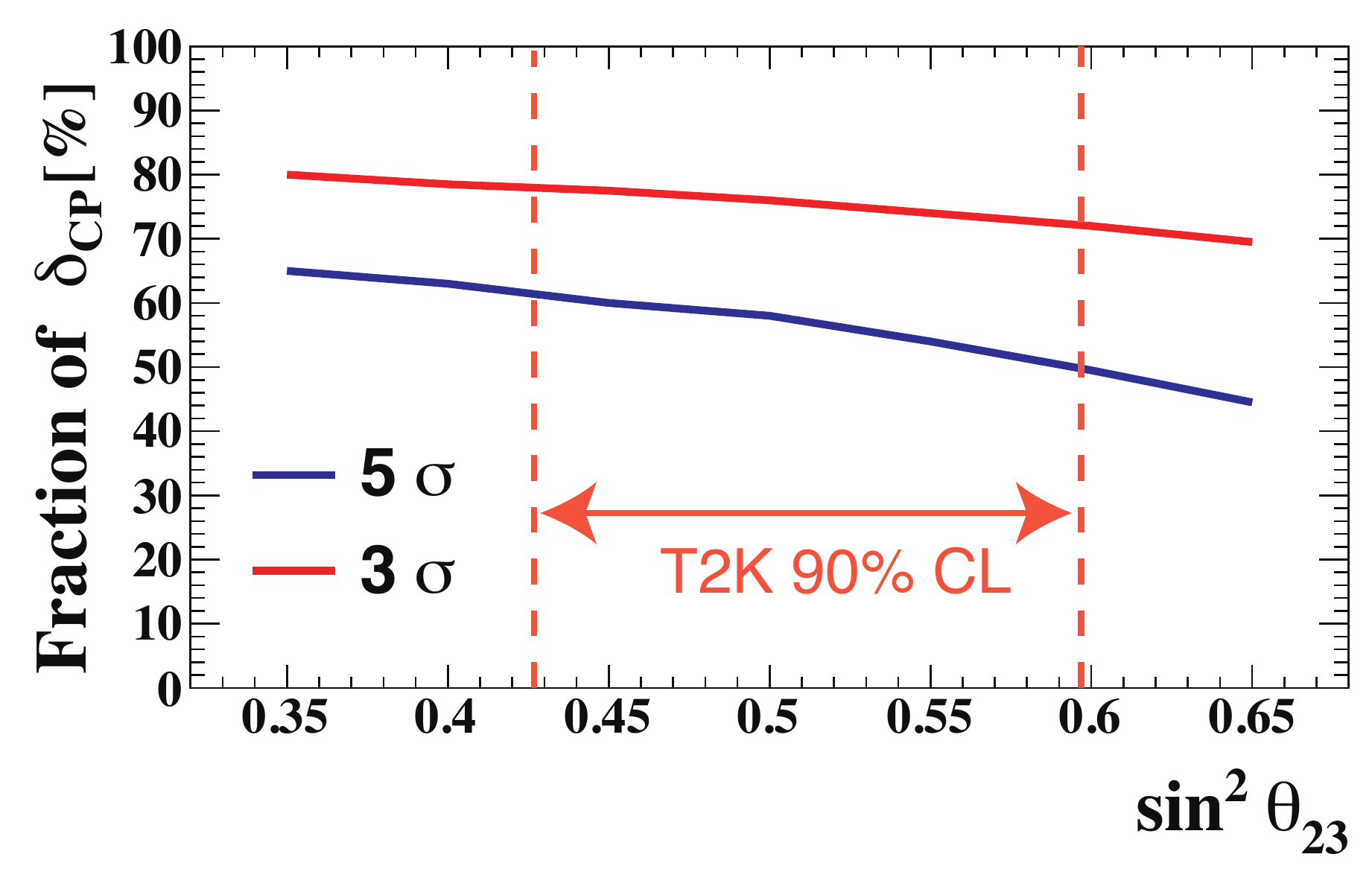}
\caption{Fraction of $\deltacp$ for which $\sin\deltacp= 0$ can be excluded with more than 3\,$\sigma$ (red) and 5\,$\sigma$ (blue) significance as a function of the true value of $\sin^2\theta_{23}$, for the normal hierarchy case.
Vertical dashed lines indicate 90\% confidence intervals of $\sin^2\theta_{23}$ from the T2K measurement in 2014~\cite{Abe:2014ugx}.
\label{fig:delta-theta23}}
\end{figure}

As the nominal value we use $\sin^2\theta_{23}=0.5$,
but the sensitivity to $CP$ violation depends on the value of $\theta_{23}$.
Figure~\ref{fig:delta-theta23} shows the fraction of $\deltacp$ for which $\sin\deltacp= 0$ is excluded with more than 3\,$\sigma$ and 5\,$\sigma$ of significance as a function of the true value of $\sin^2\theta_{23}$
with the current best knowledge of the possible $\sin^2\theta_{23}$ range by T2K collaboration~\cite{Abe:2014ugx}.

\subsection{Sensitivity to $\Delta m^2_{32}$ and $\sin^2\theta_{23}$}
\begin{figure}[tbp]
\centering
\includegraphics[width=0.95\textwidth]{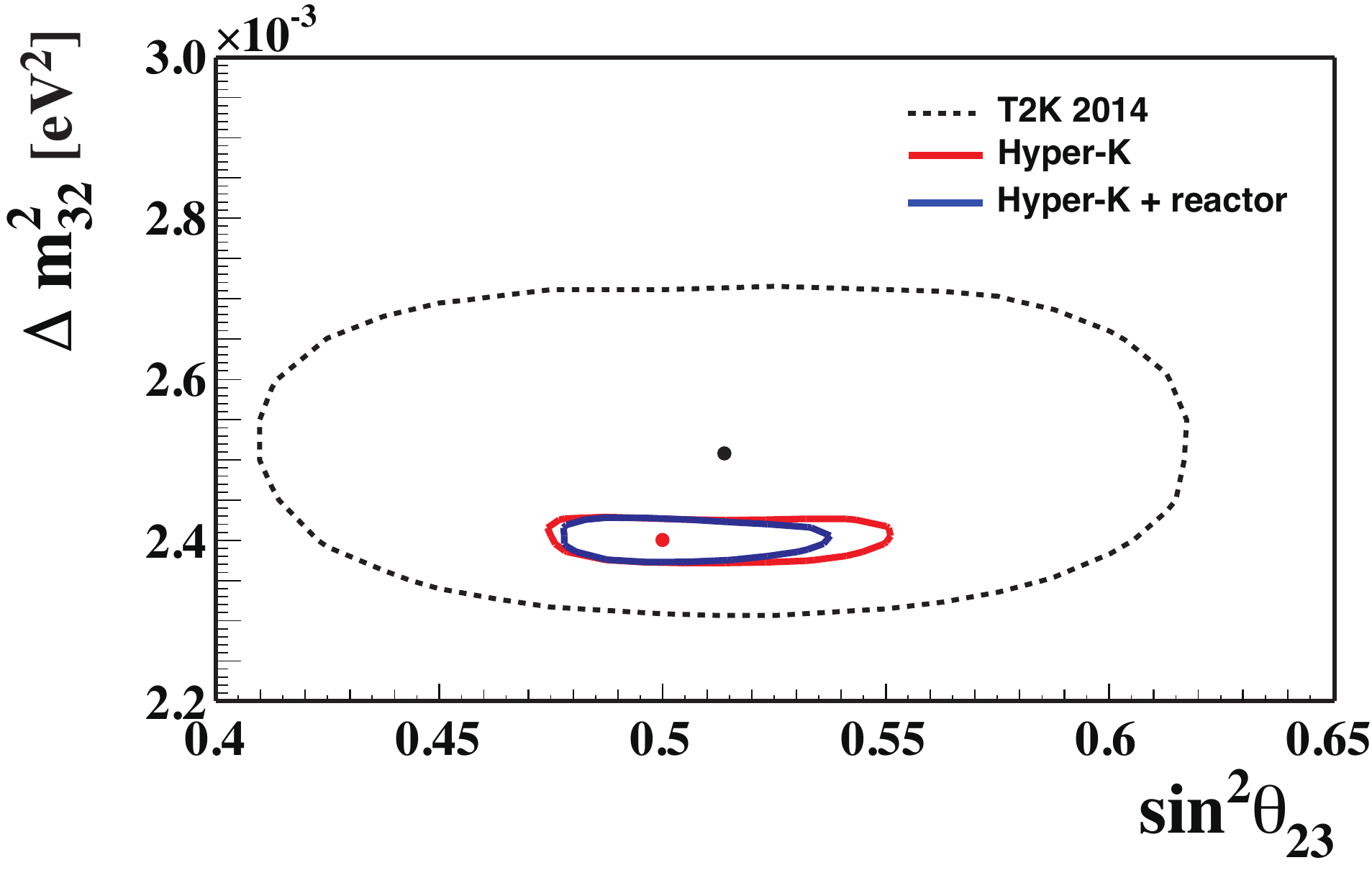}
\caption{ The 90\% CL allowed regions in the $\sin^2\theta_{23}$--$\Delta m^2_{32}$ plane.
The true values are $\sin^2\theta_{23}=0.5$ and $\Delta m^2_{32} = 2.4 \times 10^{-3}$~eV$^2$ (red point).
Effect of systematic uncertainties is included. The red (blue) line corresponds to the result with Hyper-K alone (with a reactor constraint on $\sin^22\theta_{13}$).
The dotted line is the 90\% CL contour from T2K experiment~\cite{Abe:2014ugx} with the best fit values indicated by a black point.
\label{fig:theta23-0.50}}
\end{figure}

\begin{figure}[tbp]
\centering
\includegraphics[width=0.95\textwidth]{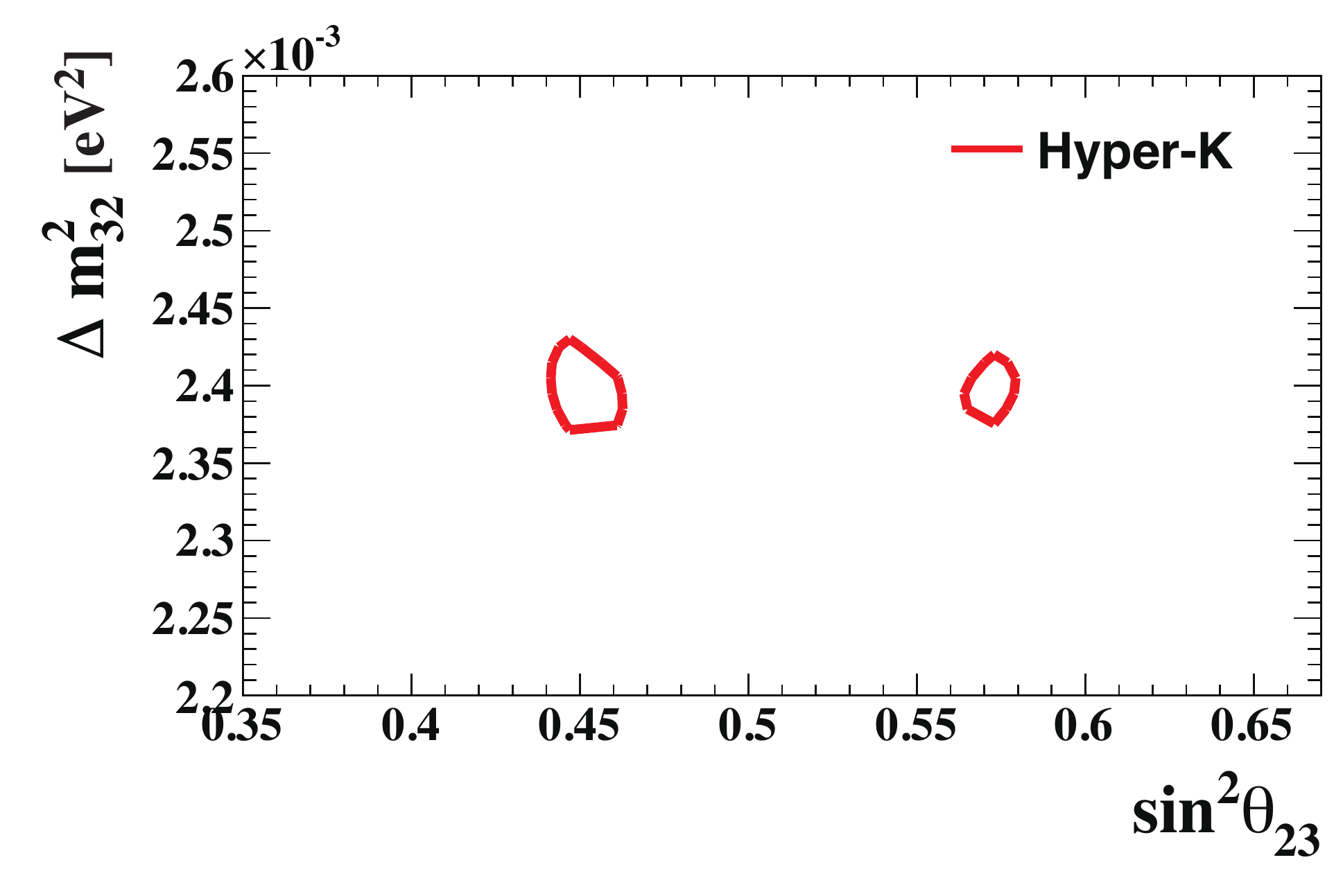}
\includegraphics[width=0.95\textwidth]{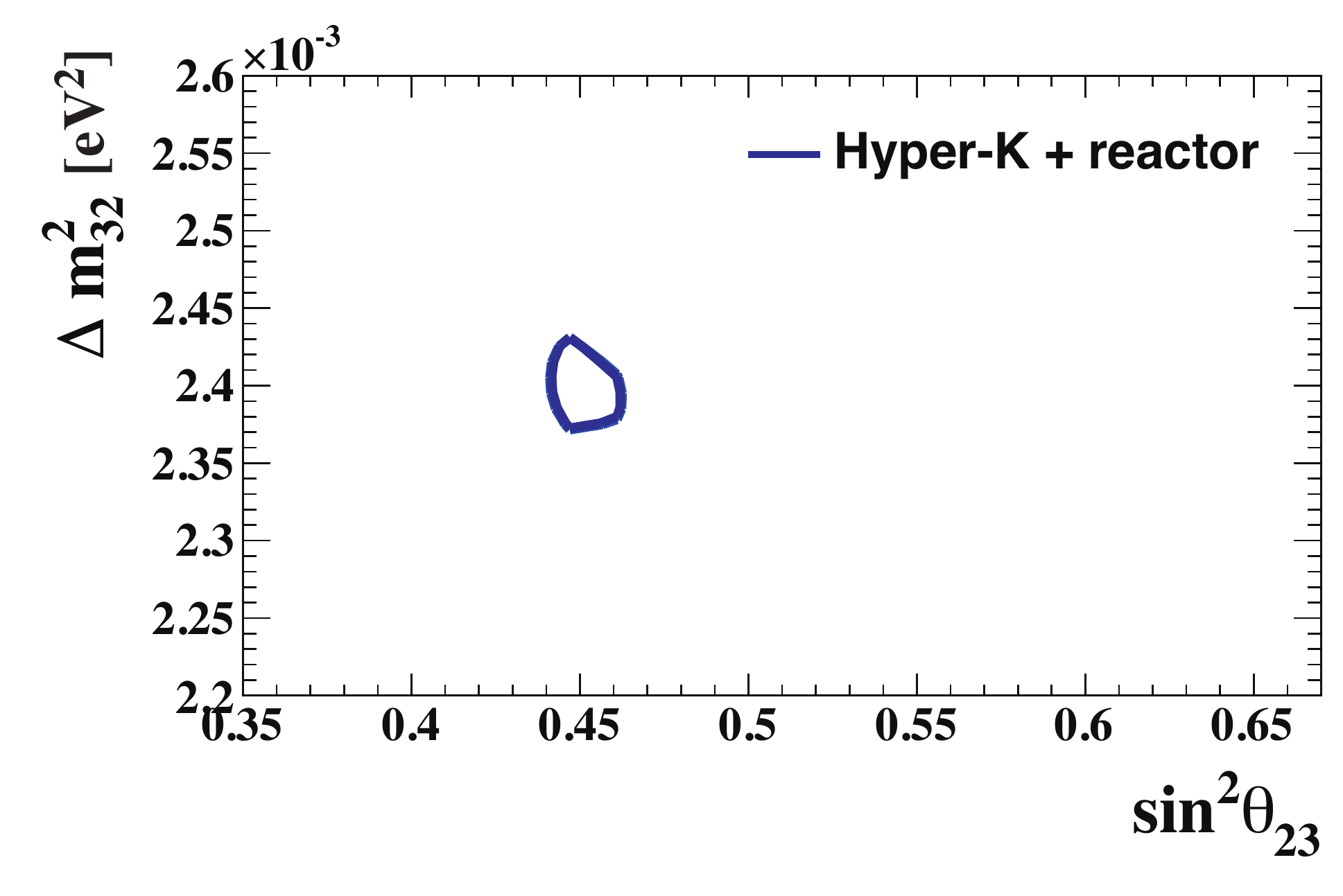}
\caption{ 90\% CL allowed regions in the $\sin^2\theta_{23}$--$\Delta m^2_{32}$ plane.
The true values are $\sin^2\theta_{23}=0.45$ and $\Delta m^2_{32} = 2.4 \times 10^{-3}$~eV$^2$.
Effect of systematic uncertainties is included.
Top: Hyper-K only. Bottom: With a reactor constraint.
\label{fig:theta23-0.45}}
\end{figure}

\begin{table}[tdp]
\caption{Expected 1$\sigma$ uncertainty of $\Delta m^2_{32}$ and $\sin^2\theta_{23}$  for true $\sin^2\theta_{23}=0.45, 0.50, 0.55$. 
Reactor constraint on $\sin^22\theta_{13}=0.1\pm 0.005$ is imposed.}
\begin{center}
\begin{tabular}{ccccccc} \hline \hline
True $\sin^2\theta_{23}$	& \multicolumn{2}{c}{$0.45$}  		& \multicolumn{2}{c}{$0.50$} 		& \multicolumn{2}{c}{$0.55$}\\ 
Parameter  				& $\Delta m^2_{32}$ (eV$^2$) 	& $\sin^2\theta_{23}$  & $\Delta m^2_{32}$	(eV$^2$)	& $\sin^2\theta_{23}$ 	& $\Delta m^2_{32}$ (eV$^2$)	& $\sin^2\theta_{23}$\\ \hline
 NH	& $1.4\times10^{-5}$		& 0.006 			& $1.4\times10^{-5}$		& 0.015				& $1.5\times10^{-5}$			& 0.009\\
 IH & $1.5\times10^{-5}$		& 0.006 			& $1.4\times10^{-5}$		& 0.015				& $1.5\times10^{-5}$			& 0.009\\
\hline \hline
\end{tabular}
\end{center}
\label{tab:23sensitivity}
\end{table}%

The result shown above is obtained with $\sin^2\theta_{23}$ and $\Delta m^2_{32}$ as free parameters as well as $\sin^22\theta_{13}$ and $\deltacp$, with a nominal parameters shown in Table~\ref{Tab:oscparam}.
The use of the $\numu$ sample in addition to $\nue$ enables us to also precisely measure $\sin^2\theta_{23}$ and $\Delta m^2_{32}$.
Figure~\ref{fig:theta23-0.50} shows the 90\% CL allowed regions for the true value of $\sin^2\theta_{23}=0.5$ together with the 90\% CL contour by T2K $\nu_\mu$ disappearance measurement~\cite{Abe:2014ugx}.
Hyper-K will be able to provide a precise measurement of $\sin^2\theta_{23}$ and $\Delta m^2_{32}$.
Figure~\ref{fig:theta23-0.45} shows the 90\% CL allowed regions on the $\sin^2\theta_{23}$-$\Delta m^2_{32}$ plane, for the true values of $\sin^2\theta_{23}=0.45$ and $\Delta m^2_{32} = 2.4 \times 10^{-3}$~eV$^2$. 
For the determination of $\theta_{23}$, a $\nu_\mu$ disappearance measurement provides precise measurement of $\sin^22\theta_{23}$.
However, when $\theta_{23}\ne \frac{\pi}{4}$, there are two possible solutions ($\theta_{23}$ and $\frac{\pi}{2}-\theta_{23}$) which give the same $\sin^22\theta_{23}$.
This is known as the octant degeneracy.
As seen from Eq.~\ref{Eq:cpv-oscprob}, $\nu_e$ appearance measurement can determine $\sin^2\theta_{23}\sin^22\theta_{13}$.
In addition, the reactor experiments provide almost pure measurement of $\sin^22\theta_{13}$. 
Thus, the combination of those complimentary measurements is known to be able to resolve this degeneracy if $\theta_{23}$ is sufficiently away from $\frac{\pi}{4}$~\cite{Fogli:1996pv,Minakata:2002jv,Hiraide:2006vh}.
As shown in Fig.~\ref{fig:theta23-0.45}, with a constraint on $\sin^22\theta_{13}$ from the reactor experiments, Hyper-K measurements can resolve the octant degeneracy and precisely determine $\sin^2\theta_{23}$.

The expected precision of $\Delta m^2_{32}$ and $\sin^2\theta_{23}$ for true $\sin^2\theta_{23}=0.45, 0.50, 0.55$ with reactor constraint on $\sin^22\theta_{13}$ is summarized in Table~\ref{tab:23sensitivity}.

\subsection{Combination with atmospheric neutrino data \label{sec:lbl-atm}}
\begin{figure}[tbp]
  \begin{center}
  \includegraphics[width=0.6\textwidth]{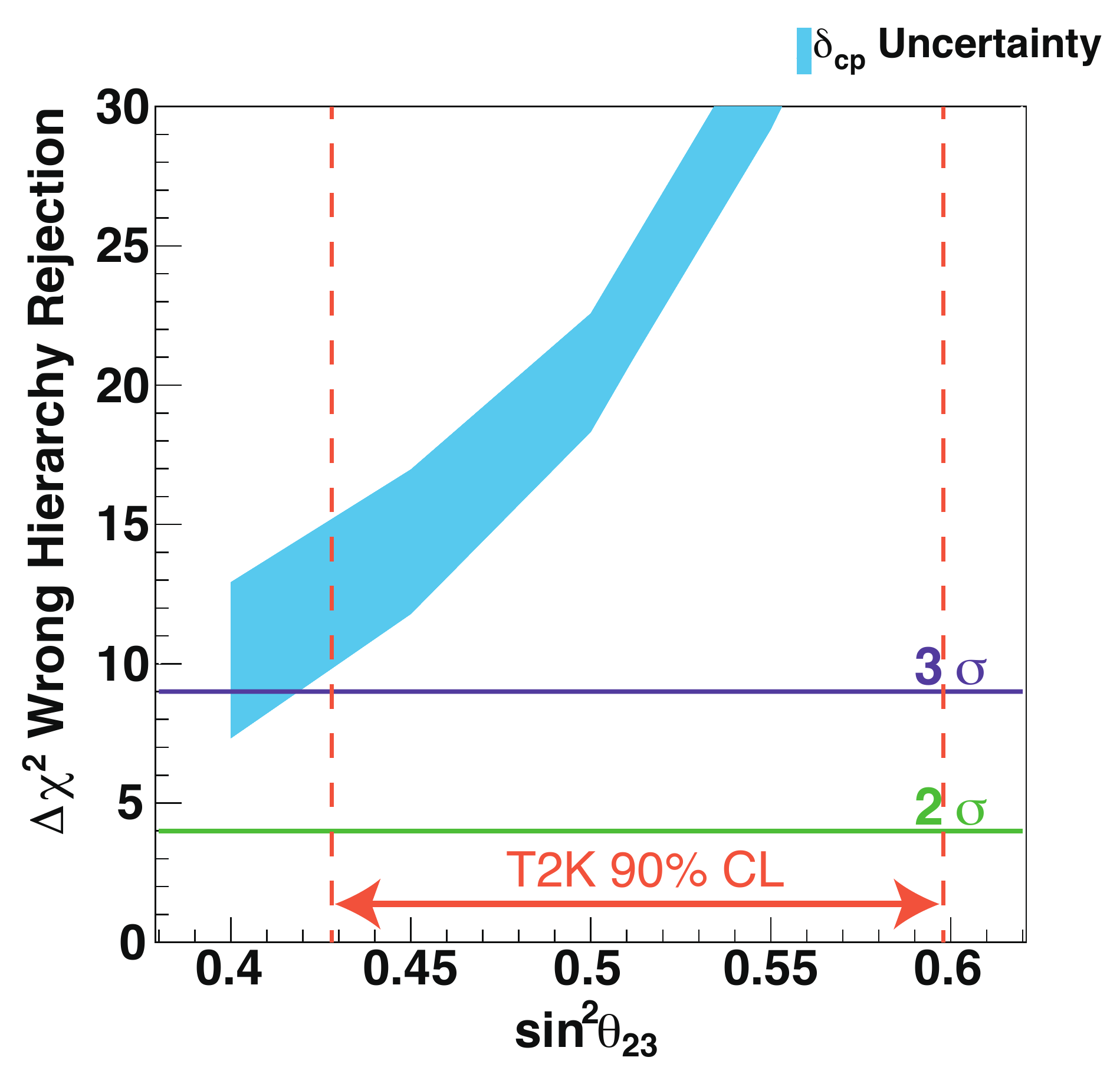}
  \includegraphics[width=0.6\textwidth]{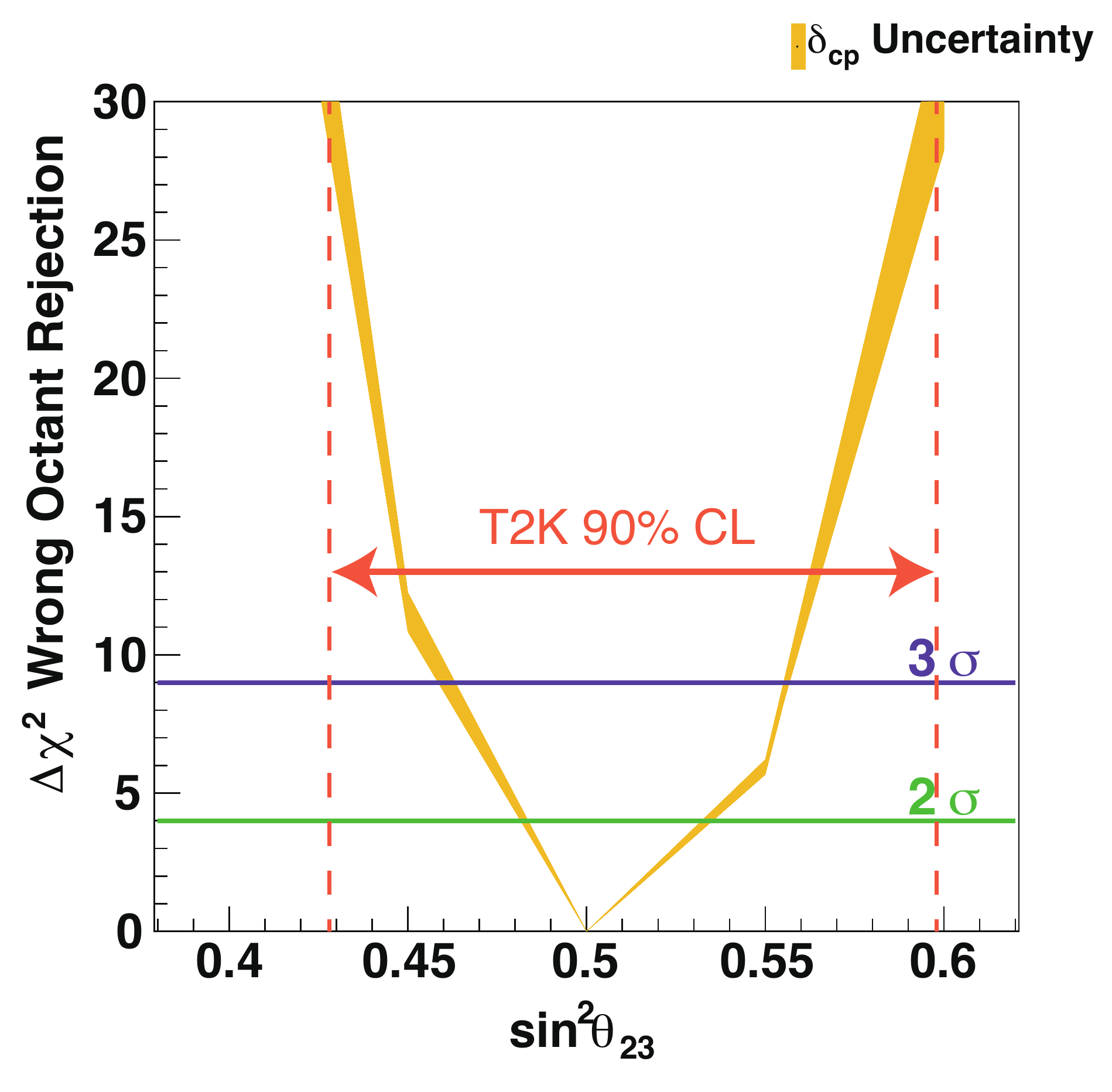}
     \caption{Atmospheric neutrino sensitivities for a ten year exposure of Hyper-K 
              assuming the mass hierarchy is normal. 
              Top: the $\Delta \chi^{2}$ discrimination of the wrong 
              hierarchy hypothesis as a function of the assumed true value of 
              $\mbox{sin}^{2} \theta_{23}$.   
              Bottom: the discrimination between the wrong octant for each value of 
               $\mbox{sin}^{2} \theta_{23}$. 
               The uncertainty from $\deltacp$ is 
              represented by the thickness of the band.
              Vertical dashed lines indicate 90\% confidence intervals of $\sin^2\theta_{23}$ from the T2K measurement in 2014~\cite{Abe:2014ugx}.
              }
     \label{fig:atmsens}
  \end{center}
\end{figure}

Atmospheric neutrinos can provide an independent and complementary information to the accelerator beam program on the study of neutrino oscillation.
For example, through the matter effect inside the Earth, a large statistics sample of atmospheric neutrinos by Hyper-K will have a good sensitivity to the mass hierarchy and $\theta_{23}$ octant.

Assuming a 10 year exposure, Hyper-K's sensitivity to the mass hierarchy and
the octant of $\theta_{23}$ by atmospheric neutrino data are shown in Fig.~\ref{fig:atmsens}.
Depending on the true value of $\theta_{23}$ the sensitivity changes considerably,
but for all currently allowed values of this parameter the mass hierarchy 
sensitivity exceeds $3\,\sigma$ independent of the assumed hierarchy. 
If $\theta_{23}$ is non-maximal, the atmospheric neutrino data can be used to discriminate 
the octant at $3\,\sigma$ if $ \sin^2 \theta_{23}<0.46$ or   $ \sin^2 \theta_{23}>0.56 $. 

In the previous sections, the mass hierarchy is assumed to be known prior to the Hyper-K measurements.
This is a reasonable assumption considering the increased opportunities, thanks to a large value of $\theta_{13}$, of ongoing and proposed projects for mass hierarchy determination.
However, even if the mass hierarchy is unknown before the start of experiment, Hyper-K itself will be able to determine it with the atmospheric neutrino measurements.

\begin{figure}[tbp]
\centering
\includegraphics[width=0.6\textwidth]{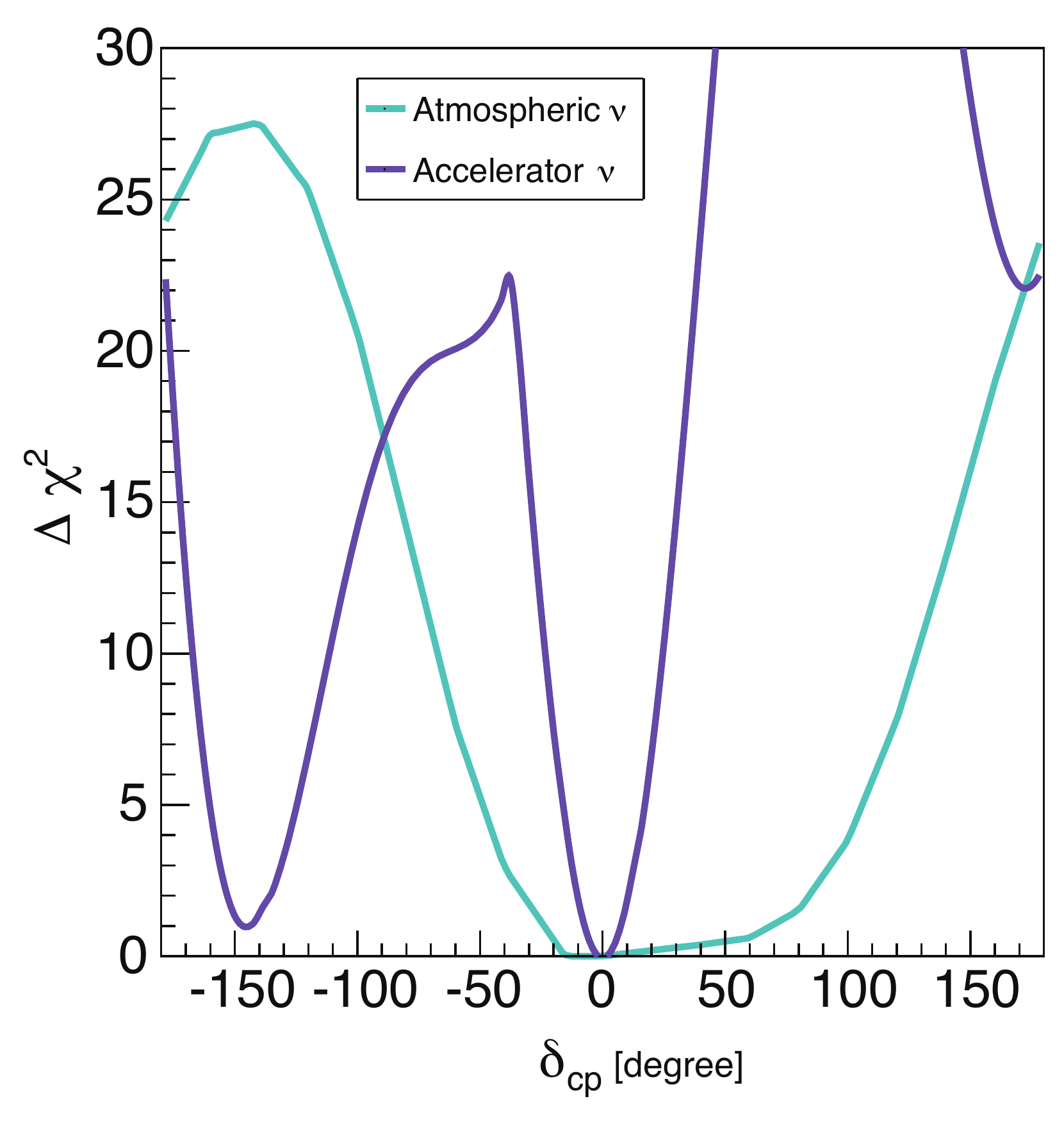}
\includegraphics[width=0.6\textwidth]{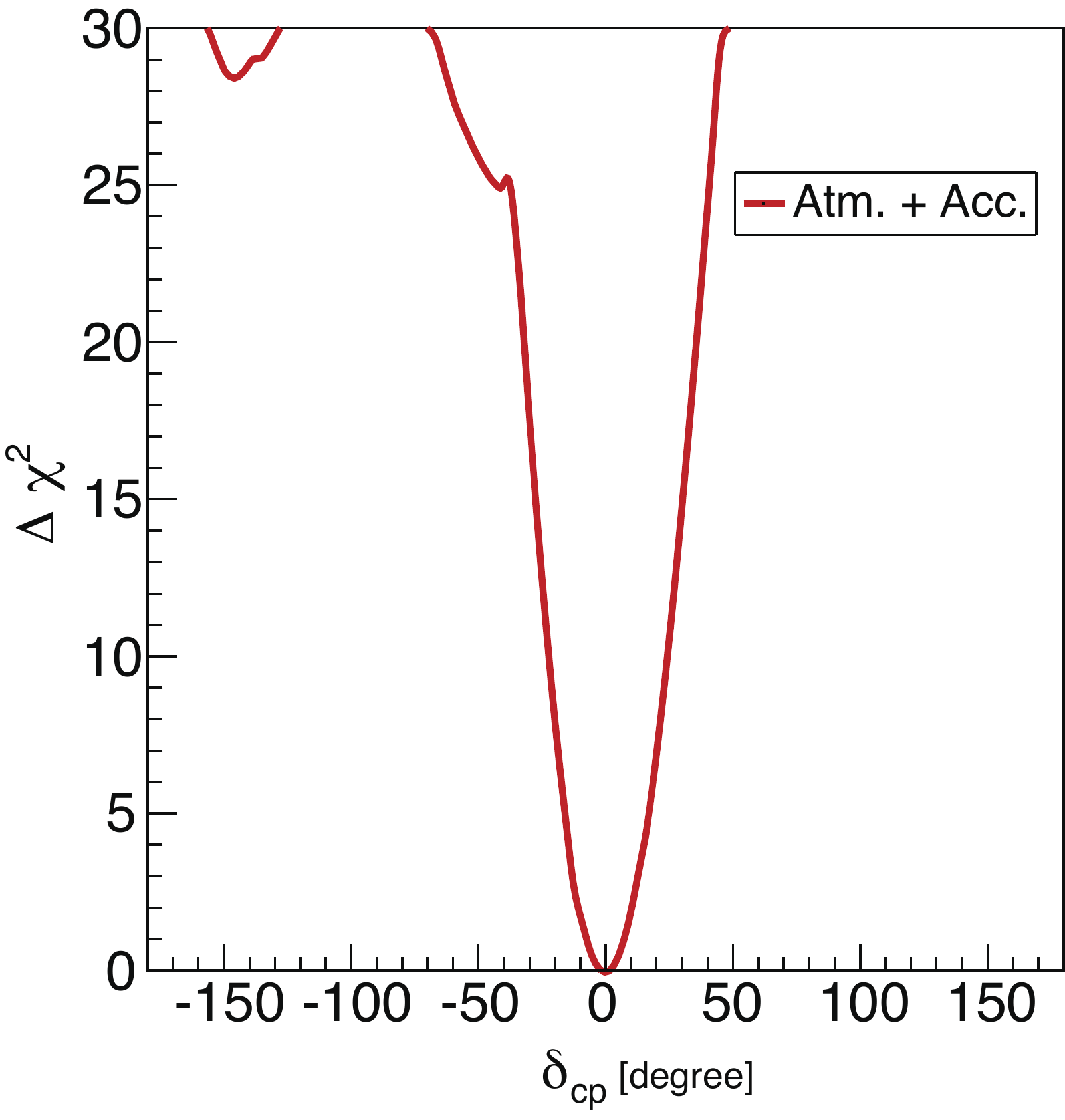}
\caption{Combination of the accelerator and atmospheric data. 
Top: Expected $\Delta \chi^2$ values for accelerator and atmospheric neutrino measurements assuming that the mass hierarchy is unknown. 
The true mass hierarchy is normal hierarchy and the true value of $\deltacp=0$.
Bottom: By combining the two measurements, the sensitivity can be enhanced. In this example study, the $\Delta \chi^2$ is simply added.
\label{fig:lbl-atm}}
\end{figure}

Because Hyper-K will observe both accelerator and atmospheric neutrinos with the same detector, the physics capability of the project can be enhanced by combining two complementary measurements.
As a demonstration of such a capability, a study has been done by simply adding
$\Delta \chi^2$ from two measurements, although in a real experiment a more sophisticated analysis is expected.
Assuming the true mass hierarchy of normal hierarchy and the true value of $\deltacp=0$,
the values of expected $\Delta \chi^2$ as a function of $\deltacp$ for each of
the accelerator and atmospheric neutrino measurements,
\textit{without} assumption of the prior mass hierarchy knowledge,
are shown in the top plot of Fig.~\ref{fig:lbl-atm}.
For the accelerator neutrino measurement, there is a second minimum near $\deltacp=150^\circ$ because of a degeneracy with mass hierarchy assumptions.
On the other hand, the atmospheric neutrino measurement can discriminate the
mass hierarchy, but the sensitivity to the $CP$ violating phase $\deltacp$ is worse than the accelerator measurement.
By adding the information from both measurements, as shown in the bottom plot of Fig.~\ref{fig:lbl-atm}, the fake solution can be eliminated and a precise measurement of $\deltacp$ will be possible.

\section{Conclusion}
The sensitivity to leptonic $CP$ asymmetry of a long baseline experiment using 
a neutrino beam directed from J-PARC to the Hyper-Kamiokande detector has been studied
based on a full simulation of beamline and detector.
With an integrated beam power of 7.5~MW$\times$10$^7$~sec,
the value of $\deltacp$ can be determined to better than 19$^\circ$ for all values of $\deltacp$
and
$CP$ violation in the lepton sector can be observed with more than 3~$\sigma$ (5~$\sigma$) significance for 76\% (58\%) of the possible values of $\deltacp$.

Using both $\nu_e$ appearance and $\nu_\mu$ disappearance data, a precise measurement of $\sin^2\theta_{23}$ will be possible.
The expected 1$\sigma$ uncertainty is 0.015(0.006) for $\sin^2\theta_{23}=0.5(0.45)$. 

\section*{Acknowledgments}
This work was supported by MEXT Grant-in-Aid for Scientific Research on Innovative Areas Number 25105004,
titled ``Unification and Development of the Neutrino Science Frontier.''
In addition, participation of individual researchers has been further supported
by funds from JSPS, Japan; the European Union ERC-207282, H2020 RISE-GA644294-JENNIFER and H2020 RISE-GA641540-SKPLUS; RSF, RFBR and MES, Russia; JSPS and RFBR under the Japan-Russia Research Cooperative Program.

\bibliographystyle{ptephy}


\end{document}